\documentclass[acmlarge]{acmart}

\usepackage{algorithm}
\usepackage{algpseudocode}
\usepackage{graphicx}
\usepackage{tikz}
\usepackage{amsmath}
\usepackage{multirow}
\usepackage{multicol}
\usepackage{enumitem}
\setitemize{topsep=2pt,parsep=0pt,partopsep=0pt,leftmargin=15pt}
\usepackage{filecontents}
\usepackage{xurl}
\usepackage{soul}
\usepackage{latexsym}
\usepackage{gensymb}
\usepackage{balance}
\usepackage{pifont}
\usepackage{anyfontsize}
\usepackage{xspace}
\usepackage{etoolbox}
\usepackage{subcaption}
\AddToHook{package/amssymb/before}{}
\usepackage{amssymb}

\newcommand{\SysName}{\mbox{SHARE}\xspace}
\definecolor{authorblue}{RGB}{0,0,0}
\newcommand{\editting}{\textcolor{authorblue}}

\algnewcommand{\Phase}[1]{\Statex \textit{$\triangleright$~#1}}

\setcopyright{cc}
\setcctype{by-nc-nd}
\acmJournal{IMWUT}
\acmYear{2026} \acmVolume{10} \acmNumber{3} \acmArticle{97}
\acmMonth{9} \acmDOI{10.1145/3832004}

\title{SHARE: Towards Head-Mounted AR with User-Centric SLAM in Shared Human–Robot Workspaces}

\author{Tianyuan Du}
\authornote{Corresponding author.}
\email{alex.du@duke.edu}
\affiliation{%
  \institution{Duke University}
  \department{Department of Electrical and Computer Engineering}
  \city{Durham}
  \state{North Carolina}
  \country{United States}
}

\author{Tianyi Hu}
\email{tianyi.hu@duke.edu}
\affiliation{%
  \institution{Duke University}
  \department{Department of Electrical and Computer Engineering}
  \city{Durham}
  \state{North Carolina}
  \country{United States}
}

\author{Hanting Ye}
\email{hanting.ye@duke.edu}
\affiliation{%
  \institution{Duke University}
  \department{Department of Electrical and Computer Engineering}
  \city{Durham}
  \state{North Carolina}
  \country{United States}
}

\author{Maria Gorlatova}
\email{maria.gorlatova@duke.edu}
\affiliation{%
  \institution{Duke University}
  \department{Department of Electrical and Computer Engineering}
  \city{Durham}
  \state{North Carolina}
  \country{United States}
}

\authorsaddresses{Authors' Contact Information: Tianyuan Du (corresponding author),
alex.du@duke.edu, Duke University, Department of Electrical and Computer Engineering,
Durham, North Carolina, United States;
Tianyi Hu, tianyi.hu@duke.edu, Duke University, Department of Electrical and Computer
Engineering, Durham, North Carolina, United States;
Hanting Ye, hanting.ye@duke.edu, Duke University, Department of Electrical and Computer
Engineering, Durham, North Carolina, United States;
Maria Gorlatova, maria.gorlatova@duke.edu, Duke University, Department of Electrical and
Computer Engineering, Durham, North Carolina, United States.}

\begin{abstract}

Human-Robot Collaboration (HRC) in shared physical spaces using Augmented Reality (AR) interfaces is powered by Simultaneous Localization and Mapping (SLAM). Existing multi-agent \editting{SLAM} systems rely on \editting{an} edge server to combine visual findings of multiple resource-constrained agents, perform computation, and schedule updates to their local maps. However, the edge treats all agents uniformly and ignores the fundamentally different latency requirements of \editting{heterogeneous} HRC agents: robots and head-mounted AR users. This uniform resource allocation often results in high lag for \editting{user} manipulation, as it does not meet the stringent latency requirements of AR. In this work, we design, implement, and evaluate \SysName, a user-centric SLAM system that strategically prioritizes AR user experience while maintaining accurate tracking performance for robots. \SysName builds a first-of-its-kind experience model for HRC agents and adaptively adjusts transmission priorities to match it. To reduce \editting{end-to-end} latency, \SysName leverages the redundancy of visual features acquired by agents in shared human-robot workspaces to reduce computation time induced by edge-based processing. Real-world deployment with commercial AR headsets and \editting{a} ground robot achieves 13.22 ms average latency for AR users (43.3\% reduction from baseline) while maintaining sub-2-centimeter tracking accuracy. User studies further reveal statistically significant improvements in user perception.
\end{abstract}



\begin{document}

\maketitle

\section{Introduction} 
\label{sec_introduction}

Augmented Reality (AR)-enabled Human-Robot Collaboration (HRC) applications facilitate seamless interaction between robots and humans in shared physical environments, ranging from collaborative assembly to domestic cleaning assistance. Achieving coordinated spatial awareness in AR-HRC scenarios requires multi-agent Simultaneous Localization and Mapping (SLAM), which leverages diverse sensory inputs (e.g., cameras and IMUs) from multiple mobile agents to jointly construct a map of an unknown environment while simultaneously localizing each agent within it.

\begin{figure}[t]    
    \vspace{-3mm}
    \includegraphics[width=.6\columnwidth]{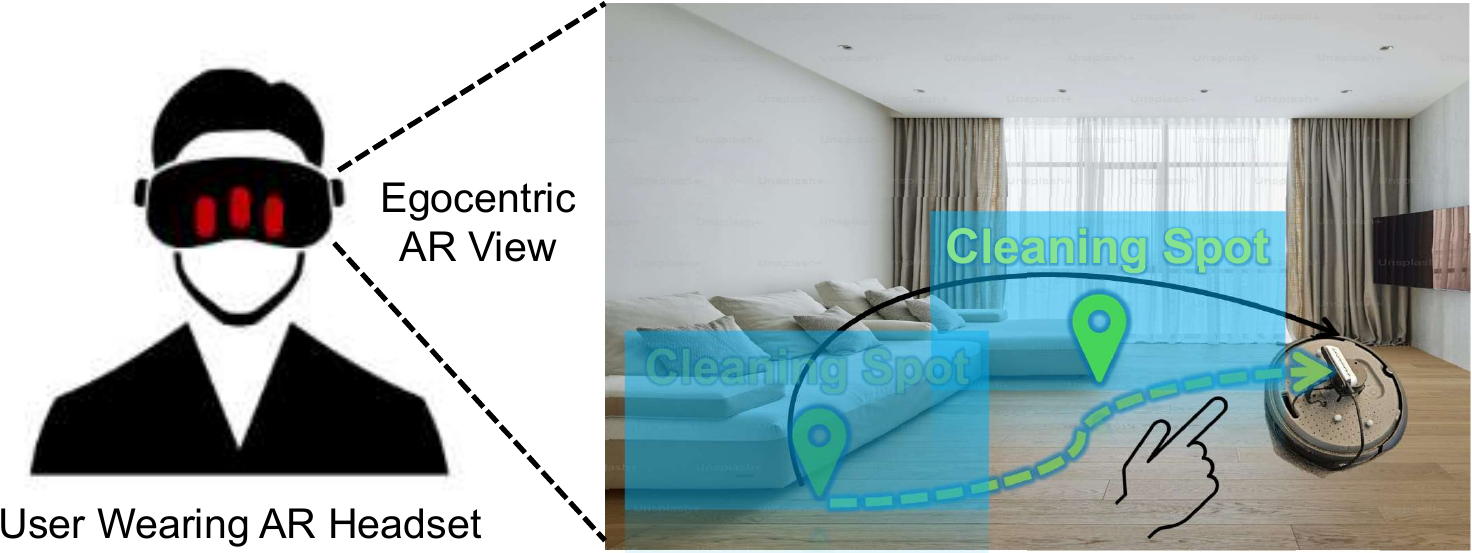}
    \vspace{-3mm}
    \caption{Latency challenge in AR-HRC. User designates a spot for the robot to clean by dragging the target marker, but the marker lags during movement due to latency.
    }
    \vspace{-3mm}
    \label{fig:teaser} 
\end{figure}

A critical bottleneck in AR-HRC that severely impairs AR user experience is \emph{latency}. AR applications impose stringent motion-to-photon latency requirements, i.e., the delay from a user's gesture control to the final visual display. Typically, this delay is required to stay below a low threshold, often set at $20~ms$, to prevent user discomfort and maintain visual coherence between virtual and real-world environments, as demonstrated in Figure~\ref{fig:teaser}. However, existing multi-agent SLAM frameworks often exceed \editting{this constraint}, with recent works reporting delays of $100-500~ms$ \cite{Zhang2024,Xu2022}. This is largely because current multi-agent SLAM systems commonly offload computationally intensive tasks to edge servers due to limited computational resources on mobile agents \cite{Ali2022,9585827,10160938,Xu2022,Chen2023, Wright2020}, and the absence of edge support for synchronizing and exchanging map data among multiple AR headsets and robots can also lead to degraded tracking accuracy. Therefore, we ask the following research question: \emph{Can we reduce the latency in current AR-powered HRC multi-agent SLAM systems while maintaining good tracking accuracy?}

To address this, we examine the causes of this performance bottleneck. Existing multi-agent SLAM systems predominantly address homogeneous agent configurations, either multiple robots~\cite{9585827,10160938,Xu2022} or multiple user devices~\cite{Dhakal2022,rob_col_vis}, whereas AR-HRC applications introduce fundamental challenges from \emph{heterogeneous coexistence}.
Thus, the key to resolving this latency concern in AR lies in the heterogeneous latency requirements between direct AR headset experiences and indirect robot control within the HRC context. 
While the AR experience is highly sensitive to latency due to the real-time nature of rendering and user interaction, users exhibit a tolerance for robot control latency that is an order of magnitude higher, up to $750~\mathrm{ms}$ to $1~\mathrm{s}$ \cite{shiwacomm,yangdelay,humanrobodelay}. 
This tolerance arises because users subconsciously map a robot's response time in decision-making and mechanical delays to that of a human, expecting delays similar to natural conversational pauses \cite{shiwacomm,robotemotim,40years,humanrobodelay}.
However, conventional multi-agent SLAM systems often treat all agents as homogeneous, failing to differentiate in resource allocation and thus causing uniformly high latency across all agents. Therefore, the first challenge we meet is: \emph{how can we realize a user-centric strategy to effectively prioritize the latency of AR applications for human users, while exploiting the looser latency requirement on the robot?}

To address this challenge, we propose differentiated quality-of-experience (QoE) policies to model varying latency tolerance across heterogeneous agents, e.g., AR headsets and robots, which differ in how latency affects user-perceived experience. We observe from prior user experience research across multiple domains, including telecommunications \cite{reichl2013logarithmic}, web browsing~\cite{zhang2008web}, and human-computer interaction \cite{mackenzie1993lag}, that users exhibit diminishing marginal sensitivity to performance improvements and that different performance factors contribute differently to perceived experience. Meanwhile, system performance metrics (e.g., Quality-of-Service (QoS) indicators such as accuracy and latency) often exhibit an exponential relationship with subjective quality perception \cite{Fiedler2010,iqx2,qoe-exp}. Inspired by this, we model an agent-specific QoE function that captures the differential priorities of AR services (latency-sensitive) versus robotic services (accuracy-critical).

However, calculating accuracy requires a ground-truth trajectory to be provided to the SLAM system, which is unrealistic, \editting{since} SLAM's objective is to estimate that trajectory. Inspired by recent advancements in embedded AI models in mobile devices, we design and deploy a lightweight deep neural network architecture on mobile agents to provide real-time accuracy estimation. Combining \editting{estimated accuracy} with measured latency, the constructed QoE model can then be used to prioritize edge-agent communication for AR headsets, while allowing relaxed latency constraints for robots.

Although QoE modeling captures metrics of latency and accuracy, these metrics are environment-dependent in the SLAM context. Even homogeneous agents, such as two users wearing \editting{identical} AR headsets, can experience different QoE \editting{due to differences in their} visual environments. Specifically, the number of features in visual frames captured by each agent significantly impacts both latency and accuracy, and subsequently the overall QoE. This indicates that latency and accuracy in the QoE model are tightly coupled through an implicit variable: \emph{visual environment dynamics}. When a mobile agent's visual sensors encounter a featureless environment, such as white walls, fewer distinguishable features are available for matching with previous frames to detect spatial changes, resulting in \editting{the} need for increased visual frames. Consequently, the second challenge we face is: \emph{How can we reasonably balance latency and accuracy to optimize the QoE of multi-agent SLAM systems in response to dynamics in visual environments?}

To tackle this challenge, a straightforward method is to set a fixed threshold or implement rule-based mechanisms for prioritization and resource allocation, i.e., once the QoE falls below a certain threshold, increase the frame rate or other computational resources to improve QoE. However, the visual environment could exhibit rapid, complex changes that render rule-based mechanisms insufficient. To address this, we propose a Proportional-Integral-Derivative (PID) \editting{scheduling} pipeline to continuously monitor environmental conditions and adjust transmission priorities across heterogeneous and homogeneous agents based on feedback from the edge server. This approach enables different agents to maintain a stable and smooth QoE, regardless of whether they are looking toward a feature-rich or a featureless environment.

After applying the methods described above to dynamically allocate resources across heterogeneous and homogeneous agents, we still find the latency of some AR applications exceeding $20~\mathrm{ms}$. We examine the causes of this performance degradation. The cause is that AR-powered HRC systems typically operate in small-scale indoor environments. In such environments, vanilla SLAM system often performs a large amount of redundant computation that does not contribute to tracking accuracy improvement. This leads to additional computational latency, especially when multiple mobile agents have highly similar views, resulting in map construction redundancy within these confined environments. Therefore, the third challenge we face is \emph{how to address inefficiency from map redundancy in AR-powered HRC SLAM scenarios?}

To solve this challenge, we propose a visual overlap calculation methodology that identifies and trims redundant map merging operations by predicting merge utility based solely on agent poses (location and orientation). We observe that in confined indoor environments, multiple agents often have their \emph{Line of Sight (LoS) intersection} within a short period of time, i.e., they all look toward the same area in the room. Visual redundancy increases during this time as the lines of sight gradually converge and decreases as the agents' viewpoints diverge. Instead of integrating all captured visual features throughout the entire sight-line intersection process, as in vanilla SLAM systems, we focus solely on the moment of LoS intersection (i.e., the point of highest visual overlap) to estimate agent locations, thereby reducing computational resource consumption. A key advantage of this approach is that only pose data (around 200 B) sent from each agent to the server are sufficient to calculate the visual overlap volume, eliminating the need to transmit full feature points (around 200 KB), and thus reducing both computational and communication overhead.

With the above design, we propose \SysName, a system implemented in real-world deployments with Meta Quest 3 AR headsets and a TurtleBot 4 ground robot, operating simultaneously within indoor environments. We collect data from five distinct trajectories totaling 483.93 meters with a motion capture system, and construct two emulation datasets based on the trajectories.
\SysName achieves an average tracking latency of 13.22 ms for AR users in a field study, representing a 43.3\% improvement over the baseline COVINS-G system and maintaining latency below the latency threshold required for AR experiences. It also exhibits 43.1\% and 39.4\% latency improvement on emulation datasets. In addition, \SysName maintains sub-2-cm absolute trajectory error and sub-1-cm translational relative pose error for all agents, which is comparable to state-of-the-art SLAM systems. \editting{To validate real-world usability, we conduct a user study with 20 participants performing robot control tasks through AR interfaces.} The results show statistically significant improvements in both robot responsiveness and movement predictability compared to baselines, with \editting{65\%} of the participants rating our system as their most preferred solution. \editting{The system is open-sourced at https://github.com/tyduduke/SHARE.}

\begin{figure}[t]
    \includegraphics[width=.8\columnwidth]{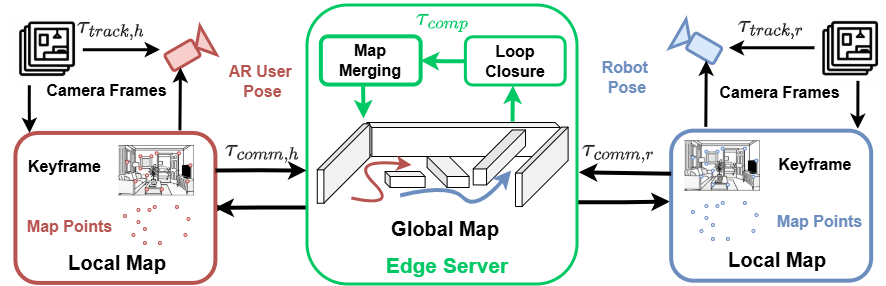}
    \caption{Structure of edge-assisted SLAM for heterogeneous agents. Agents build individual local maps and \editting{collaboratively construct} a global map. 
    }
    \label{fig:edge-slam-spec}
\end{figure}
\vspace{-1mm}
\section{Primer on Multi-agent SLAM}
\label{sec_background}
Multi-agent SLAM systems involve multiple mobile agents collaboratively constructing a map while simultaneously localizing themselves within it as they explore an unknown environment. Such systems can deliver strong spatial awareness capabilities across a wide range of applications, from autonomous vehicles to advanced AR and human-robot collaboration. Multi-agent SLAM could be either distributed or centralized, with the latter approach recognized as advantageous for accuracy \cite{kimeramulti,ravic}. In a centralized SLAM implementation, individual agents perform local sensing and computation, while an edge server coordinates global optimization processes \cite{9585827,10160938,Xu2022,Zhang2024,ccmdslam,Schmuck2019}. 
The architecture is shown in Figure \ref{fig:edge-slam-spec}.

\textbf{Mobile agents}, such as robots and AR devices, employ a SLAM frontend pipeline to process visual data. In the frontend, the camera continuously captures raw image data at a fixed rate, producing \textit{camera frames}. Since not all camera frames are suitable for localization and mapping, the SLAM frontend selectively processes camera frames based on criteria such as motion magnitude, converting qualifying camera frames into \textit{SLAM frames} for pose estimation. Among these SLAM frames, the system further identifies frames with significant viewpoint changes or rich geometric details as \textit{keyframes}, which contain sufficient distinctive visual features that serve as \textit{map points}. All the keyframes and map points form the \textit{local map} maintained by the frontend used to track the agent's spatial position and orientation, commonly referred to as \textit{pose}, which is represented as a six Degrees of Freedom (6-DoF) transformation matrix. The latency of the entire agent pose estimation process (tracking) is denoted as $\tau_{track}$.

\textbf{Edge server} receives captured keyframes and map points from multiple agents and is used to maintain the \textit{global map} across all agents. The process begins with \textit{loop closure detection}, which examines new keyframes for visual overlap with existing ones. This occurs when an agent revisits previously mapped locations (intra-agent loops) or when multiple agents observe common environmental features (inter-agent loops). The edge server invokes a map merging if it detects a loop closure. This involves integrating spatial information from multiple agents into a unified coordinate framework, enabling the construction of a comprehensive environmental representation that exceeds the observational capacity of individual agents. The total computation latency incurred when keyframes are processed at the edge server for loop closure and map merging is denoted as $\tau_{comp}$.

\textbf{Communication in-between.} As shown in Figure \ref{fig:edge-slam-spec}, the total latency of bidirectional data transmission between mobile agents and the edge server is denoted as $\tau_{comm}$. The agent transmits the corresponding keyframe along with its map point to the edge server when there is a substantial displacement between its latest estimated pose and the previous pose. On the edge server, the conventional method uses the global map to update the agent's local map, sends the new local map to the agent, and the agent uses the updated local map to calculate its pose \cite{Zhang2024,Xu2022,Ali2020}. However, COVINS and COVINS-G \cite{9585827,10160938}, a state-of-the-art open-source visual SLAM framework, use a more efficient way by sending only a pose correction matrix instead of the local map to agent, achieving lower bandwidth and latency overhead.

\section{Related Work}
\label{sec_relatedwork}

\textbf{Edge-Assisted Multi-Agent SLAM.} The computational demands of real-time SLAM motivate research into edge-assisted architectures, such as Edge-SLAM~\cite{Ali2022,Cao2022}, which distribute processing between mobile devices and edge servers. \editting{Early cloud-offloaded SLAM systems such as C2TAM~\cite{Riazuelo2014} established the paradigm of offloading map optimization to remote servers, though at the cost of transmitting the full global map back to each agent. Subsequent centralized collaborative frameworks, including CVI-SLAM~\cite{Karrer2018} and CCM-SLAM~\cite{Schmuck2019}, refined this architecture by having agents run local odometry while a central server handles global optimization, but exclusively targeted homogeneous robotic teams.}
COVINS \cite{9585827} and COVINS-G \cite{10160938}
extend this approach to collaborative multi-agent scenarios. Furthermore, SwarmMap~\cite{Xu2022} addresses visual SLAM for multi-agent collaboration in large open areas, \editting{CloudSLAM~\cite{Wright2020} applies edge offloading to vehicular SLAM demonstrating domain-specific latency concerns,} while Map++~\cite{Zhang2024} explores efficiency improvements in map merging.
However, these works are mainly based on homogeneous agents and focus on improving the scalability of SLAM for large-scale environments, often overlooking the unique challenges posed by AR-specific adaptations in real-world settings. In this work, we \editting{are the first to} address the stringent latency requirements of AR by dynamically balancing the differing latency tolerances of heterogeneous human and robot agents and exploiting redundancy opportunities in small indoor environments.

\textbf{SLAM for Head-Mounted AR.} 
Early SLAM systems for AR optimize on-device computation through parallel tracking and mapping threads~\cite{Klein2007} and dynamic accuracy-efficiency balancing~\cite{Park2017}, but due to device limitations, they operate on stationary equipment, such as a personal computer. With the development of head-mounted AR devices, more recent approaches either address pervasiveness through sensor fusion, such as RoVAR~\cite{dasari2023rovar}, or address computational limitations through edge-assisted offloading, such as SLAM-Share~\cite{Dhakal2022}. \editting{Further works study the multi-user AR communication problem: Ran et al.~\cite{Ran2020SPAR} reduce transmission overhead between homogeneous AR users through content-aware selective keyframe sharing, while Chen et al.~\cite{Chen2023CommMag} characterize latency as the central bottleneck across edge-cloud and peer-to-peer architectures for localization-based multi-user AR.} However, these SLAM solutions for head-mounted AR focus solely on supporting human users. In contrast, our work leverages agent heterogeneity, specifically the coexistence of heterogeneous agents such as robots and humans in AR-based collaborative environments, to address critical latency concerns. To accommodate the diverse needs of different agent types, we propose a QoE modeling mechanism to capture the distinct requirements of each agent type and employ a PID scheduler to dynamically allocate resources between homogeneous (multiple human users) and heterogeneous (human-robot) agent configurations based on QoE.

\vspace{-1mm}
\section{When Head-Mounted AR Meets Multi-Agent SLAM}
\label{sec_challenges}

In this section, we describe what we encounter when deploying multi-agent SLAM in practical AR scenarios. Although SLAM can provide accurate pose tracking for AR devices to render virtual content, its latency results in drift in the perceived spatial locations of rendered virtual content as perceived by users.

\vspace{-2mm}
\subsection{Latency Challenges of Multi-Agent SLAM Adoption in Head-Mounted AR}
\subsubsection{Harsh motion-to-photon latency requirements from head-mounted AR}
Motion-to-photon latency, defined as the cumulative delay from motion sensing through tracking to visual display, represents a critical performance metric for AR systems. \editting{Head-mounted displays are uniquely latency-sensitive among interactive computing systems because unlike earth-fixed displays such as monitors and projectors, the entire visual scene is rigidly yoked to the user's head. This coupling means that any rendering delay causes the displayed scene to lag behind the user's actual head position, producing a discrepancy between where the visual scene appears and where the vestibulo-ocular reflex (VOR), a gaze-stabilizing neural circuit in the brainstem, expects it to be~\cite{stauffert2020latency, mtplatencyneareye}. This discrepancy has no equivalent in conventional display systems and is the reason latency detection thresholds in head-mounted displays are substantially lower than that in any other display class, with psycho-physical experiments reporting just-noticeable differences of less than 17~ms~\cite{adelstein2003head, mania2004perceptual, jerald2010relating}. These findings establish 20~ms as a perceptually derived upper bound for motion-to-photon latency, beyond which VOR-induced image slip becomes reliably detectable and causes perceptual instability during typical head movements~\cite{adelstein2003head, mania2004perceptual, stauffert2020latency, warburton2023measuring}.}

\begin{figure}[t]
    \centering
    \begin{subfigure}{0.48\textwidth}
        \centering
        \includegraphics[width=\linewidth]{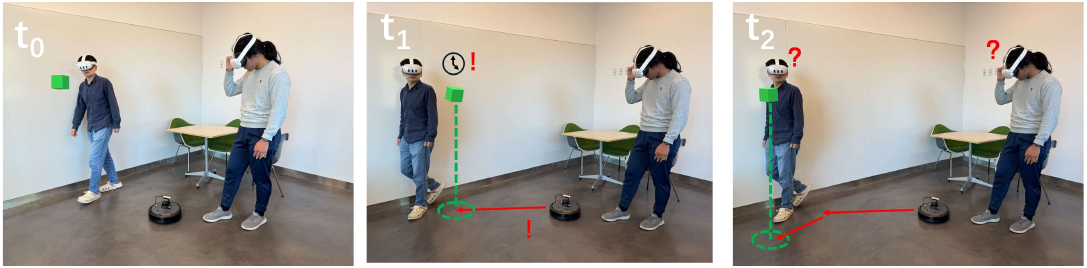}
        \caption{$t_1 - t_0 < \tau_{AR} < t_2 - t_0$}
        \label{fig:latency-impact-1}
    \end{subfigure}
    \hfill 
    \begin{subfigure}{0.48\textwidth}
        \centering
        \includegraphics[width=\linewidth]{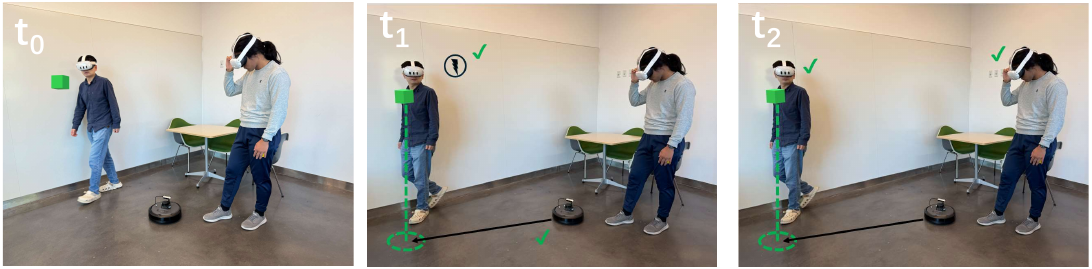}
        \caption{$\tau_{AR} < t_1 - t_0$}
        \label{fig:latency-impact-2}
    \end{subfigure}
    \vspace{-3mm}
    \caption{Effect of motion-to-photon latency ($\tau_{AR}$) on AR rendering. Time points $t_0$, $t_1$, and $t_2$ denote the user's start, displacement, and an idle period after it. In (a), failure to update the cube's position within $t_1 - t_0$, causing it to drift from correct location. At $t_2$, the cube is re-rendered in the correct location, but confusion on virtual contents' implication and on robot's planned trajectory had already occurred. In (b), cube location update is done within $t_1 - t_0$, preventing confusion of both kinds.}
    \label{fig:delay_latencyimpact} 
    \vspace{-3mm}
\end{figure}

\editting{A demonstration of latency in AR is shown in} Figure~\ref{fig:delay_latencyimpact}, where a virtual object acts as an interface between the user and the robot. In Figure~\ref{fig:delay_latencyimpact}(a), when the user moves in the real world and their pose changes, high motion-to-photon latency prevents the AR device from updating the AR object's position (its spatial coordinates) in accordance with the user's new pose. As a result, the AR object appears to maintain its original relative distance from the user, causing it to drift from the correct pose. This failure of virtual objects to update their positions synchronously with the user's head movements creates a phenomenon known as \textit{judder} or \textit{swim}~\cite{slocum2023going,lutwak2023user}. 

\editting{In AR, the physical world provides a simultaneous zero-latency reference, rendering any lag in virtual content rendering immediately apparent as spatial misalignment and scaling errors by the product of head angular velocity and end-to-end latency~\cite{AzumaBishop1994}, producing displacements of several centimeters at arm's length even at moderate head-rotation rates~\cite{Holloway1997}. This geometric constraint independently corroborates the perceptual 20~ms threshold and reinforces its application specifically to head-mounted AR.}

\vspace{-2mm}
\subsubsection{What's the role of SLAM in head-mounted AR's latency?} The motion-to-photon latency for AR users mainly stems from three components: \textit{sensor collection latency}, \textit{SLAM estimation latency}, \textit{visual rendering and display latency}~\cite{stauffert2020latency, warburton2023measuring}.
The sensor collection latency and the visual rendering and display latency are either hardware-limited (depending on inertial and visual sensor types) or determined by the AR device choice (depending on whether a video-see-through or optical-see-through solution is used), and are therefore outside the scope of this work.
\editting{
Therefore, they consume a fixed portion of the 20 ms budget regardless of software optimization. Consequently, the SLAM estimation component is allocated strictly less than 20 ms, since any SLAM latency at or exceeding this bound guarantees violation of the total motion-to-photon constraint, independent of all other system choices. This motivates treating 20 ms as a hard ceiling for SLAM latency.}
In AR applications, SLAM introduces two types of latency: 1) When the user's range of motion is small and the environment detected by the AR device does not change significantly, the AR device can estimate the user's pose and render the AR content locally. The latency experienced by the user in this case is the tracking latency of the mobile agent, i.e., $\tau_{track}$. 2) When the user's range of motion is large or the environment detected by the AR device changes considerably, the AR device cannot accurately estimate the user's pose on its own. In this case, the latest keyframe captured by the AR device must be uploaded, and the user's pose must be updated with the help of the global map constructed from the perspectives of other users and robots. The latency experienced by the user is then the sum of the communication and computation latencies of the entire agent–edge–agent process in the edge-assisted SLAM system, i.e., $\tau_{track} + \tau_{comp} + \tau_{comm}$. As shown in Figure \ref{fig:delay_latencyimpact}(b), when SLAM latency is reduced, AR objects rendered according to the updated pose maintain consistency in the user's perception.

\begin{figure}[t]
    \begin{minipage}[c]{0.4\columnwidth}
        \centering
        \includegraphics[width=\textwidth]{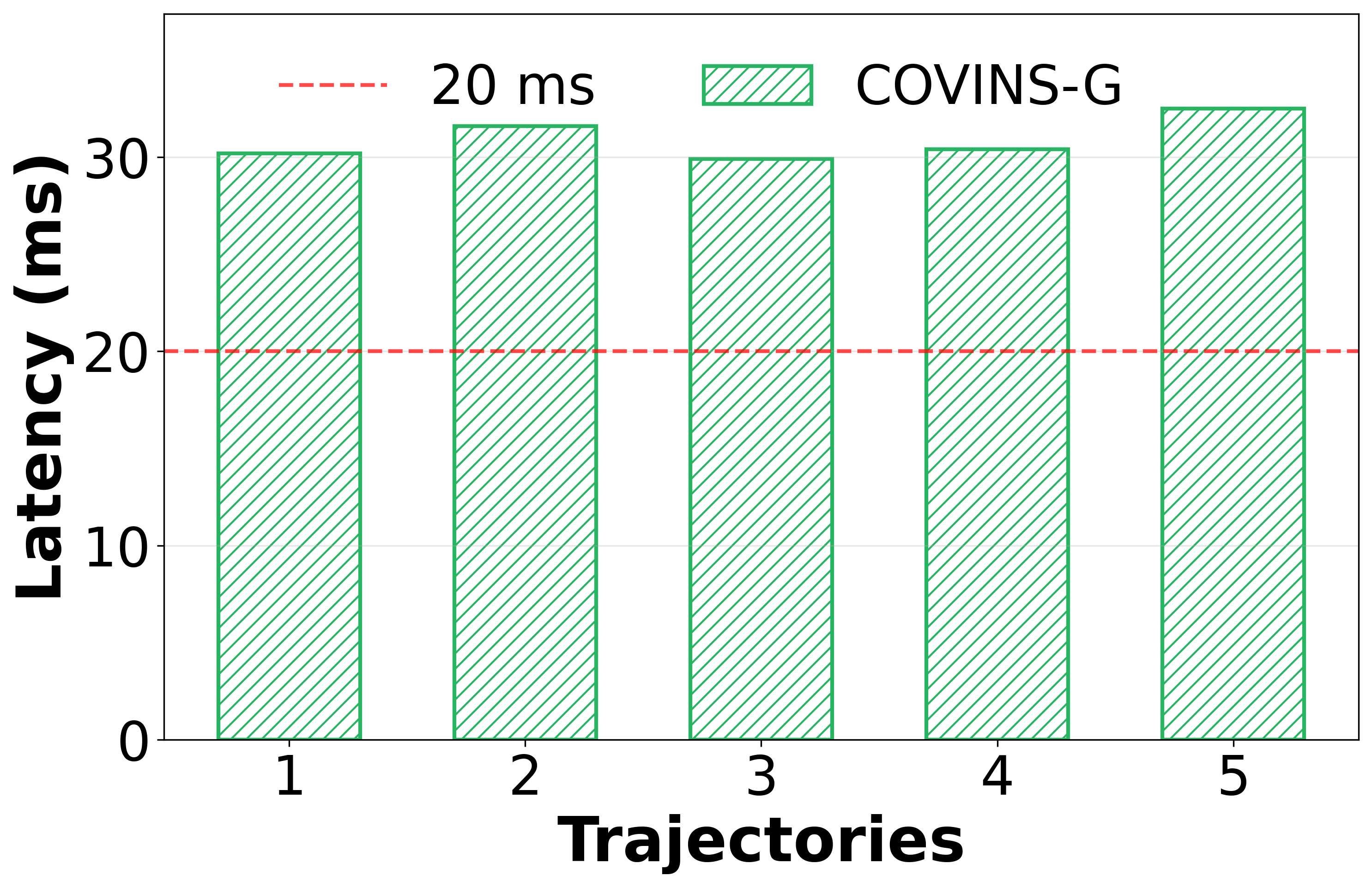}
        \vspace{-8mm}
        \caption{COVINS-G latency on our field study data. The 20 ms threshold is the required upper limit for Augmented Reality.}
        \label{fig:delay_analysis}
    \end{minipage}
    \hfill
    \begin{minipage}[c]{0.55\columnwidth}
        \centering
        \includegraphics[width=.85\textwidth]{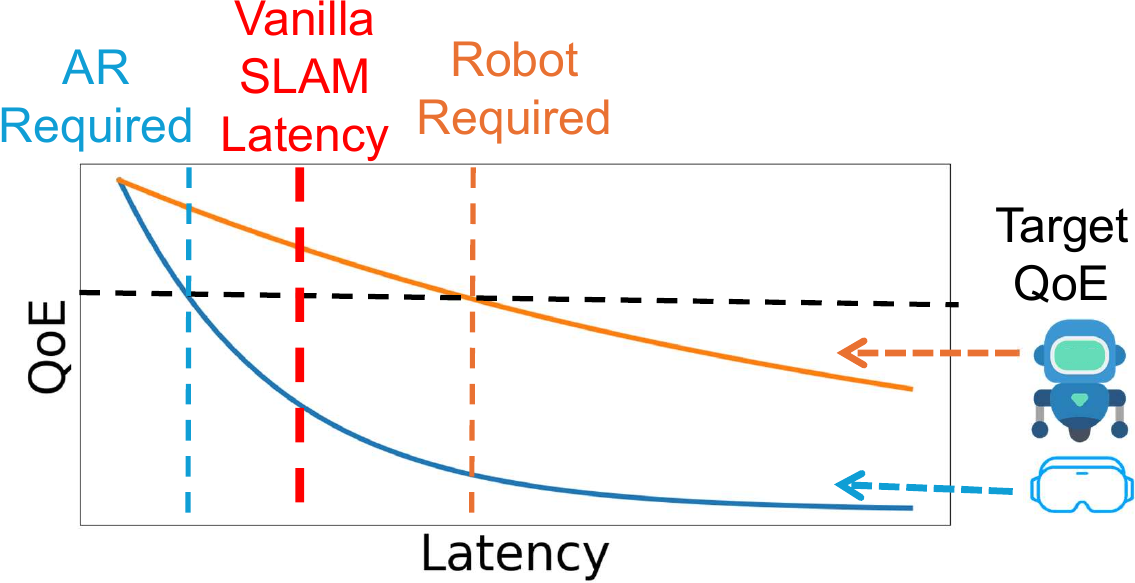}
        \vspace{-3mm}
        \caption{QoE modeling. Compared to vanilla SLAM that causes equal latency between agents, we capture the different latency requirements between agents and target equal QoE.}
        \label{fig:qoe-modeling}
    \end{minipage}
    \vspace{-3mm}
\end{figure}

\subsection{\editting{Key Observations for AR-HRC SLAM}}

Due to the distributed nature and communication requirements of multi-agent SLAM, existing systems such as SwarmMap \cite{Xu2022} and Map++\cite{Zhang2024} report delays of tracking, $\tau_{track}$, to be 50 to 100 ms and delays of map processing, $\tau_{comp}$, of 300 ms to 600 ms. 
To address harsh latency challenges in head-mounted AR, we take a deeper look into multi-agent systems in the HRC context to identify latency optimization opportunities that exist across heterogeneous agents, homogeneous agents, and the edge server. Below, we present our key observations that can be leveraged to reduce communication and computation latency.

\vspace{-2mm}
\subsubsection{Observation 1: Heterogeneous latency requirements between AR applications and robots for human experience}
We observe a fundamental distinction in users' localization requirements between AR experience and robot control. \editting{This distinction arises from the fundamental difference in how latency manifests perceptually: while other mobile agents (e.g., robots) experience latency as delayed state updates affecting their control loops, AR devices directly expose latency-induced spatial inconsistencies to human perception. However, traditional multi-agent SLAM frameworks do not differentiate between AR devices and other mobile agents.
As suggested in Figure~\ref{fig:qoe-modeling}, this observation indicates that implementing differentiated quality-of-experience (QoE) policies to prioritize communication for AR headsets while relaxing latency constraints for robots could satisfy both applications}.

\vspace{-2mm}
\subsubsection{Observation 2: Dynamic tracking conditions among homogeneous agents}
\begin{figure}[t]
\centering
\begin{minipage}[c]{0.5\columnwidth}
    \centering
    \includegraphics[width=\textwidth]{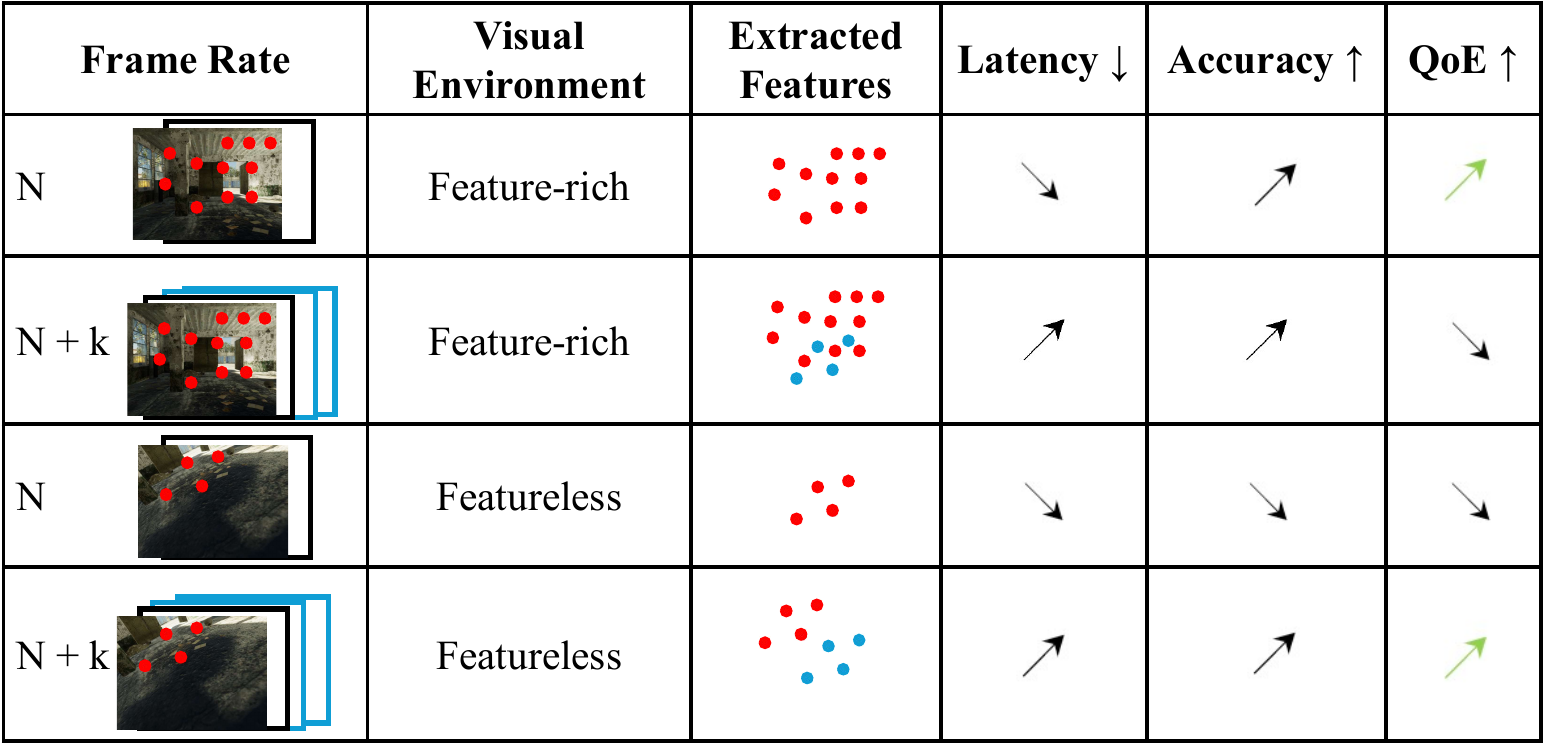}
    \caption{Dynamic visual environment observation: up arrow indicates resulting increase, down arrow indicates resulting decrease. In feature-rich environments, more visual frames create redundancy. In featureless cases, limited visual frames cause poor accuracy.}
    \label{fig:observation2}
\end{minipage}%
\hfill
\begin{minipage}[c]{0.45\columnwidth}
    \centering
    \includegraphics[width=\textwidth]{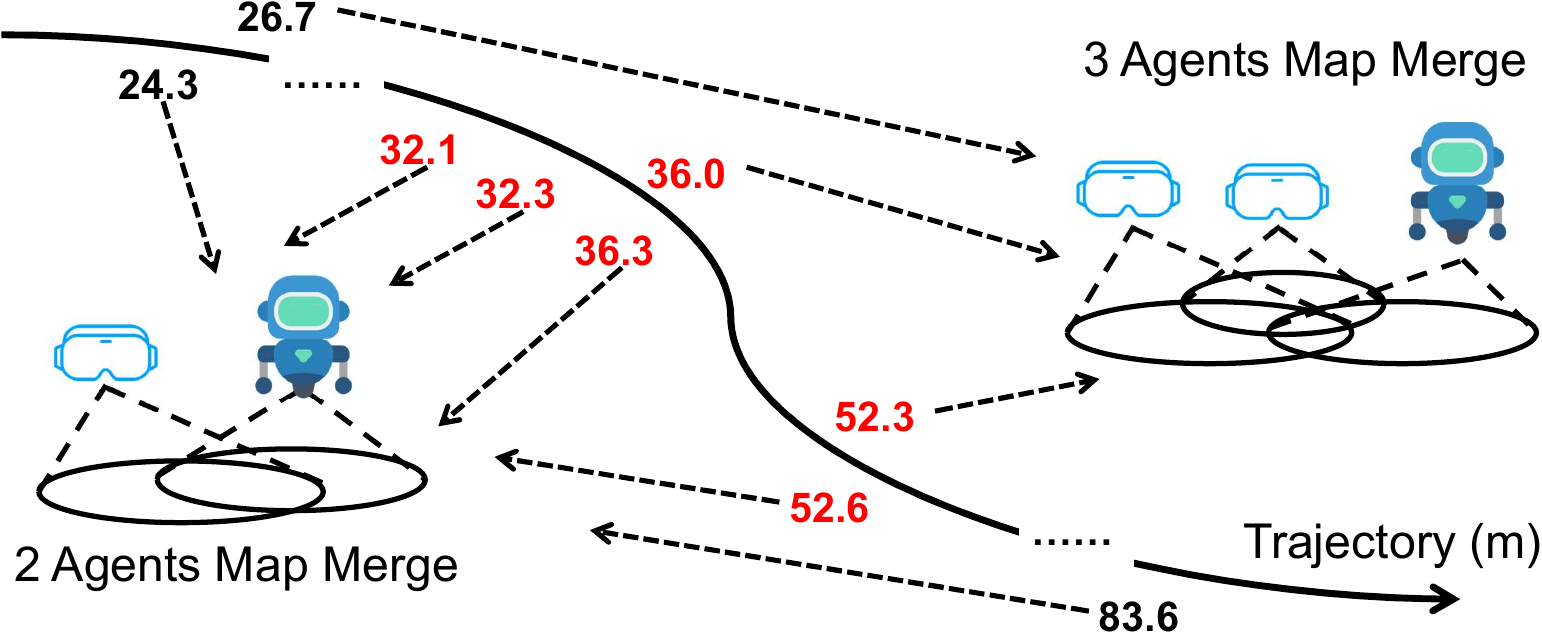}
    \vspace{2mm}
    \caption{Observation of redundancy in map merging. Values reflect the length of trajectory before selected merges. Map merging is aggregated over certain windows, suggesting redundancy.}
    \label{fig:observation3}
\end{minipage}
\end{figure}

Traditional multi-agent SLAM systems treat homogeneous agents, e.g., human-to-human, as functionally equivalent and interchangeable entities. However, SLAM tracking performance can vary significantly across homogeneous agents due to differences in the visual information they capture, reflecting \emph{environmental dynamics.} This observation indicates the potential for resource allocation based on visual conditions. As demonstrated in Figure \ref{fig:observation2}, latency increases when redundant visual frames are captured under favorable tracking conditions (adequate features, slow motion), while accuracy drops when limited visual frames are captured under challenging visual conditions (scarce features, fast motion). This sheds light on the need for an adaptive resource allocation approach based on real-time assessments of each agent's visual conditions. Such an approach should prioritize communication resources for agents experiencing challenging visual conditions, while reducing resource allocation to agents in favorable tracking circumstances, thereby maintaining overall tracking performance.

\label{fig:combined}

\vspace{-2mm}
\subsubsection{Observation 3: Inefficient computation of map merging in shared human–robot workspaces}
Unlike autonomous driving, which primarily deals with open-world SLAM in unstructured outdoor environments involving scalability and long-term consistency \cite{SLAM-auto-drive-review}, we focus on AR use cases that apply SLAM in indoor, constrained, personal spaces. These spatially constrained environments usually have well-defined boundaries and limited structural complexity. Moreover, when multiple agents are placed in a spatially constrained region, they generate map merge requests to the edge server for segments that exhibit high structural similarity. Figure \ref{fig:observation3} demonstrates all the map merges when our baseline SLAM system with backend optimization (COVINS-G) is running on one trajectory of our collected data. Aggregated merging is repeatedly observed, where map merging can happen three times within as little as 1 meter of total movement distance. For example, the red-marked distances in Figure \ref{fig:observation3} show merges occurring frequently in a short duration between 32.1~m and 32.3~m, 36.0~m and 36.3~m, and 52.3~m and 52.6~m.
Therefore, the resulting frequent merging operations process incrementally similar geometric information, leading to only marginal accuracy improvements while significantly contributing to computation latency, $\tau_{comp}$. This observation suggests that developing selective map merging strategies could reduce computation latency while maintaining tracking accuracy for household SLAM applications.

\vspace{-2mm}
\section{System Design}
\label{sec_system}

\begin{figure}[]     
\includegraphics[width=.8\columnwidth]{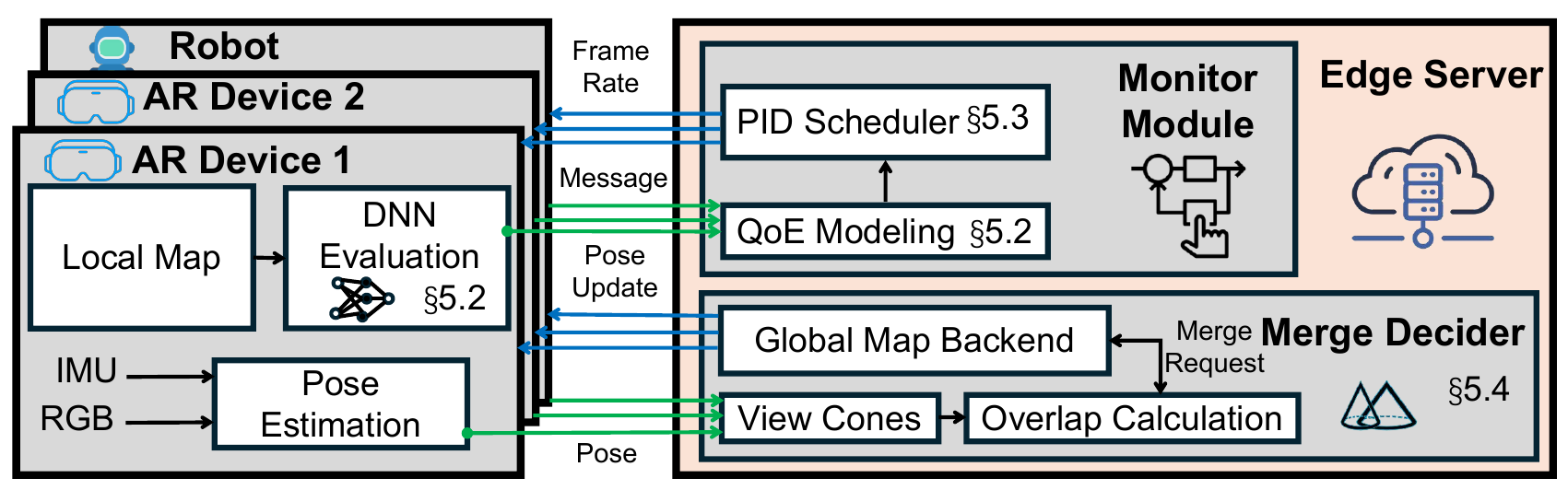}   
\caption{Main system diagram. Each message contains the error estimation from DNN evaluation and real-time latency, calculated by adding Round Trip Time (RTT) and keyframe processing time.}
\vspace{-7mm}
\label{fig:main-system} 
\end{figure}

\subsection{System Overview}
\SysName addresses the latency requirements in multi-agent SLAM systems through three primary components. 

\textbf{QoE Modeling (Section~\ref{sec_system_qoe})} is employed to quantify tracking performance for individual agents. The QoE assessment integrates error estimation metrics with latency measurements to evaluate tracking quality. This quantification enables differentiated resource allocation policies, allowing the system to prioritize communication for AR devices while applying relaxed constraints to robotic agents (Observation 1), thereby accommodating the heterogeneous latency requirements across agent types. DNN-based evaluation module is implemented to evaluate all keyframes. The evaluation module analyzes visual features to estimate localization errors, generating parameters that inform QoE modeling. This evaluation enables assessment of tracking accuracy without requiring ground-truth data, facilitating resource management adaptation to tracking conditions.

\textbf{PID Scheduler (Section~\ref{sec_system_pid})} dynamically adjusts frame transmission rates according to QoE values. The scheduler modulates frame rates to allocate communication resources based on tracking difficulty, increasing frame rates for agents experiencing challenging visual conditions while reducing rates for agents under favorable circumstances. This adaptive scheduling addresses the dynamic tracking variations among homogeneous agents (Observation~2), preventing latency increases from redundant transmissions and accuracy degradation from insufficient frame updates.

\textbf{Visual Overlap Analysis (Section~\ref{sec_system_overlap})} is based on view cone overlap between agents to determine spatial relationships. Applying overlap thresholds, the module selectively filters map merge requests, reducing redundant merging operations that occur within limited movement ranges. This approach minimizes computational overhead in shared human–robot workspaces (Observation~3) while maintaining tracking accuracy for household SLAM applications \editting{at the same time}.

These components are distributed across the system architecture as follows: the DNN evaluation module operates in the frontend to provide real-time quality assessment; the QoE modeling function and PID scheduler comprise the Monitor Module for adaptive resource management; and the visual overlap analysis functions within the Merge Decider coordinate with the global map backend taken from the implementation of COVINS-G \cite{10160938}. Details of the system design are demonstrated in Figure~\ref{fig:main-system}.

\vspace{-2mm}
\subsection{Agent-specific QoE modeling}
\label{sec_system_qoe}

To capture and utilize this difference in latency requirements, we develop a Quality of Experience (QoE) modeling method for different services.

Effective resource scheduling for tracking-dependent systems requires modeling user perception of tracking quality: how users perceive the tracking performance when interacting with them. To establish a principled approach to this modeling challenge, we examine how QoE has been formulated across domains where user perception of system performance is critical. In web server load balancing, queuing analysis revealed logarithmic correlation between queue lengths and QoE~\cite{zhang2008web}. In human-computer interaction, MacKenzie and Ware~\cite{mackenzie1993lag} established logarithmic perceptual scaling effects even for small delays. In telecommunications services, Reichl et al.~\cite{reichl2010logarithmic,reichl2013logarithmic} established that user-perceived quality follows logarithmic relationships with Quality of Service metrics representing positive resource parameters such as bandwidth or throughput, grounded in the psychophysical Weber-Fechner law. Complementing these findings, the IQX hypothesis~\cite{Fiedler2010,iqx2,hossfeld2008testing} formalizes the exponential relationship between influence factors and QoE. 

These findings converge on a common principle that \textbf{users exhibit diminishing marginal sensitivity to performance improvements, and different performance factors contribute differentially to experience.} The foundational work by Keeney and Raiffa~\cite{keeney1993decisions} on Multi-Attribute Utility Theory (MAUT) provides theoretical justification for weighted logarithmic formulations that naturally capture such diminishing returns in multi-factor quality perception.

We adapt these established QoE modeling principles to the human-robot collaboration context, where the SLAM tracking quality directly impacts task completion, safety, and user confidence. In this domain, latency determines responsiveness of collaborative actions, while tracking accuracy affects spatial awareness and manipulation precision. To the best of our knowledge, this work represents the first attempt to formulate QoE-driven resource management for SLAM systems in HRC scenarios.

\textit{Agent-specific QoE Modeling.}
We adopt an agent-specific QoE model that captures heterogeneous service priorities by explicitly distinguishing agent-centric and marginally contributing performance determinants for different agents (Robot and AR). QoE is defined as a function of latency and accuracy, which are normalized and parameterized differently depending on the agent type:
\begin{equation}
\text{QoE}_i = e^{-\beta_i \cdot P_i} \cdot \left(1 - e^{-\gamma_i \cdot Q_i}\right),
\label{eq:qoe_unified}
\end{equation}
where the parameters' implications can be explained below:

\textbf{Agent-centric factor penalty ($P_i$).}
The agent-centric factor penalty captures the dominant QoE bottleneck for a given agent and enforces steep QoE degradation when this bottleneck is violated.

\begin{itemize}
    \item \textit{AR services}: Latency is the dominant factor due to human perceptual sensitivity. We define $P_i = d - d_0$, where $d$ is the motion-to-photon or system latency in milliseconds, and $d_0$ is the 20 ms threshold. This choice reflects that even small latency increases can significantly degrade user experience. The parameter $\beta_i$ controls the severity of QoE degradation with respect to latency, enforcing the strength of QoE collapse when latency deteriorates, visually reflected by the steepness of the curve in Figure~\ref{fig:qoe-modeling}.
    
    \item \textit{Robot services}: Task correctness and control stability are the key concerns. We define $P_i = a_0 - a$, where $a$ denotes task accuracy and $a_0$ is the minimum acceptable accuracy threshold, heuristically set as 1 cm. This formulation reflects the fact that users will experience high QoE drop when robots fail to achieve their assigned task. The parameter $\beta_i$ controls the severity of QoE degradation with respect to accuracy, enforcing the strength of QoE collapse when accuracy deteriorates.
\end{itemize}

\textbf{Marginal performance contribution ($Q_i$).}
The marginal performance contribution models auxiliary benefits that improve QoE but exhibit diminishing marginal returns once the primary requirement is satisfied.

\begin{itemize}
    \item \textit{AR services}: Accuracy contribution has diminishing marginal returns. We define $Q_i = a$, reflecting that higher visual or spatial accuracy improves immersion and task performance, but cannot compensate for excessive latency. The parameter $\gamma_i$ determines the rate at which accuracy improvements contribute to QoE. The term $(1 - e^{-\gamma_i \cdot Q_i})$ ensures diminishing returns.
    
    \item \textit{Robot services}: Latency has diminishing marginal returns since robots can compensate for moderate delays through control buffering. We define $Q_i = (d + \theta_i)^{-1}$, where $\theta_i$  is a normalization constant that bounds latency sensitivity and reflects agent-specific tolerance. This inverse formulation captures the asymptotic improvement in QoE as latency decreases, without allowing latency to dominate accuracy-driven QoE. The parameter $\gamma_i$ determines the rate at which latency improvements contribute to QoE, with the term $(1 - e^{-\gamma_i \cdot Q_i})$ ensuring diminishing returns, reflecting the observation that beyond a certain point, improvements in accuracy yield progressively smaller benefits, as reflected in the damping effect of curves in Figure~\ref{fig:qoe-modeling}.
\end{itemize}

\vspace{-2mm}
\subsubsection{Deep Neural Network-based SLAM Accuracy Estimation}
\label{sec_system_dnn}

Modeling QoE requires the accuracy term $a$ to be estimated in real time. However, accurate assessment of SLAM localization accuracy presents a fundamental challenge, as ground-truth trajectory information from Motion Capture (MoCap) systems or similar mechanisms is unavailable in real-world settings. To address this limitation, we employ a lightweight deep neural network architecture for real-time SLAM error estimation, building upon established methodologies that utilize keyframe analysis for visual SLAM error prediction~\cite{seesys}. We develop a streamlined architecture based on MobileNetV2, specifically optimized for low-latency inference in resource-constrained environments~\cite{Sandler_2018_CVPR}. \editting{The full training pipeline, including label generation scripts, dataset preparation, and model training code, is released as part of the open-source codebase accompanying this paper.}

\editting{
\noindent\textbf{Training label generation.}
Since ground-truth localization error is unavailable at runtime, we generate per-frame error labels offline using a custom monocular SLAM pipeline built on pyslam~\cite{Freda2025pySLAM}. For each frame in the training dataset, pyslam runs ORB feature detection and tracking with DBoW3-based loop closure to produce an estimated camera pose. The corresponding ground-truth pose is obtained from the SenseTime SLAM dataset~\cite{sensetimedataset} synchronized to the image timestamps, and the per-frame localization error is computed by comparing the two. Simultaneously, pyslam exposes its internal tracking state at each successfully tracked frame, from which the Point Spatial Distribution (PSD) matrix and the eight scalar features are extracted and cached alongside the error label, forming the complete input--label pairs used for supervised training.}

\begin{figure}[t]
    \centering
    \includegraphics[width=0.7\linewidth]{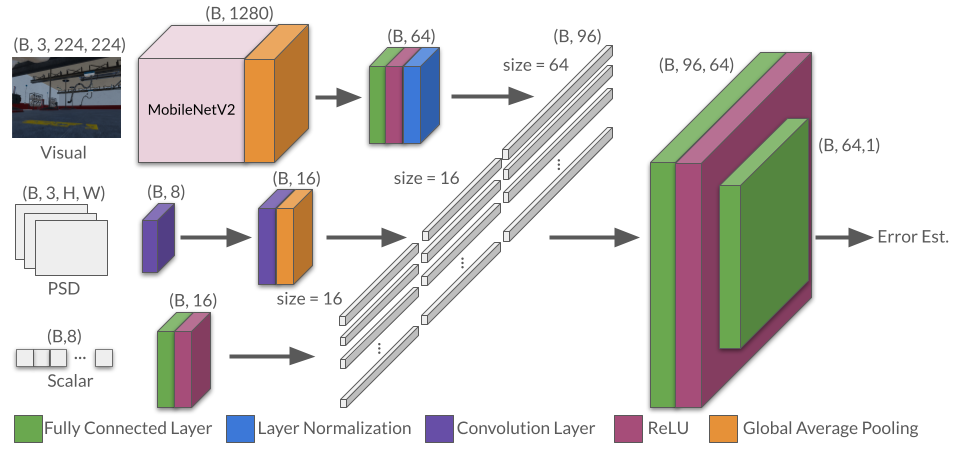}
    \caption{\editting{Architecture of the DNN-based SLAM error estimator. B refers to batch size, and H and W represent the spatial height and width of the PSD grid. Three parallel branches process visual keyframe data, PSD matrices, and scalar tracking features, respectively. Their outputs are concatenated into a 96-dimensional joint representation and passed through a two-layer fully connected network to produce the predicted per-frame localization error.}}
    \label{fig:dnn-arch}
\end{figure}

\editting{
\noindent\textbf{Neural network architecture.}
As illustrated in Figure~\ref{fig:dnn-arch}, the network processes three complementary input modalities through dedicated branches, whose outputs are fused to produce the predicted tracking error.}

\editting{
The \textit{visual branch} processes each keyframe through a pre-trained MobileNetV2 backbone with the classification head removed. Global average pooling over the final feature map yields a 1280-dimensional representation, which a single fully connected layer projects to 64 dimensions, followed by ReLU activation and layer normalization.}

\editting{
The \textit{PSD branch} encodes the spatial arrangement of inlier feature points, which substantially impacts tracking robustness~\cite{seesys}. Inlier map points are encoded in a three-channel spatial grid capturing keypoint response strength, the inlier indicator, and estimated depth. Two sequential convolutional layers (8 and 16 filters, $3{\times}3$ kernels, stride~2) with ReLU activations, followed by global average pooling, reduce this grid to a 16-dimensional representation.}

\editting{
The \textit{scalar branch} captures SLAM system state characteristics that cannot be encoded spatially. An 8-element vector comprising image-level statistics (brightness, contrast, entropy, Laplacian variance) and tracking-quality indicators (detected keypoints, successful matches, active map points, and aggregate tracking quality score) is processed through a single fully connected layer with ReLU activation to produce a 16-dimensional representation.}

\editting{
The three branch outputs are concatenated into a 96-dimensional joint representation, which is passed through a two-layer fully connected network and ReLU activation, producing the scalar predicted tracking error.}

\subsection{PID scheduling pipeline}
\label{sec_system_pid}

\begin{algorithm}[t]
\caption{\editting{\SysName PID Scheduling System}}
\begin{algorithmic}[1]
\Function{PID}{$\text{agent},\ Q$}
    \State $s \gets \text{state}[\text{agent}]$ \Comment{load per-agent controller state}
    \State $Q^{*} \gets \text{qoe\_target}[\text{agent}]$
    \State $t \gets \Call{GetCurrentTime}{\,}$
    \State $\Delta t \gets t - s.t_{\text{prev}}$
    \State $e \gets Q^{*} - Q$

    \Statex \textbf{Proportional term:}
    \State $P \gets K_p \cdot e$ \Comment{compute proportional term from current error}

    \Statex \textbf{Integral term:}
    \State $s.I \gets \Call{Clip}{s.I + K_i \cdot e \cdot \Delta t,\ -I_{\max},\ I_{\max}}$ \Comment{accumulate error with clamping}
    \State $I \gets s.I$

    \Statex \textbf{Derivative term:}
    \State $d_{\text{raw}} \gets (e - s.e_{\text{prev}}) / \Delta t$ \Comment{estimate error change rate}
    \State $d \gets \alpha \cdot d_{\text{raw}} + (1 - \alpha) \cdot s.d_{\text{prev}}$ \Comment{filter the derivative estimate}
    \State $D \gets K_d \cdot d$

    \Statex \textbf{Output update:}
    \State $u \gets P + I + D$ \Comment{combine PID terms}
    \State $R \gets \Call{Clip}{s.R_{\text{prev}} + u,\ R_{\min},\ R_{\max}}$ \Comment{apply bounded rate update}
    \State $s.e_{\text{prev}} \gets e$;\quad $s.d_{\text{prev}} \gets d$;\quad $s.R_{\text{prev}} \gets R$;\quad $s.t_{\text{prev}} \gets t$ \Comment{store state for the next update}
    \State \Return $R$
\EndFunction
\Statex
\Procedure{MainControlLoop}{\,}
    \ForAll{evaluation messages received}
        \State $\text{agent} \gets \Call{GetAgentFromMessage}{\,}$
        \State $Q \gets \Call{CalculateQoE}{\text{agent.delay},\ \text{agent.accuracy}}$
        \State $R \gets \Call{PID}{\text{agent},\ Q}$
    \EndFor
\EndProcedure
\end{algorithmic}
\end{algorithm}

In addition to the different requirements among heterogeneous agents, another observation is derived from the dynamic visual environment at different times of operation, which affects SLAM performance. To address this, we employ a Proportional-Integral-Derivative (PID) scheduling framework for QoE optimization, utilizing frame transmission rate modulation as the primary control mechanism. The PID scheduler treats the QoE functions as target setpoints and dynamically adjusts frame transmission rates to maintain desired performance levels across heterogeneous agents.

The control system architecture treats each agent's QoE deviation from its target value as an error signal, applying proportional, integral, and derivative corrections to determine optimal frame transmission rates. This approach enables the system to balance immediate QoE requirements (proportional term), compensate for persistent performance deficits (integral term), and anticipate future resource needs based on QoE trends (derivative term), thereby achieving robust multi-agent resource allocation in dynamic communication environments. The error calculation is defined as:
\begin{equation}
err(t) = \text{QoE}_{\text{target}} - \text{QoE}_{\text{current}}(t),
\end{equation}
where $\text{QoE}_{\text{target}}$ represents the desired performance level for all agents, and $\text{QoE}_{\text{current}}(t)$ is computed using the agent-specific QoE models. The scheduler output combines three complementary terms through the standard PID formulation:
\begin{equation}
u(t) = K_p \cdot err(t) + K_i \cdot \int_0^t err(\tau) d\tau + K_d \cdot \frac{derr(t)}{dt},
\end{equation}
where the proportional term $K_p \cdot err(t)$ provides immediate response proportional to current QoE deviation, the integral term $K_i \cdot \int_0^t err(\tau) d\tau$ eliminates steady-state error by accumulating historical performance deficits, and the derivative term $K_d \cdot \frac{derr(t)}{dt}$ anticipates future trends based on error rate of change. To mitigate high-frequency noise in QoE measurements, the derivative term incorporates exponential filtering and anti-windup mechanisms. 

The PID output is mapped to frame transmission rate adjustments through:
\begin{equation}
R(t) = R_{\text{prev}} + u(t),
\end{equation}
where $R(t)$ represents the adjusted frame transmission rate for the agent. The continuous nature of PID scheduling enables fine-grained resource allocation adjustments, providing higher responsiveness compared to discrete threshold-based approaches employed in prior work~\cite{Xu2022}. Rather than binary decisions based on predetermined QoE thresholds, the PID framework facilitates smooth, incremental modifications to frame transmission rates, minimizing the effect of abrupt environmental transformations.

\begin{figure}[t]
\centering
\begin{subfigure}{0.16\columnwidth}
    \centering
    \includegraphics[width=\textwidth]{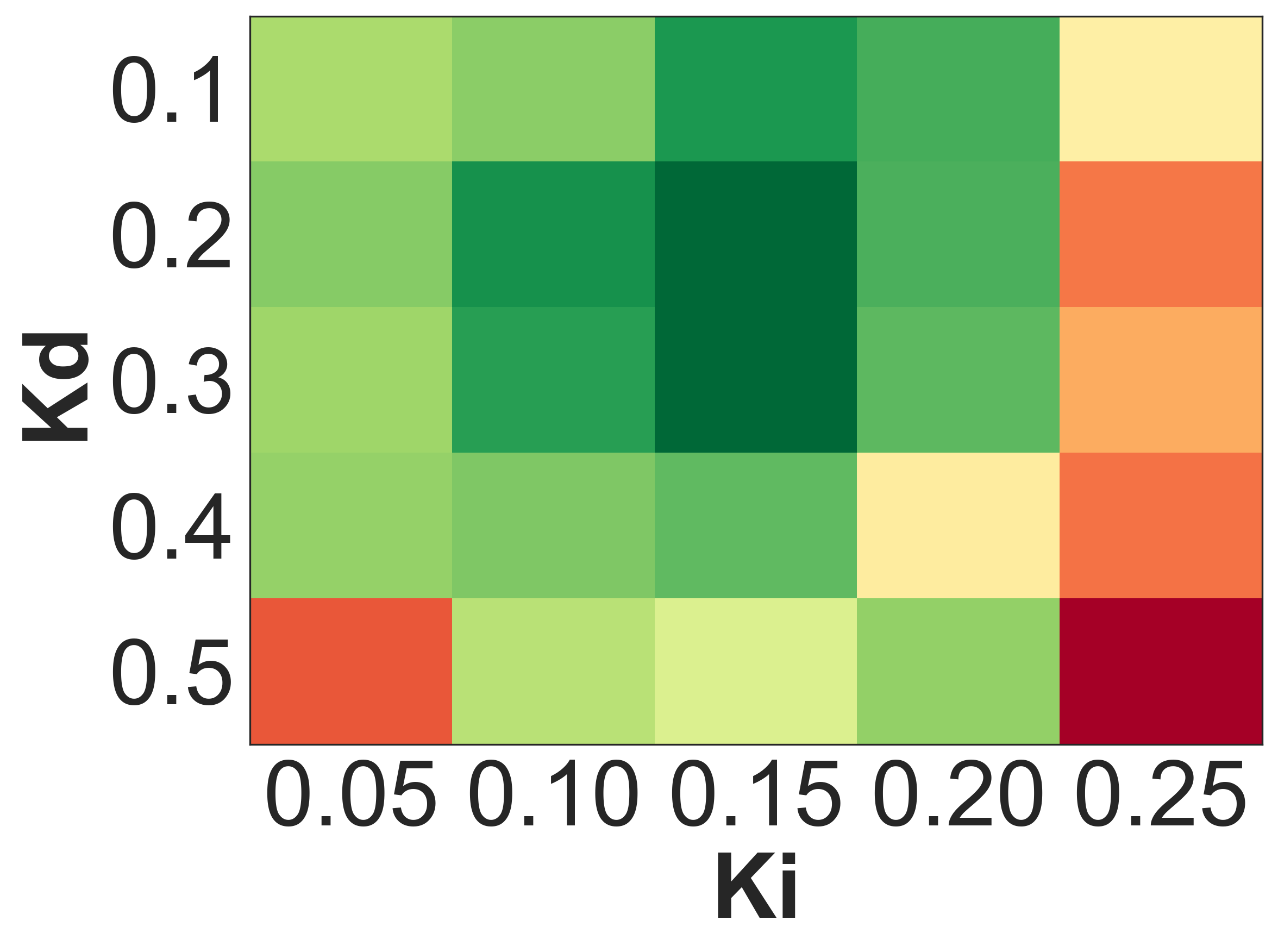}
    \label{fig:heatmap_kp08_acc}
\end{subfigure}%
\hspace{-1mm}%
\begin{subfigure}{0.16\columnwidth}
    \centering
    \includegraphics[width=\textwidth]{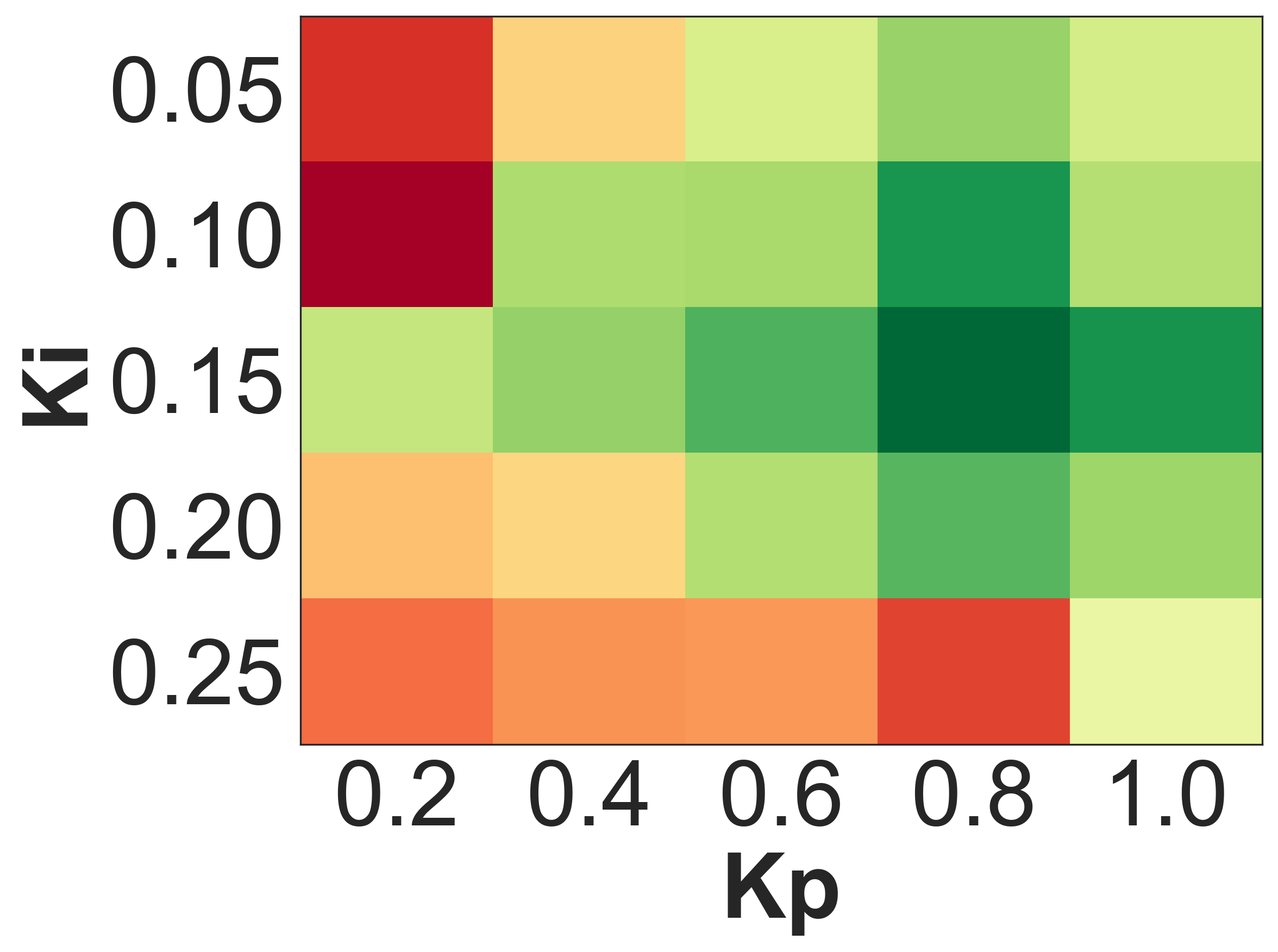}
    \label{fig:heatmap_kd02_acc}
\end{subfigure}%
\hspace{-1mm}%
\begin{subfigure}{0.16\columnwidth}
    \centering
    \includegraphics[width=\textwidth]
    {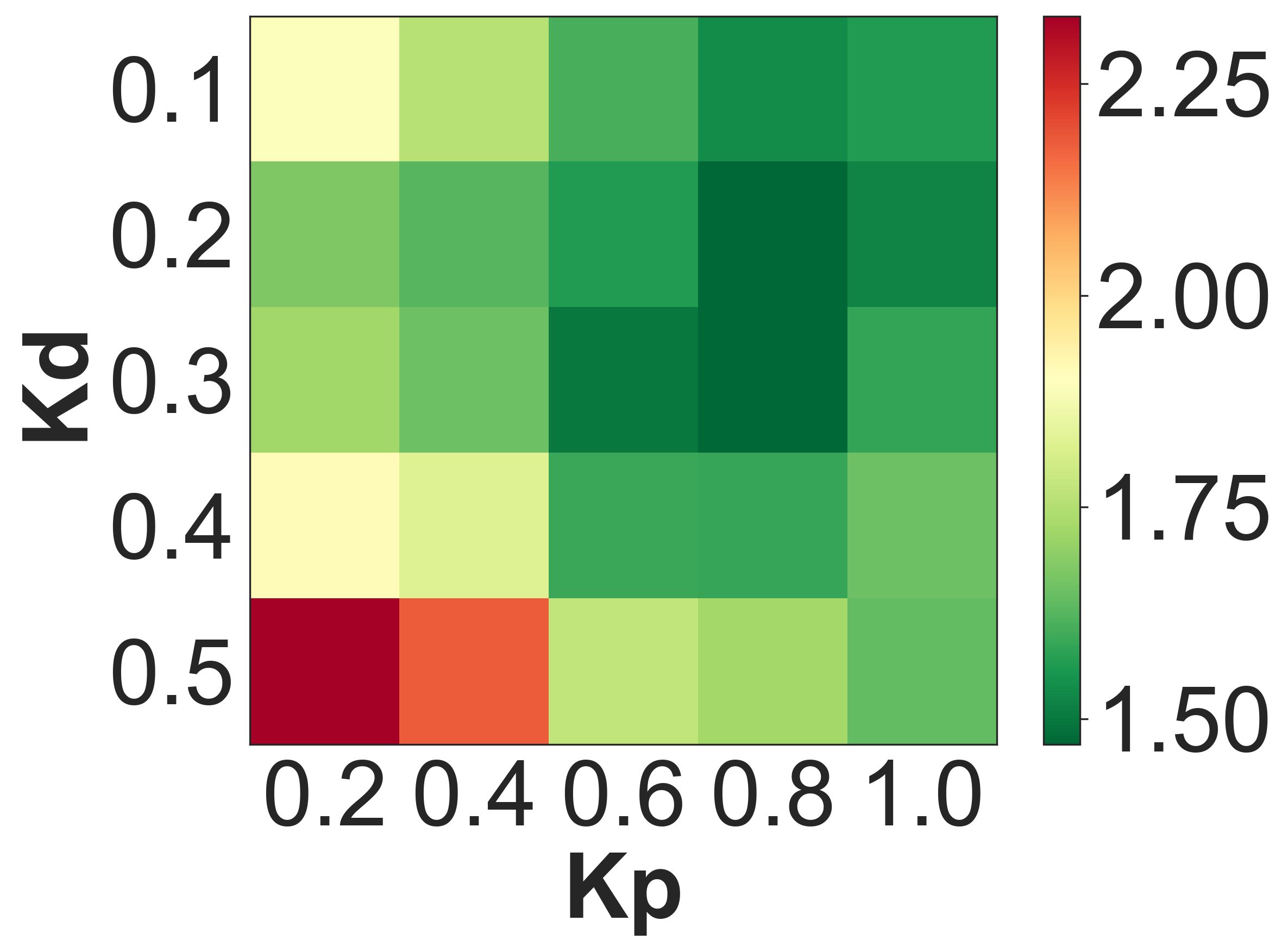}
    \label{fig:heatmap_ki015_acc}
\end{subfigure}
\hspace{-1mm}%
\begin{subfigure}{0.16\columnwidth}
    \centering
    \includegraphics[width=\textwidth]{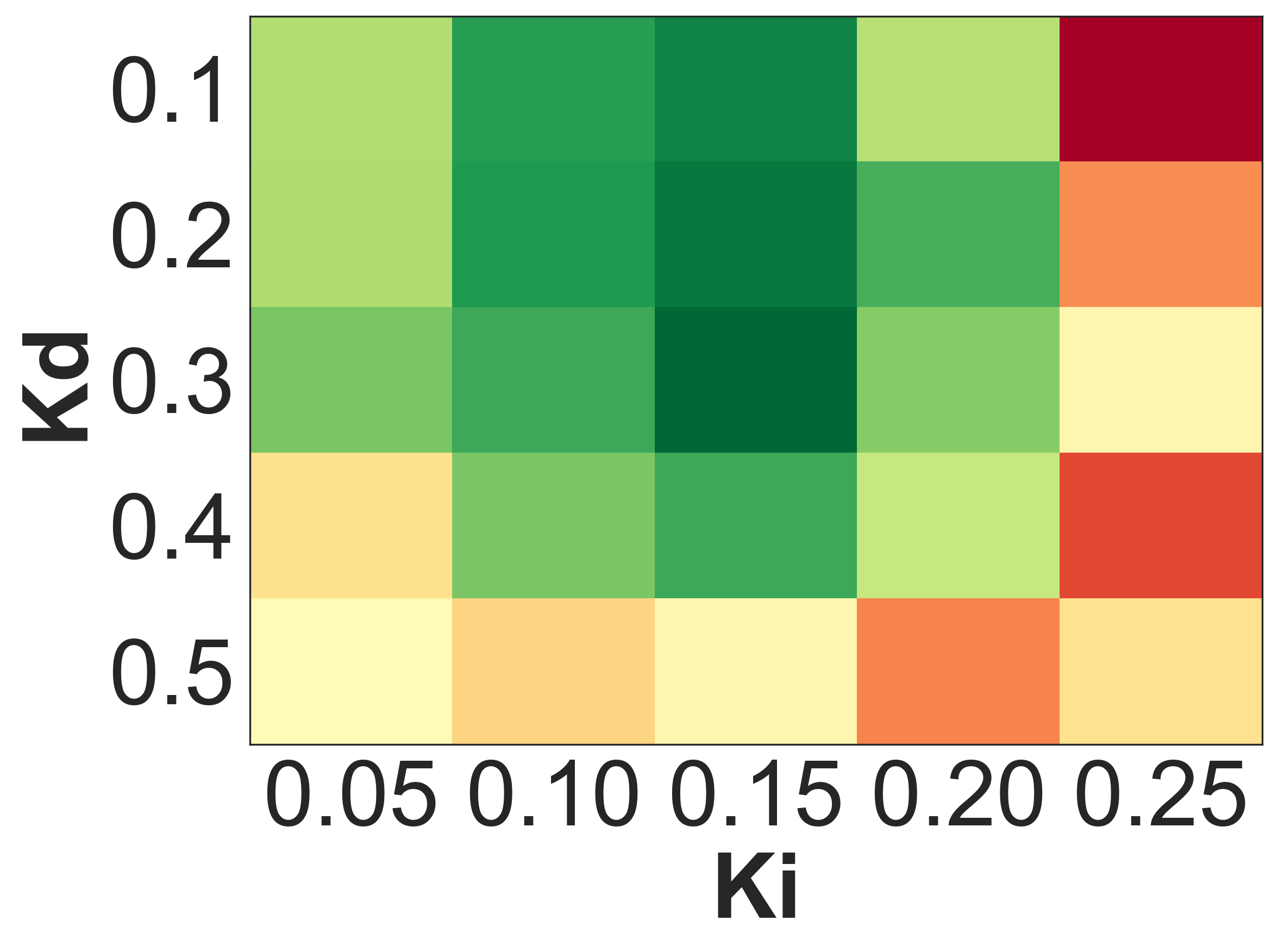}
    \label{fig:heatmap_kp08_lat}
\end{subfigure}%
\hspace{-1mm}%
\begin{subfigure}{0.16\columnwidth}
    \centering
    \includegraphics[width=\textwidth]
    {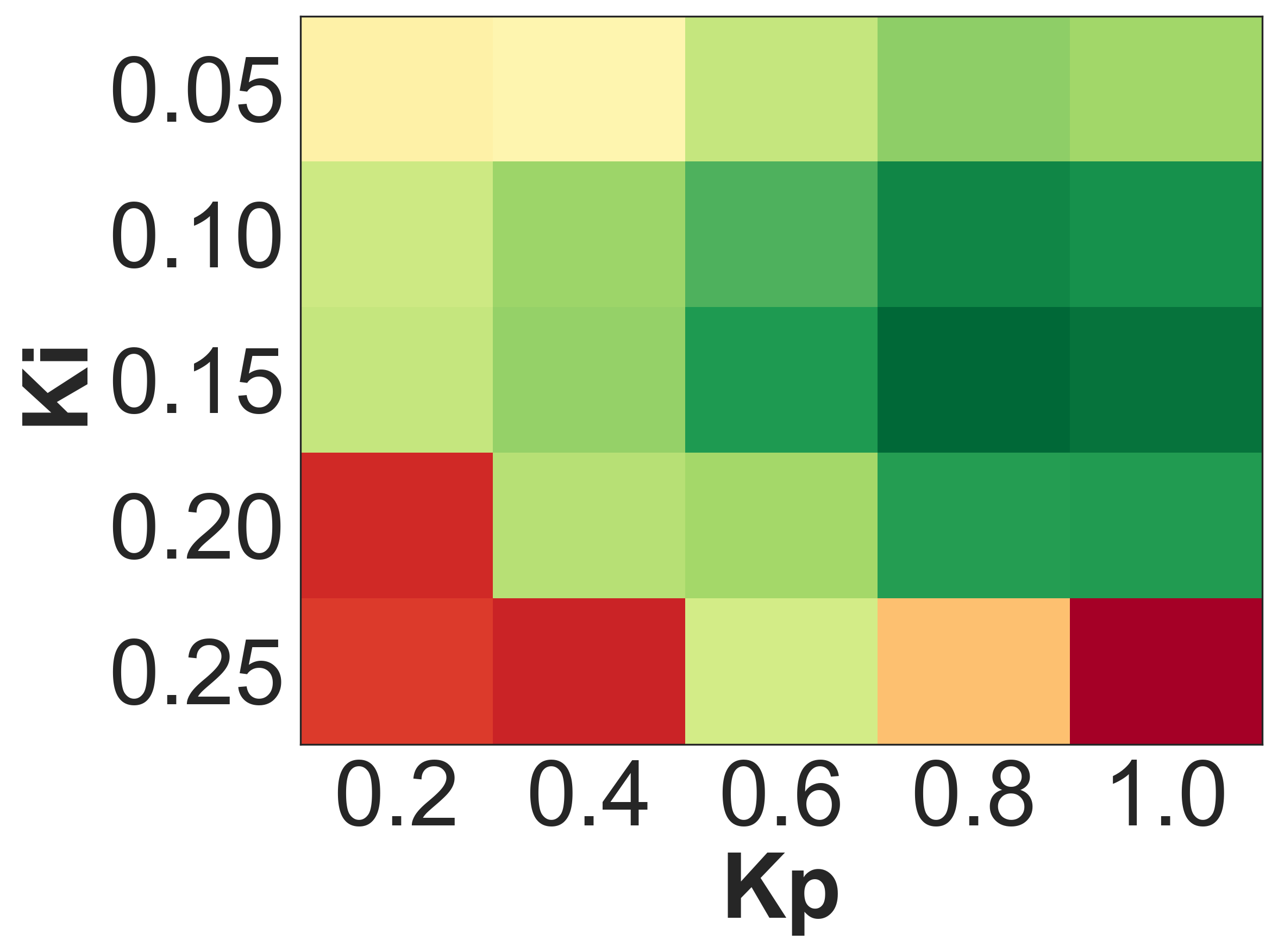}
    \label{fig:heatmap_kd02_lat}
\end{subfigure}%
\hspace{-1mm}%
\begin{subfigure}{0.16\columnwidth}
    \centering
    \includegraphics[width=\textwidth]{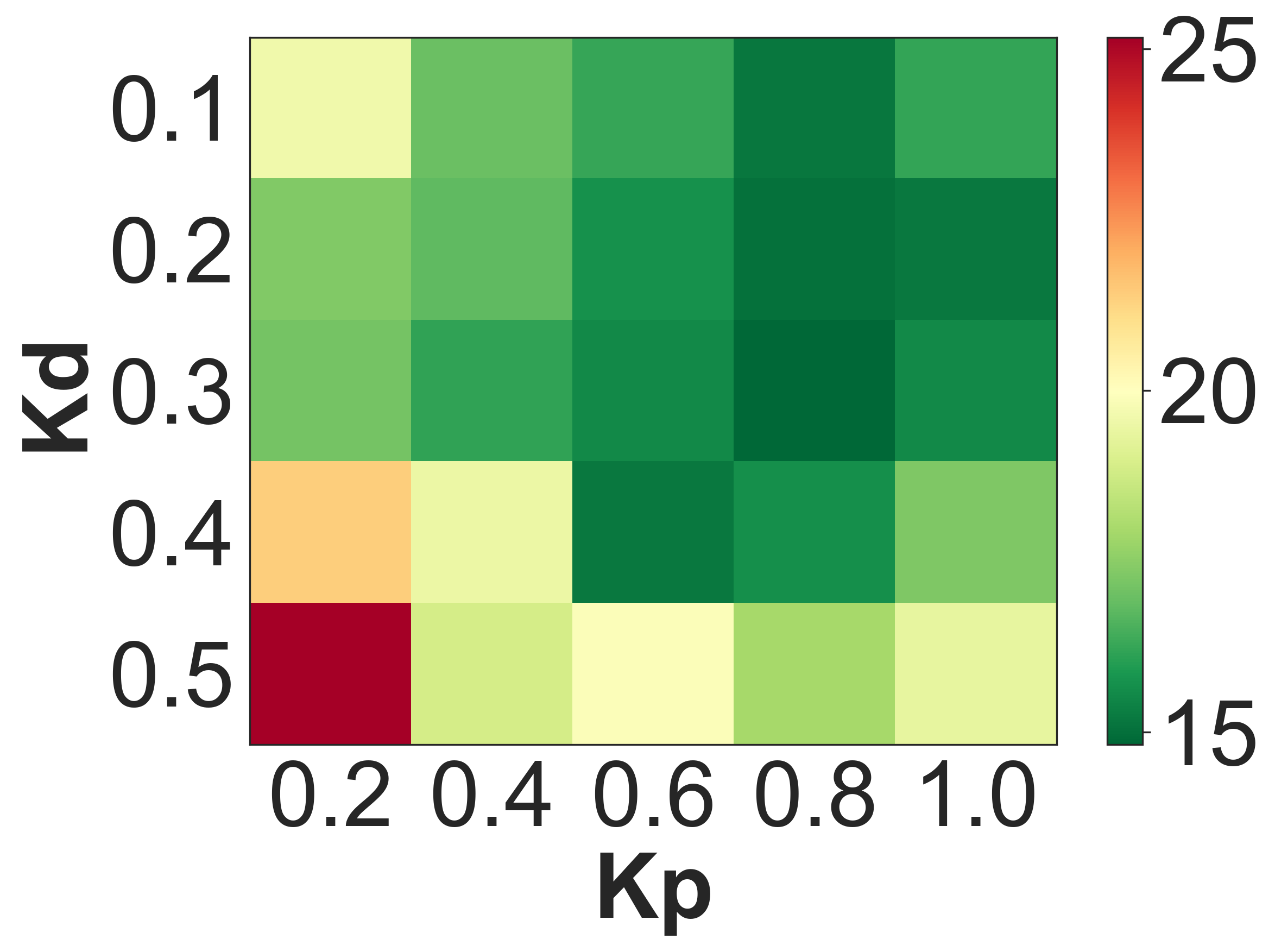}
    \label{fig:heatmap_ki015_lat}
\end{subfigure}%

\vspace{-5mm}
\caption{PID scheduler parameter grid search results arranged left to right. For each parameter setting, left shows tracking error (cm) and right shows latency (ms). The three pairs from left to right fix $K_p=0.8$, $K_i=0.15$, $K_d=0.2$ respectively, representing parameters that optimize scheduling by balancing emphasis on current, accumulated, and projected error.}
\label{fig:pid_gridsearch}
\end{figure}

The parameters of the PID scheduler are determined by conducting a grid search on a separate dataset that is collected in the same environment as the experiments, but not used for evaluating the system. The optimal combination of parameters is selected based on latency and accuracy, as demonstrated in Figure~\ref{fig:pid_gridsearch}. The grid search results reveal the following insights: (1) The proportional term is dominant, suggesting that the scheduler prioritizes immediate response to current QoE fluctuations. Large proportional gain ensures rapid correction of instantaneous QoE drops such as sudden movements, which smaller values could overlook. (2) The integral term is small but non-negligible, indicating that the scheduler accounts for accumulated effects over time, such as map aging. The presence of a modest, non-zero integral term is consistent with the memory characteristics of SLAM systems, while refraining from excessive corrective actions on historical errors, such as drifts. (3) The derivative term is present at a moderate level, indicating that the scheduler has predictive responsiveness to emerging QoE trends caused by increased latency or decreased accuracy. A high derivative gain would make the scheduler excessively sensitive and cause instability in response to sensor noise.

\subsection{Visual overlap calculation methodology}
\label{sec_system_overlap}
The deployment characteristics of \SysName in small-scale indoor environments for human-robot collaboration introduce specific optimization opportunities for map merging operations.

\subsubsection{Pose-based Overlap Volume Calculation}
Traditional map merge triggering mechanisms rely on counting keypoints in overlapped views, which requires computation using two frames and introduces complexity. We employ a calculation scheme that uses only the 6-DOF poses of two cameras, enabling rapid overlap assessment with limited information.

The camera frustum model \cite{frustum1,frustum2} represents the visible volume for each agent as a truncated pyramid defined by horizontal field of view $\theta_h$, vertical field of view $\theta_v$, near clipping distance $d_{near}$, and far clipping distance $d_{far}$. Agent poses are represented as 6-DOF transformations combining 3D position vectors $\mathbf{p} = [x, y, z]^T$ and orientation quaternions $\mathbf{q} = [q_x, q_y, q_z, q_w]^T$, which produce a 4×4 homogeneous transformation matrix $\mathbf{T}$ encoding the complete spatial relationship. After that, point-in-frustum testing determines whether arbitrary 3D points lie within the visible volume of a given camera. For a candidate point $\mathbf{x} = [x, y, z]^T$ and camera transformation $\mathbf{T}$, the point is transformed to the camera coordinate system as $\mathbf{x}_{cam} = \mathbf{T}^{-1} [\mathbf{x}^T, 1]^T$. The transformed point is then evaluated against frustum constraints:
\begin{equation}
\arctan\left(\frac{|x_{\text{cam}}|}{|z_{\text{cam}}|}\right) \leq \frac{\theta_{\text{h}}}{2}, \quad
\arctan\left(\frac{|y_{\text{cam}}|}{|z_{\text{cam}}|}\right) \leq \frac{\theta_{\text{v}}}{2}.
\end{equation}

Volume discretization employs regular 3D grid sampling to approximate the continuous overlap volume through discrete point counting. The sampling domain is established by computing a bounding box encompassing all agent positions with expansion margins based on maximum camera range $d_{far}$. For agents with poses $\{\mathbf{p}_1, \mathbf{p}_2, \ldots, \mathbf{p}_n\}$, the sampling space is $\mathcal{S} = \{(x, y, z) : (x,y,z) \in [\mathbf{p}_{min}, \mathbf{p}_{max}]\}$, where $\mathbf{p}_{min} = \min_i(\mathbf{p}_i)$ and $\mathbf{p}_{max} = \max_i(\mathbf{p}_i)$, sampled at intervals of $\delta$. For cameras $i$ and $j$, let $\mathcal{V}_i$ and $\mathcal{V}_j$ denote the sets of points visible to each camera. The overlap volume is approximated by sampling points visible to both cameras and scaling by the volume element $
V_{overlap} = |\mathcal{V}_i \cap \mathcal{V}_j| \cdot \delta^3.$

The intersection-over-union ratio quantifies the relative overlap magnitude as $R_{IoU} = V_{overlap}/V_{union}$, where $V_{union} = |\mathcal{V}_i \cup \mathcal{V}_j| \cdot \delta^3$. This pose-based overlap calculation enables efficient spatial redundancy assessment without requiring explicit map reconstruction, facilitating real-time overlap monitoring during multi-agent SLAM operations. The discretization resolution $\delta$ provides a trade-off between computational efficiency and volume estimation accuracy.

\begin{figure}[t]
    \centering
    \begin{subfigure}{0.48\textwidth}
        \centering  
        \includegraphics[width=\linewidth]{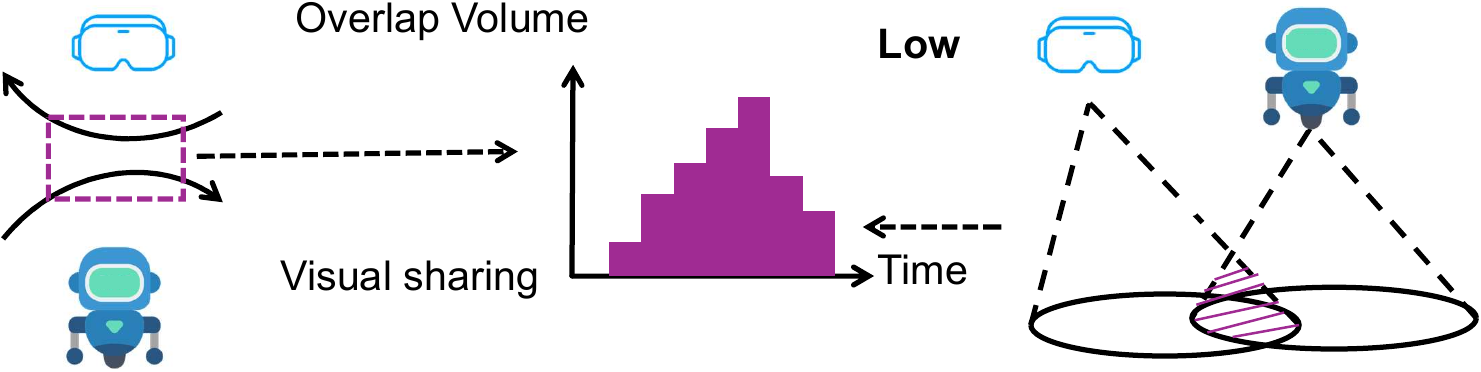}
        \caption{Conventional map merging.}
        \label{fig:map-merge-1} 
    \end{subfigure}
    \hfill
    \begin{subfigure}{0.48\textwidth}
        \centering
        \includegraphics[width=\linewidth]{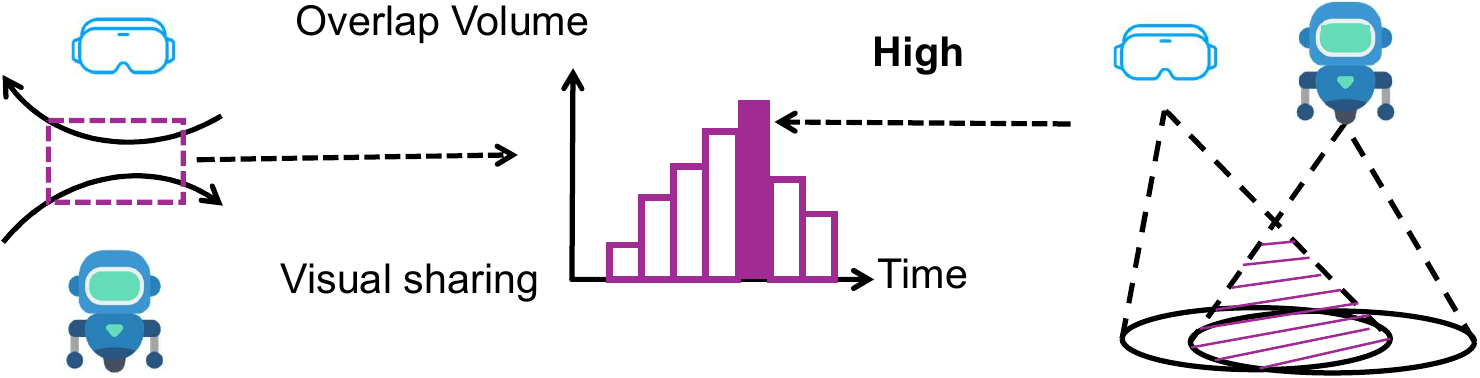}
        \caption{Proposed map merging.}
        \label{fig:map-merge-2} 
    \end{subfigure}
    \vspace{-4mm}
    \caption{Map merging redundancy removal. Map merging is triggered when agents share visual information. The conventional method incurs repeated map merging (left), while our method only merges on high overlap, achieving reduced redundancy.}
    \label{fig:map-merge}
    \vspace{-3mm}
\end{figure}

\subsubsection{Spatial Redundancy Reduction}
To address computational inefficiency in constrained indoor environments, we propose a selective map merging framework based on overlap volume analysis that distinguishes between essential and redundant merge operations, as suggested in Figure~\ref{fig:map-merge}. Map merges are conducted to align maps of multiple agents and provide inter-agent loop closure, which eliminates drift from local tracking. Therefore, when drifts are not significant, map merges contribute limited accuracy improvements and may be skipped to reduce latency. The overlap volume calculation quantifies the spatial intersection between agents as a geometric measure of information redundancy. 

Map merge is triggered when agents approach each other and their cameras' fields of view intersect. If the approach continues, map merges are repeatedly triggered, and if they separate symmetrically to the trajectory they approach, the same map merges occur again, resulting in increased redundancy. As shown in Figure~\ref{fig:map-merge}, overlap volume transitions from increasing to decreasing during this process, with the higher volume indicating the closer distance between agents, hence suggesting more shared visual features for better map merging, and should be less likely to be skipped. This is achieved through probabilistic thresholding on overlap volume, assigning lower probability of skipping map merges with closer agent distances, which are equally computationally heavy but more informative than their counterparts.

To implement this strategy, a buffer maintains overlap volumes from completed merge operations, denoted as $\{V_1, V_2, \ldots, V_N\}$, with historical mean overlap volume computed as $\bar{V} = \frac{1}{N} \sum_{i=1}^{N} V_i$. When a new merge request is generated with overlap volume $V_{new}$, the system computes a skip probability that reflects the likelihood of the request representing a local optimum for merging. The skip probability $P_{skip}$ is defined as a linear function of the ratio between the merge volume $V_{new}$ and the historical mean:
\begin{equation}
P_{skip} = \max\left(0, 1 - \alpha_{m} \cdot \frac{V_{new}}{\bar{V}}\right),
\end{equation}
where $\alpha_{m}$ is a scaling parameter that controls the sensitivity of the redundancy removal mechanism and $\bar{V}$ is the mean volume in the buffer. This probabilistic approach enables the system to maintain map accuracy while reducing computational load in scenarios where agents frequently encounter overlapping regions.

\section{Implementation}
\label{sec_implementation}

We implement \SysName with 3 components: two AR user agents, a
robot agent, and an edge server built upon the COVINS-G multi-agent SLAM framework \cite{9585827}, which follows a front-end to back-end structure. The frontend is implemented on all agents. It consists of sensing, the communication scheme of COVINS-G, and a pose estimation module. The pose estimation module is \editting{custom-}designed by combining ORB-SLAM3 Visual-Inertial Odometry~\cite{Campos2021}, IMU-PARSAC algorithm~\cite{rdvio}, and Extended Kalman Filter~\cite{ekf} to enhance its robustness in feature extraction. The backend is implemented on the server. Additionally, we incorporate our QoE modeling and PID scheduling in Python, visual overlap calculation and map merging in C++, on both agents and the server side. 
The DNN evaluator in each agent is trained using pySLAM ~\cite{Freda2025pySLAM} and SenseTime visual SLAM dataset \cite{sensetimedataset}.

\begin{figure}[t]
\centering
\begin{minipage}[b]{0.45\columnwidth}
    \centering
    \begin{subfigure}[c]{.55\linewidth}
       \includegraphics[width=\textwidth]{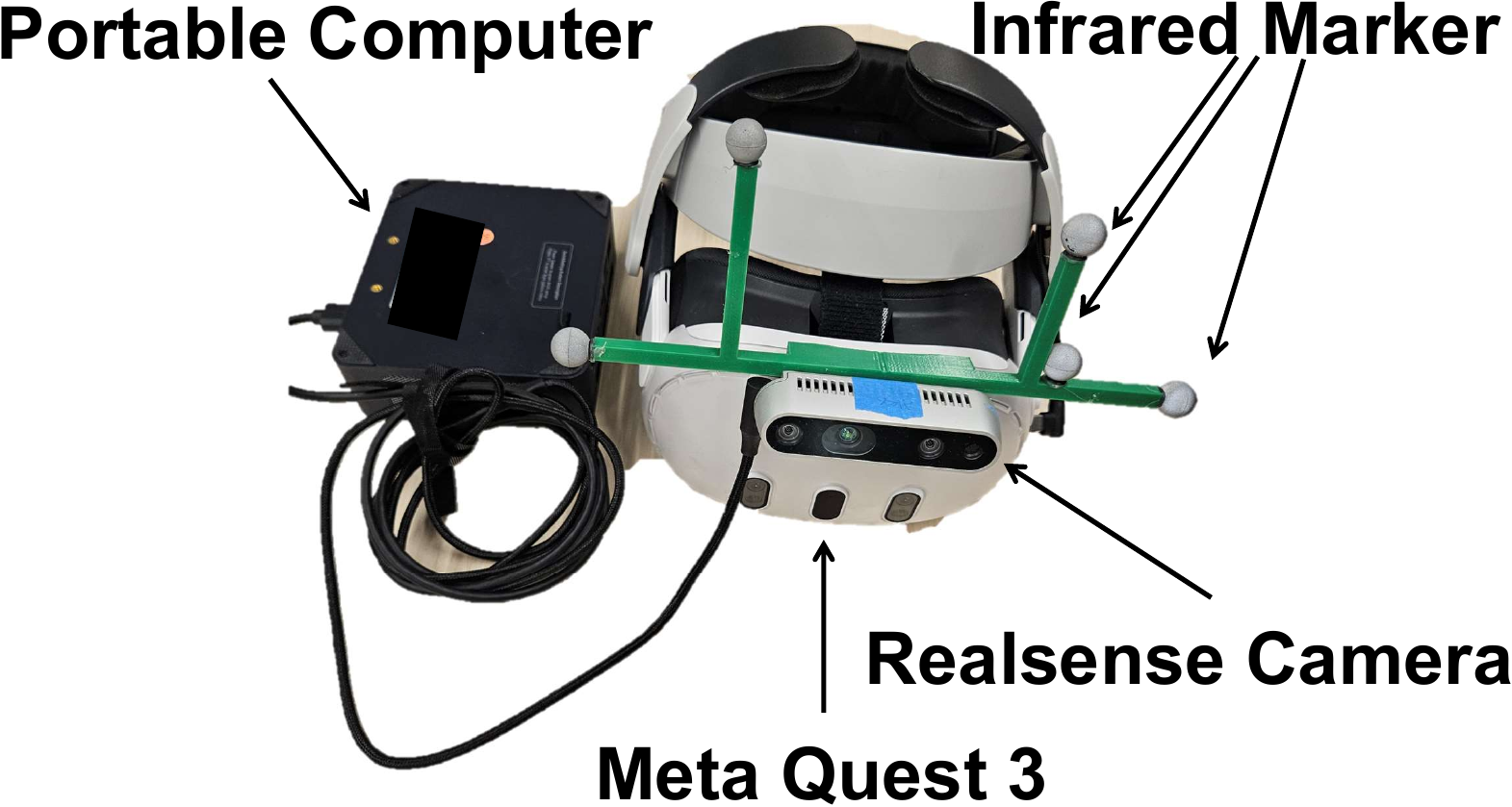}
    \subcaption{AR headset (Meta Quest 3) and added camera.}
    \label{headset-pic} 
    \end{subfigure}
    \hfill
    \begin{subfigure}[c]{.4\linewidth}
          \includegraphics[width=\textwidth]{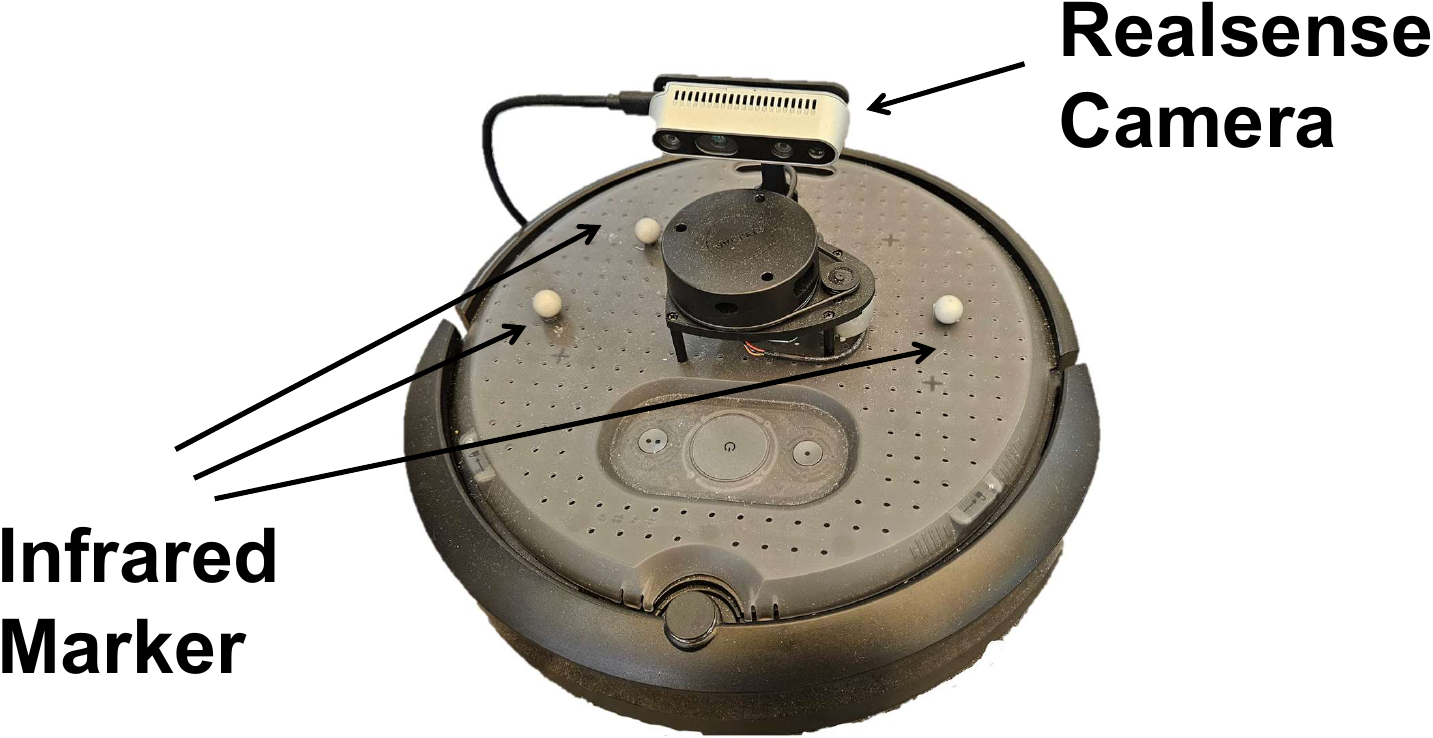}
        \subcaption{Turtlebot 4.}
        \label{robot-pic} 
    \end{subfigure}
    \vspace{-3mm}
    \caption{Devices used in this study.}
    \label{fig_headset_robot}
\end{minipage}%
\hfill
\begin{minipage}[b]{0.55\columnwidth}
    \centering
    \includegraphics[width=\textwidth]{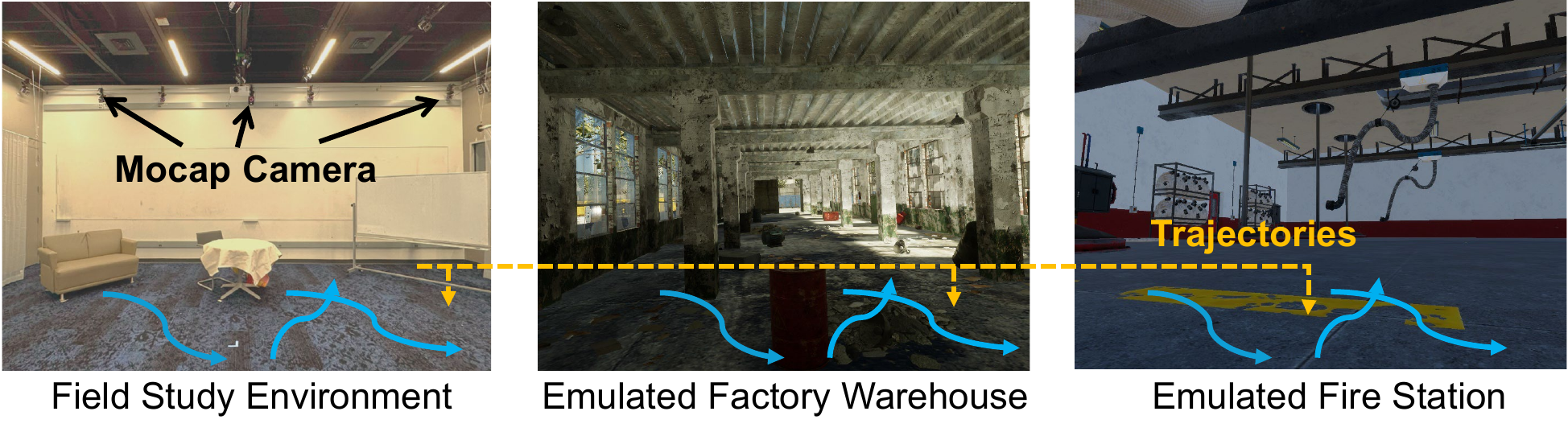}
    \vspace{-3mm}
    \captionof{figure}{Experiment environments used to construct datasets using same trajectories collected from field study.}
    \label{fig:emulator_scenes}
\end{minipage}
\vspace{-5mm}
\end{figure}

\begin{figure}[t]
    \includegraphics[width=.6\columnwidth]{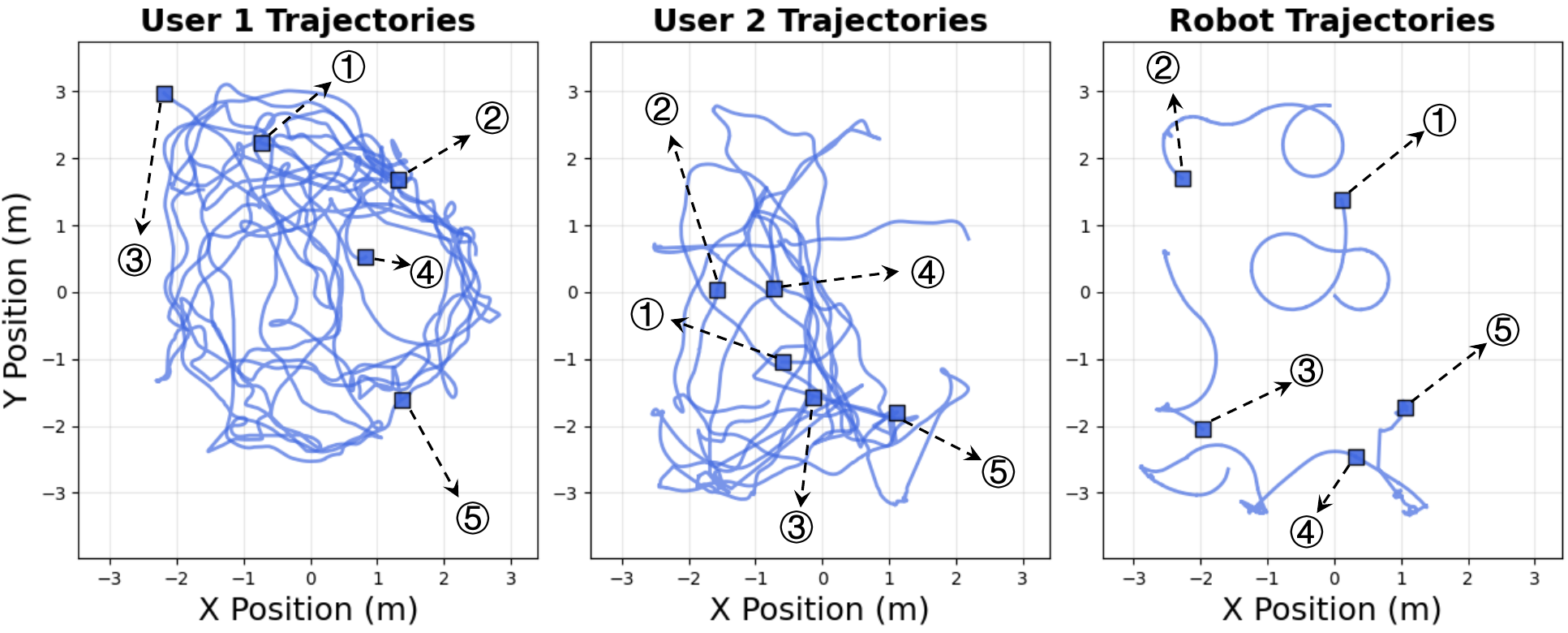}
    \vspace{-3mm}
    \caption{Demonstration of trajectories across all agents. For User 1, total length of trajectories \textcircled{1} - \textcircled{5} are 82.56 m, 33.39 m, 28.48 m, 54.02 m, 37.67 m. For User 2, total length of trajectories \textcircled{1} - \textcircled{5} are 17.21 m, 43.87 m, 35.66 m, 40.77 m, 34.17 m. For Robot, total length of trajectories \textcircled{1} - \textcircled{5} are 10.06 m, 10.88 m, 8.16 m, 23.91 m, 23.09 m.}
    \label{fig:traj-gt}
    \vspace{-5mm}
\end{figure}

\textbf{AR users.} 
We use the Meta Quest 3 as the AR headset worn by each AR user. To address the restriction on direct access to integrated cameras and IMU data of commercial AR headsets due to privacy considerations, we mount a RealSense D435i camera onto the Quest 3 and connect it to a BOSGAME Mini PC \cite{bosgame}, featuring an Intel N100 processor, as demonstrated in Figure~\ref{headset-pic}. The Intel N100 processor offers processing capabilities comparable to those of Meta Quest 3 with 4 cores and a 6 MB cache. \editting{The additionally mounted camera is placed to allow camera access on commercial, closed-source headsets, providing camera intrinsics necessary for feature matching in SLAM operations. This is a common approach in various prior works \cite{Drako2020,Drako2021,Tome2023,outcamera1,outcamera2}.}

\textbf{Robot.}
We use TurtleBot 4 (Figure~\ref{robot-pic}) as the robot collaborator in \SysName. 
We choose this robot for its wide applicability in HRC research~\cite{turtlebot1,turtlebot2,turtlebot3} in addition to its common application as a cleaning robot that applies to our use case.
Turtlebot 4 runs ROS 2 navigation components ~\cite{roshumble} based on the SLAM functionality we design and embed within an LXC subsystem running Ubuntu 20.04 to ensure compatibility with the backend.

\textbf{Edge server.}
For the edge server that hosts our backend processing and coordination of all agents' SLAM system, along with the real-time QoE values, PID scheduling and overlap calculations, we use a Lenovo ThinkPad T14 running on Ubuntu 20.04. It is equipped with an Intel i7 CPU with 4 cores and 8 threads, a maximum frequency of 4.9 GHz and an 8 MB cache. 

\textbf{Networking.}
All agents and the edge server are connected to the same Wi-Fi network operating on the 5.0 GHz frequency band. 
Temporal synchronization across multiple heterogeneous devices is critical for timing recovery and communication delay analysis. We implement the network time protocol over the Wi-Fi network, with the motion capture workstation serving as the authoritative time server \cite{ntp_pub}. This synchronization approach achieves millisecond-level timing accuracy through systematic calculation of clock offsets and round-trip delays, ensuring temporal alignment of sensor data streams across all participating agents.

\section{Evaluation}
\label{sec_evaluation}

\subsection{Experiment Setup}

\textbf{Dataset collected from field studies.} To evaluate \SysName, we collect real-world experimental data under controlled laboratory conditions. The data collection setup consists of two AR users equipped with Meta Quest~3 headsets and one TurtleBot~4 ground robot, all operating within a 9~m~$\times$~6~m~$\times$~4~m motion capture space instrumented with 22 ceiling-mounted Vicon cameras~\cite{Mocap}. We attach infrared reflective markers to all devices, enabling the collection of ground-truth pose measurements with millimeter-level accuracy for precise tracking evaluation. After synchronized pose sequence optimization~\cite{Hu2025XRRealityCheck}, we compute the rigid transformations required to map motion capture rigid body frames onto the device reference frames.

The experimental protocol requires simultaneous operation of all three agents, each starting from arbitrary initial positions and following unconstrained trajectories within the capture volume. The robot is teleoperated by an AR user (one of the agents) using wireless navigation commands, while both the operator and an additional human observer move freely within the environment. The motion capture system continuously records ground-truth pose trajectories for all agents. Additionally, sensor data and associated timestamps are logged from each device to evaluate both localization accuracy and system latency. We collect five distinct trajectory sequences, each involving all three agents operating concurrently within the motion capture space demonstrated in Figure~\ref{fig:emulator_scenes}. The proposed system operates at 30 frames per second, and for each trajectory, the agents move for 60 seconds. The total data collection yields 25,657 frames across all agents. Trajectories are demonstrated in Figure \ref{fig:traj-gt}. Each sequence represents a complete multi-agent SLAM session with varying trajectory characteristics and agent interaction patterns. The total lengths per sequence are summarized in Figure~\ref{fig:traj-gt}.

\textbf{Synthesized datasets.}
To test the robustness of \SysName
in new environments, the dataset collected solely from the lab environment is insufficient. However, there is a scarcity of large-scale visual SLAM datasets (with accurate
ground-truth pose data) that cover a diverse range of realistic, commonly encountered visual environments. To address this limitation, we employ an emulator-based
synthetic data generation approach \cite{AReval}, as illustrated in Figure \ref{fig:emulator_scenes}. This approach substitutes synthetic visual scenes for real camera imagery while preserving the original IMU data from the physical experiments within easily modifiable virtual environments. 
To produce realistic virtual environments for our synthetic datasets, we use the Unity game engine's High Definition Render Pipeline. We place virtual cameras in these environments, where the virtual camera intrinsics are set for simulation environments, referencing established prior works \cite{AReval,VREval2,VREval3}. The virtual cameras move along the same trajectories collected from the field study and collect visual frames in the same frame rate. Visual frames are then combined with real IMU information from the trajectory collection process to form the synthesized dataset. 

The virtual environments used in the evaluation are shown in Figure \ref{fig:emulator_scenes}. They are specifically designed to assess multi-agent SLAM application scenarios and to provide a diverse spectrum of visual features. For each environment, we construct a dataset equivalent in size to the data collected from the field study, containing the identical trajectories. The first dataset, \textit{Warehouse}, is based on an emulated factory warehouse environment. It includes lighting from multiple directions, pillars and walls with various features, and a sufficient number of real-world objects. This dataset aims to cover the usage of robots in construction and logistics support, as well as navigation in dynamically lit environments. The second dataset, \textit{Firestation}, is based on an emulated fire station facility. The environment is more spread out compared to the \textit{Warehouse} environment, with walls and floors featuring fewer visual cues, which makes feature extraction more challenging. This dataset is intended to model scenarios involving robots used in first-response operations.

\textbf{Goal and metrics.} The goal of the evaluation is to demonstrate that our proposed pipeline is able to achieve 1) low tracking latency that enables its compatibility with real-time AR, we use \emph{delay (ms)} as the metric; and 2) high accuracy for tracking.

As discussed in the prior sections, the latency of interest in AR is measured from a certain frame to the best estimate of pose from the frame, as demonstrated in Figure.\ref{fig:delay_latencyimpact}. This latency is calculated by averaging latency from local pose estimate of all frames and edge corrections from applicable frames. Regarding accuracy, we employ \emph{Absolute Trajectory Error (ATE)} and \emph{Relative Pose Error (RPE)}. ATE measures the global consistency of a trajectory by comparing the estimated trajectory against the ground-truth trajectory after aligning them in a common reference frame. RPE quantifies the local drift in between consecutive poses, measuring how much displacement between two estimated poses differs from the corresponding ground-truth displacement. RPE is calculated through root mean square error on both translation and rotation perspective or pose drifts.

\textbf{Baselines.} While several multi-agent SLAM systems \cite{Xu2022, Dhakal2022, dasari2023rovar} have been proposed, establishing a fair comparison with our proposed system presents challenges due to fundamental differences in system design, implementation availability, and hardware requirements. Both \emph{SwarmMap}~\cite{Xu2022} and \emph{SLAM-Share}~\cite{Dhakal2022} utilize powerful and expensive GPU acceleration (a 512-core Volta GPU costing \$1300 in \emph{SwarmMap}, and a Tesla V100 GPU costing \$450 in \emph{SLAM-Share}\editting{, priced at the time of submission of this paper}) for both front-end keypoint detection on mobile devices and global map optimization on edge servers. Due to the limited cost of the AR headset (e.g., Meta Quest 3 at \$500), such powerful GPU resources are not currently available in commercial AR headsets. Note that our developed \SysName framework could also benefit from GPU acceleration once such resources become available on headsets in the future. In addition, \emph{RoVAR}~\cite{dasari2023rovar} diverges significantly in its approach by fusing a new sensor modality, \editting{Radio-Frequency tracking with Ultra-wideband}, with the original visual output, which is beyond the scope of this work.

Given the aforementioned considerations, we compare the performance of the following three edge-assisted solutions: (1) \emph{Our \SysName}. (2) \emph{Streamed ORB-SLAM3:} implemented by enforcing a real-time stream of raw visual and inertial data to the server and running three separate single-agent ORB-SLAM3 instances, serving as a trivial baseline of tracking without multi-agent collaboration or edge offloading. (3) \emph{COVINS-G:} the COVINS-G system without modification \cite{10160938}, serving as the foundational reference implementation that provides standard multi-agent collaborative SLAM functionality without any adaptive resource management or agent-specific optimizations. COVINS-G is used because it is an open-source, state-of-the-art, centralized system.

\vspace{-2mm}
\subsection{Results Analysis}
\label{sec_performance_field}
We evaluate the performance of \SysName using data collected from both field studies and emulated environments.

\begin{figure}[t]
	\centering
	\subcaptionbox{Field study users.\label{fig:lat_field_users}}[0.32\textwidth]{%
		\includegraphics[width=0.32\textwidth]{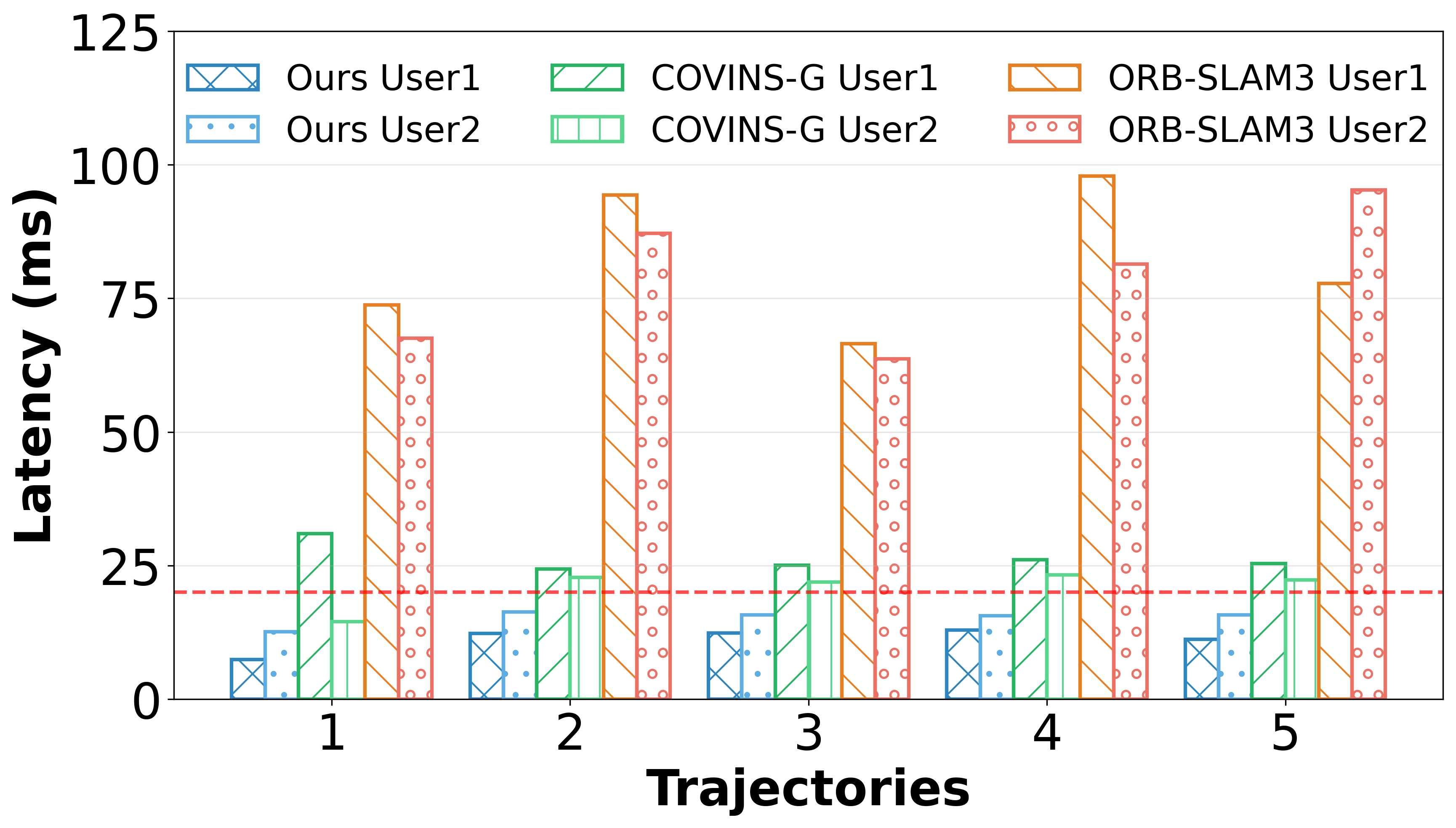}%
	}%
	\hfill%
	\subcaptionbox{Warehouse (Emulation) users.\label{fig:lat_warehouse_users}}[0.32\textwidth]{%
		\includegraphics[width=0.32\textwidth]{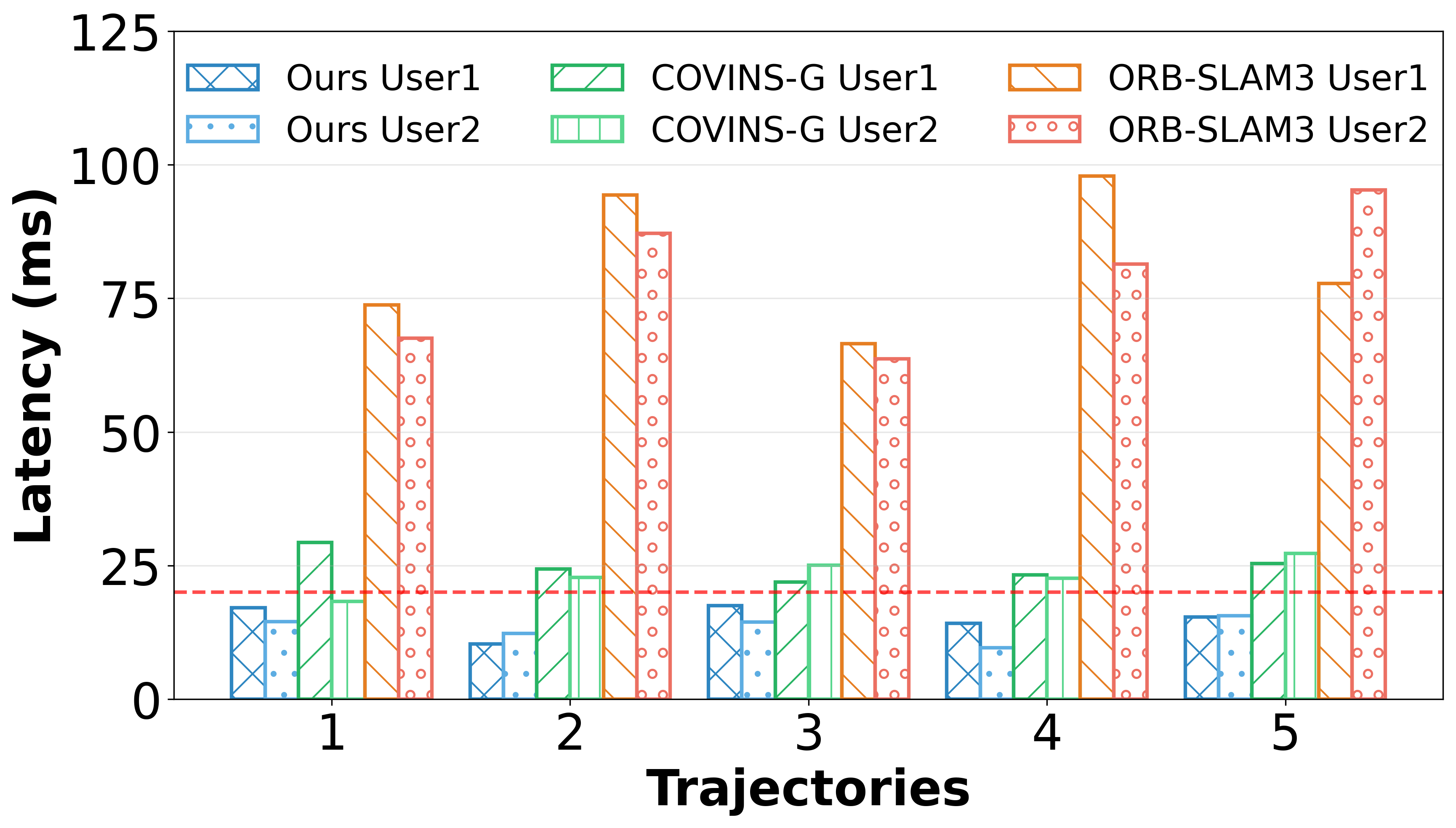}%
	}%
	\hfill%
	\subcaptionbox{Firestation (Emulation) users.\label{fig:lat_firestation_users}}[0.32\textwidth]{%
		\includegraphics[width=0.32\textwidth]{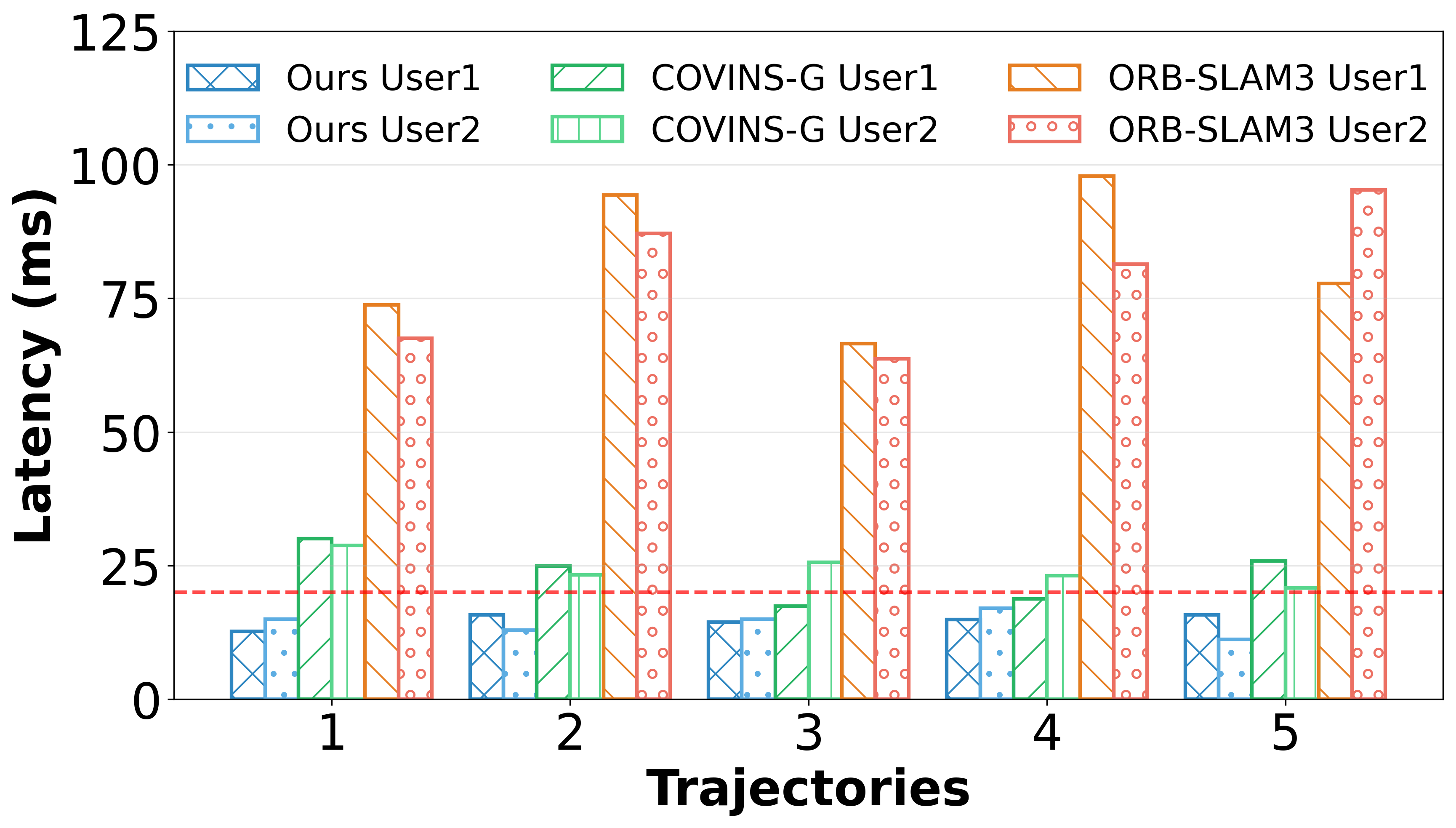}%
	}
	
	\medskip%
	
	\subcaptionbox{Field study robot.\label{fig:lat_field_robot}}[0.32\textwidth]{%
		\includegraphics[width=0.32\textwidth]{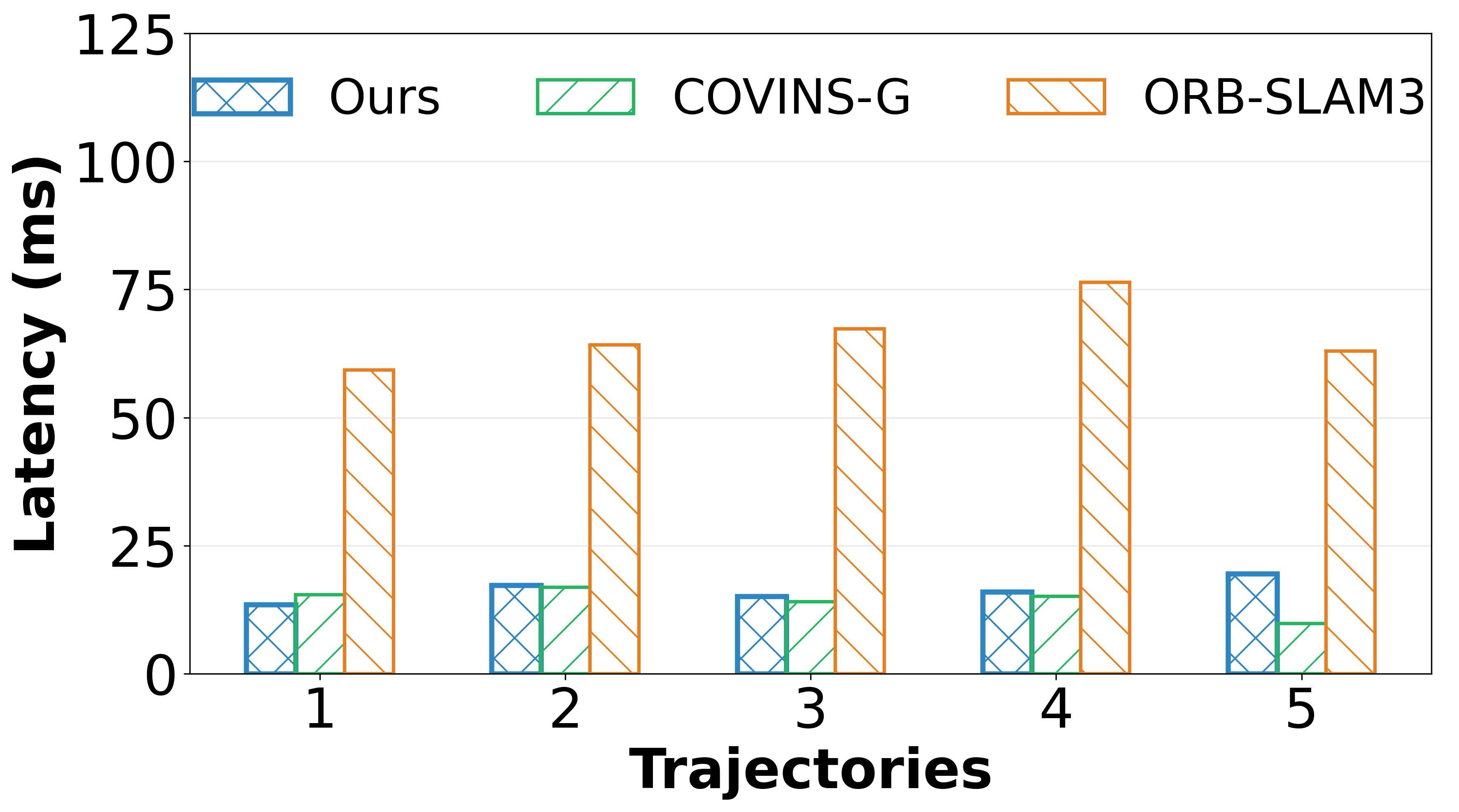}%
	}%
	\hfill%
	\subcaptionbox{Warehouse (Emulation) robot.\label{fig:lat_warehouse_robot}}[0.32\textwidth]{%
		\includegraphics[width=0.32\textwidth]{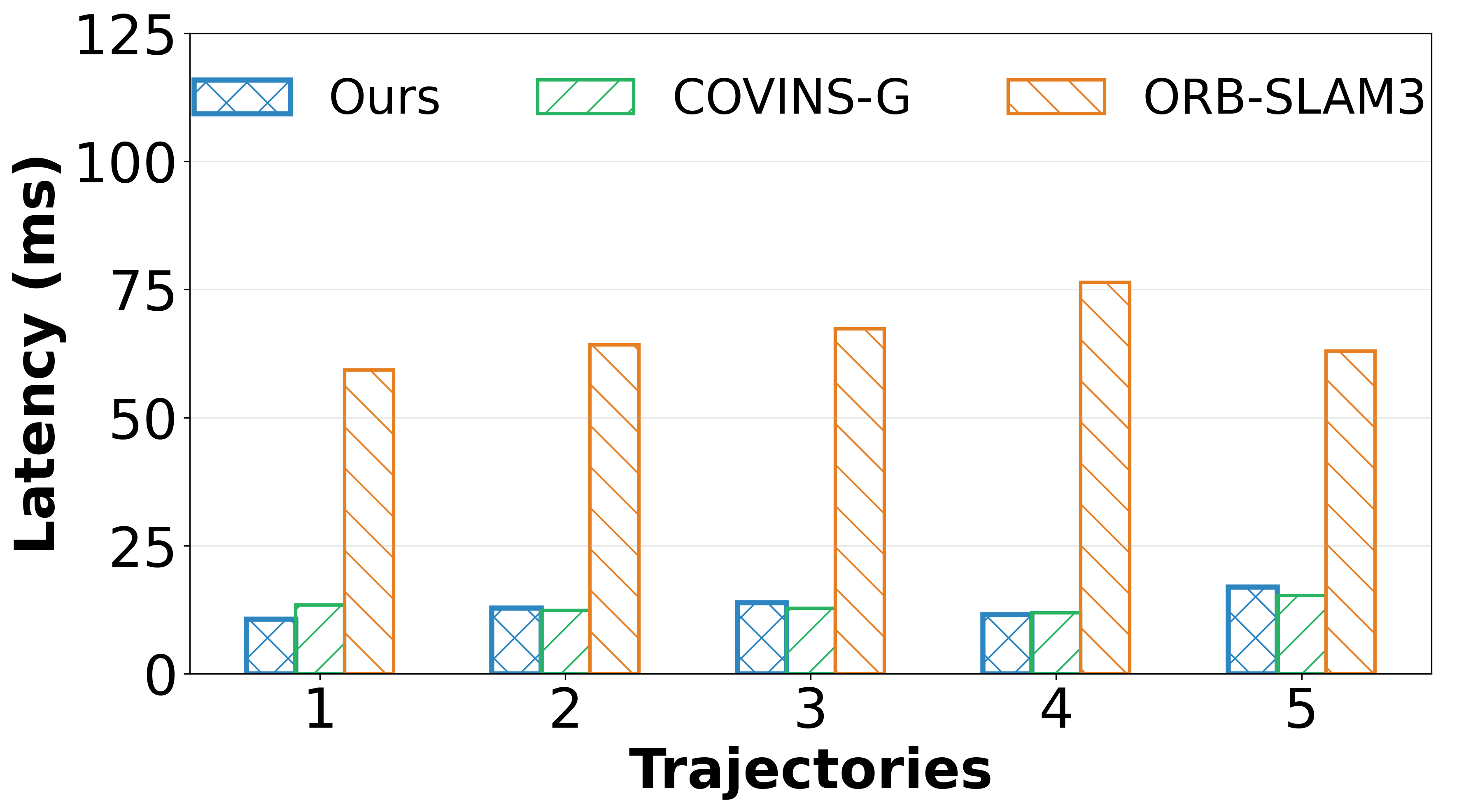}%
	}%
	\hfill%
	\subcaptionbox{Firestation (Emulation) robot.\label{fig:lat_firestation_robot}}[0.32\textwidth]{%
		\includegraphics[width=0.32\textwidth]{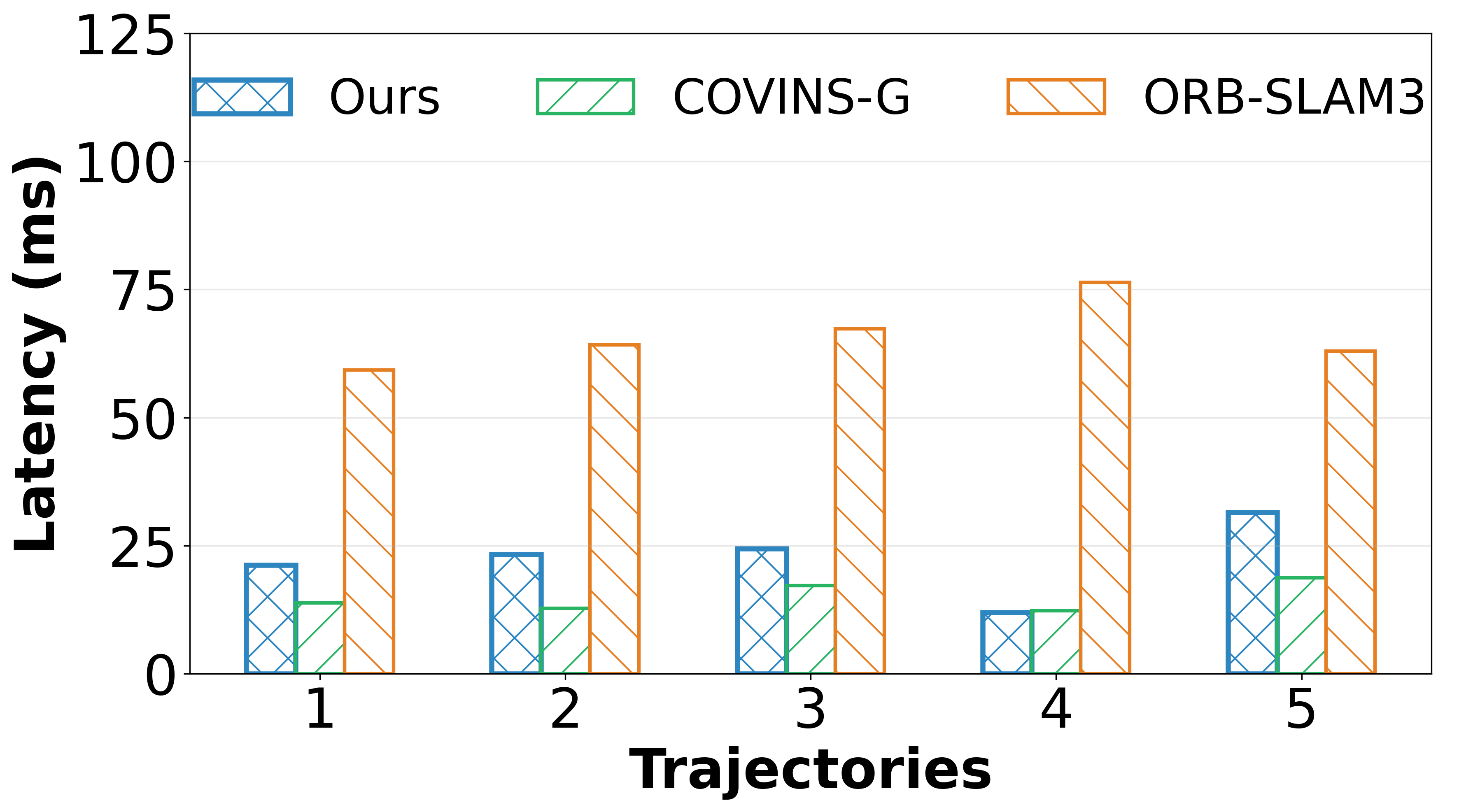}%
	}
	
	\caption{Latency comparison between \SysName, COVINS-G, and ORB-SLAM3 across three environments. Red line marks 20 ms AR limit. Lower is better.}
	\label{fig:latency_full}
\end{figure}

\textbf{Latency Analysis.}
The latency results demonstrate consistent performance improvements across the field study dataset and the synthesized datasets, as shown in Figure~\ref{fig:latency_full}. In the field study, \SysName maintains latency under the 20 ms AR threshold across all trajectories and human users, averaging 13.22~ms and achieving a 43.3\% reduction compared to COVINS-G. In contrast, COVINS-G exceeds this threshold in all cases except for user 2 in trajectory \textcircled{1}, indicating its limited suitability for AR applications. \editting{Both user 1 and user 2 maintain consistent sub-threshold latency under \SysName across all five trajectories despite their differing trajectory lengths, demonstrating that the PID scheduler effectively adapts resource allocation to sustain AR latency requirements regardless of the operational distance covered by individual users.} The synthesized datasets validate these findings: \SysName reduces latency by 43.1\% in the warehouse scene and 39.4\% in the fire station scene, consistently maintaining performance below 20~ms, yet COVINS-G is only able to achieve this threshold for user 2 under the trajectory \textcircled{3}, rendering it insufficient for the latency requirement.

For the robot, \SysName exhibits marginally higher latency in most cases, with increases below 10\% across field study trajectories (except trajectory~\textcircled{5}). In the emulated environments, this pattern persists, where robot cameras positioned lower in the virtual scene encounter incorrectly rendered virtual objects at scene joints and corners, contributing to modest latency variations. These results validate the system's strategic communication resource allocation, prioritizing AR latency requirements while maintaining robot latency around 25~ms, which is acceptable for most robot tasks. The variability in latency trends across the same trajectories over the three distinct datasets indicates that visual environment characteristics influence tracking performance. \editting{Overall, \SysName's gains over COVINS-G are most pronounced under operationally realistic conditions with longer trajectories, feature-sparse environments, and sustained multi-agent interaction, which are precisely the conditions that characterize target deployment scenarios.}

In the cases of both human users and robots, \SysName and COVINS-G both significantly outperformed ORB-SLAM3 with raw frame streaming. This reflects the efficiency advantage of strategic offloading on communication and computation. Importantly, the system attains the aforementioned low latency despite relying on a modestly provisioned edge server with only 4 cores, as described in Section~\ref{sec_implementation}. This result underscores the efficiency and robustness of \SysName.

\textbf{Accuracy Analysis.}
The accuracy metrics demonstrate consistent performance across all datasets, as shown in Figure~\ref{fig:ate_full}. In line with the latency results, both \SysName and COVINS-G outperform ORB-SLAM3 with raw frame streaming, reflecting that collaborative map construction contributes to significant improvement in SLAM accuracy. Comparing the two, \SysName achieves errors below 2~cm for both users and robots in all cases. In the field study, \SysName outperforms COVINS-G in 13 out of 15 trajectory-agent combinations, achieving the lowest ATE on trajectories \textcircled{1} through \textcircled{4} for human users and trajectories \textcircled{2} through \textcircled{5} for robots. \editting{The limited cases where COVINS-G achieves lower ATE arise on short trajectories with restricted viewpoint diversity, where adaptive frame dropping occasionally discards keyframes carrying non-redundant geometric information; this effect diminishes as trajectory length grows and redundant observations accumulate.} The emulated datasets corroborate these findings. \editting{In the fire station scene, which contains fewer visual features, the PID scheduler compensates by increasing frame transmission rates, a behavior absent in COVINS-G, explaining \SysName's stronger relative accuracy gains there.} The consistency of ATE across field study and emulated environments, both maintaining sub-2~cm error, validates the system's stability across diverse settings. \editting{A concrete direction for improvement is to make the frame dropping policy more conservative in the early phase of short trajectories where keyframe diversity is most critical, transitioning to more aggressive dropping once sufficient map density has been established.}

\vspace{-2mm}
\subsection{Ablation Study}
\label{sec_ablation}
To evaluate the effectiveness of individual modules, we compare \SysName against two implementations tailored to isolate specific components:

\editting{
(1) \emph{COVINS-G and Linear Frame Dropping (Linear Baseline):} It extends the original COVINS-G system by incorporating a linear frame dropping mechanism designed to mitigate communication overhead under varying latency conditions. This approach follows
simple adaptive methodologies established in prior work~\cite{Damigos2024CommunicationAware} that do not consider agent heterogeneity. Specifically, no frames are dropped when the measured end-to-end latency $d$ is at or below the 20\,ms AR threshold $d_0$. When latency rises above this threshold, the drop rate increases proportionally as $(d - d_0) / (d_{\max} - d_0)$, reaching its maximum of 1.0 at $d_{\max} = 1{,}000$\,ms, and is applied uniformly across all agents without differentiation.
}
(2) \emph{\SysName with Equal QoE (Equal Baseline):} It implements equivalent QoE allocation across all participating agents, treating both human agents and robot agents with identical priority and resource allocation policies on the original COVINS-G system. This baseline specifically removes agent discrimination mechanisms in the original design of \SysName. 
(3) \emph{\SysName with Random PID Parameters (Random Baseline):} It demonstrates the system without tuning the PID scheduler, implemented by averaging performance from 10 randomly selected parameter configurations of the PID system. This baseline demonstrates the functionality of PID scheduling. The ablation study is based on the dataset collected from the field study.

\begin{figure*}[t]
	\centering
	\subcaptionbox{Field study users.\label{fig:ate_field_users}}[0.32\textwidth]{%
		\includegraphics[width=0.32\textwidth]{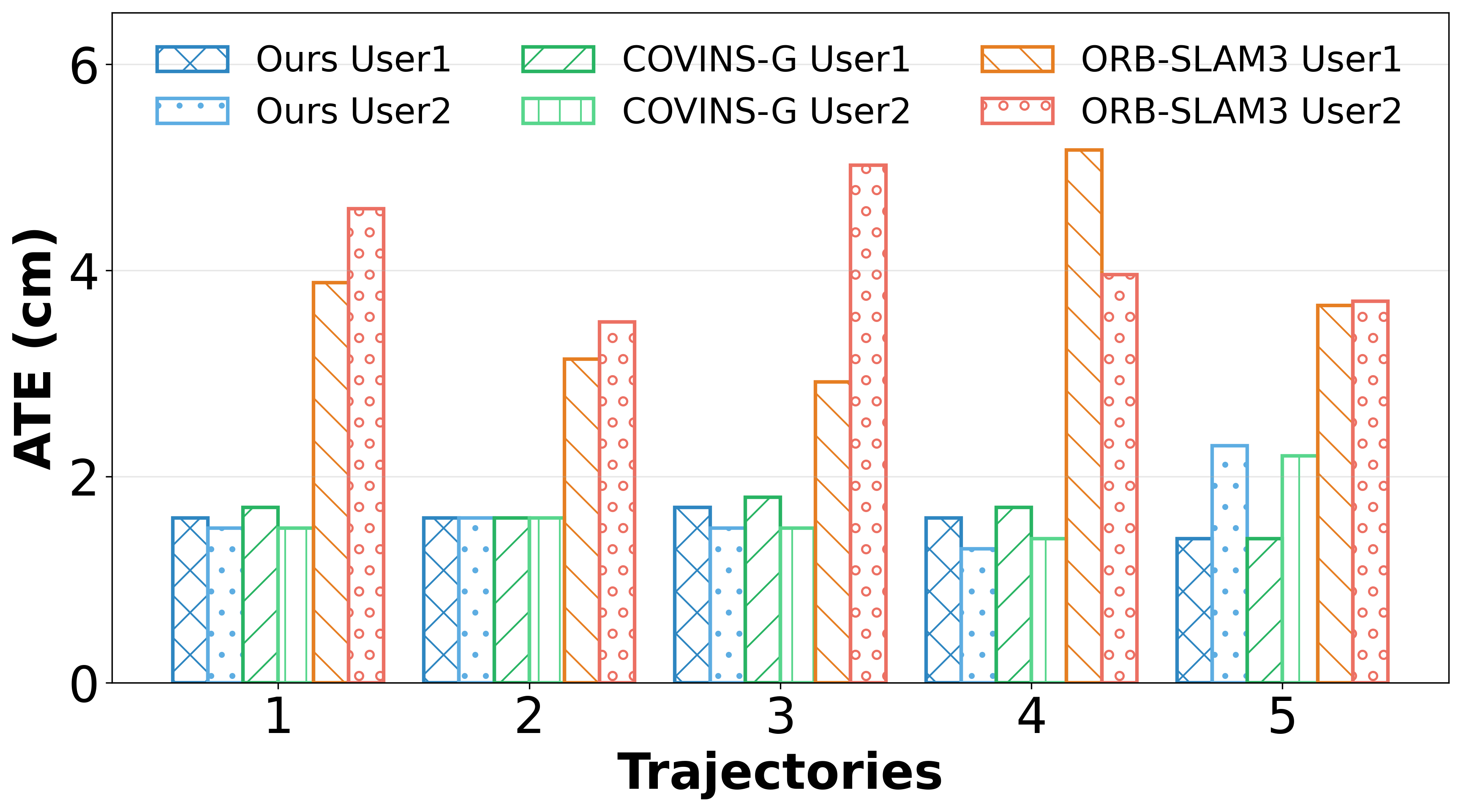}%
	}%
	\hfill%
	\subcaptionbox{Warehouse (Emulation) users.\label{fig:ate_warehouse_users}}[0.32\textwidth]{%
		\includegraphics[width=0.32\textwidth]{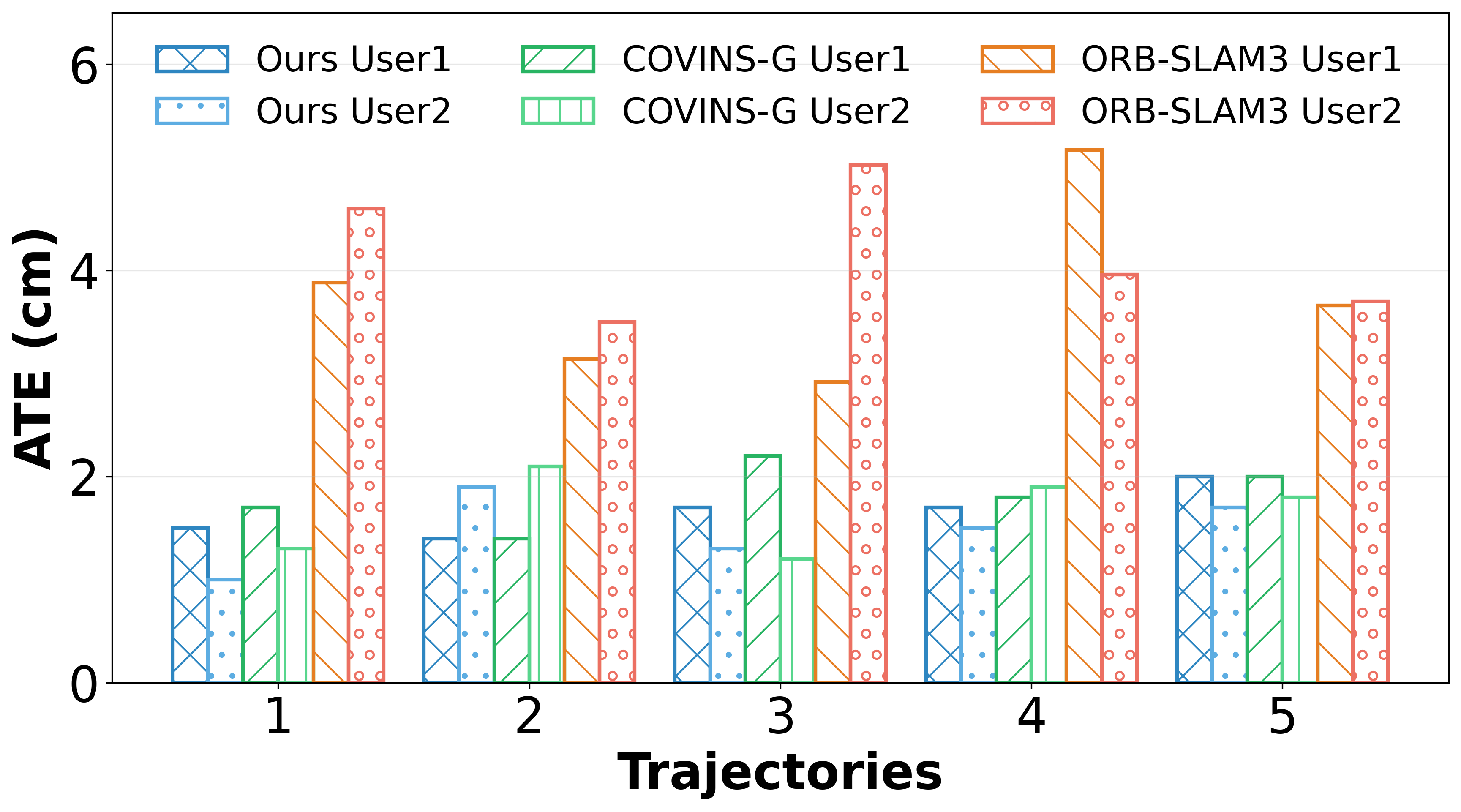}%
	}%
	\hfill%
	\subcaptionbox{Firestation (Emulation) users.\label{fig:ate_firestation_users}}[0.32\textwidth]{%
		\includegraphics[width=0.32\textwidth]{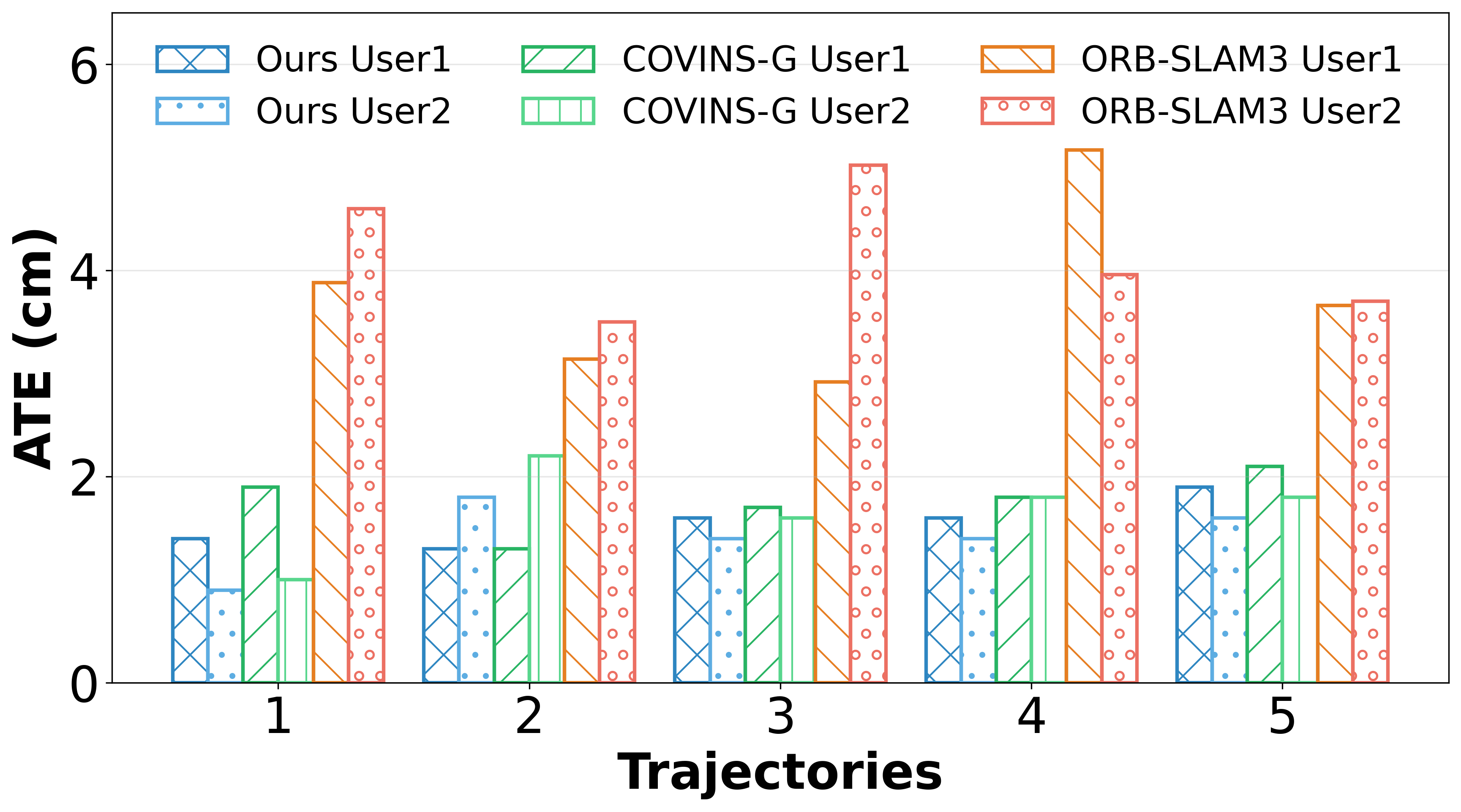}%
	}
	
	\medskip%
	
	\subcaptionbox{Field study robot.\label{fig:ate_field_robot}}[0.32\textwidth]{%
		\includegraphics[width=0.32\textwidth]{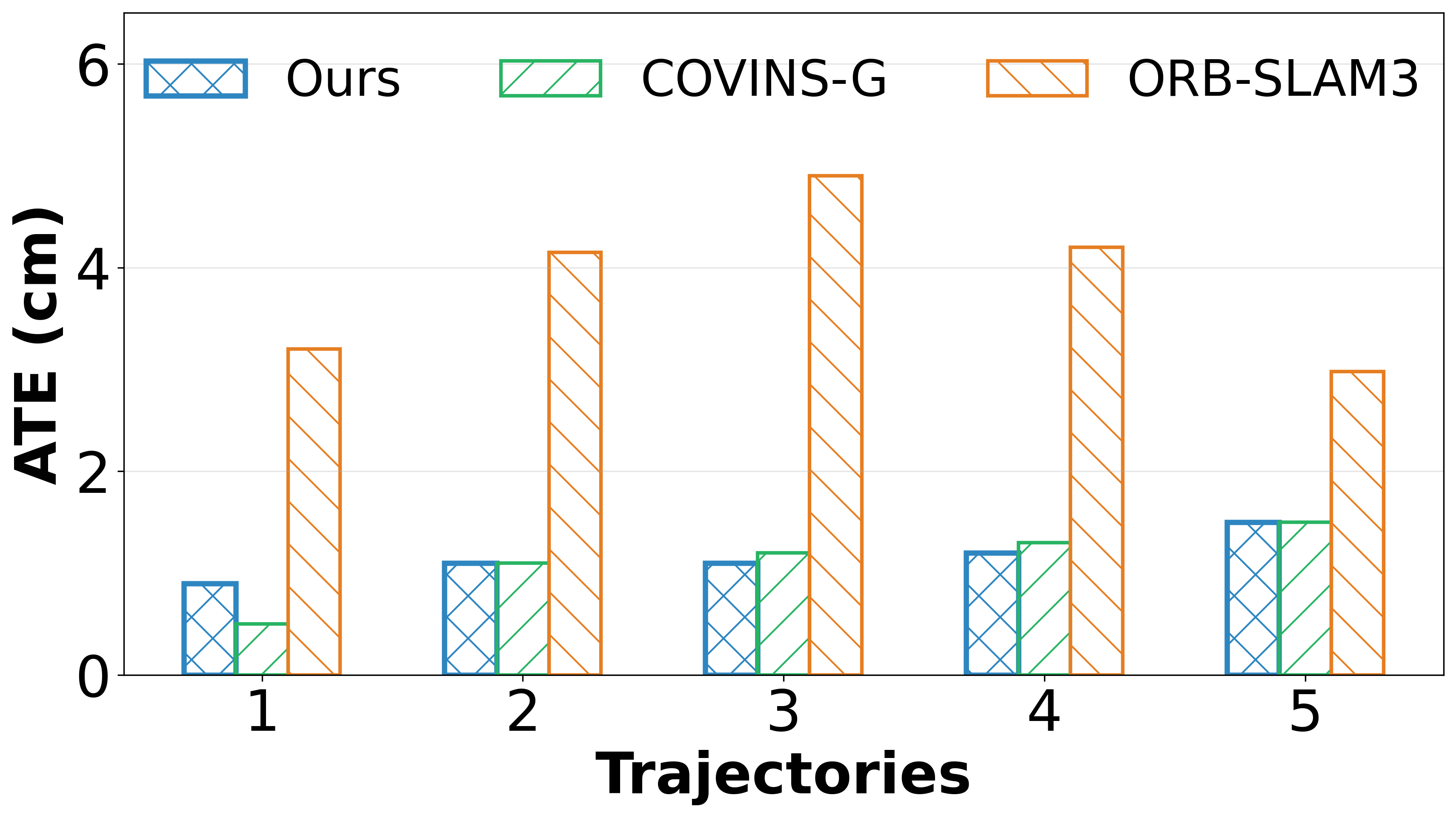}%
	}%
	\hfill%
	\subcaptionbox{Warehouse (Emulation) robot.\label{fig:ate_warehouse_robot}}[0.32\textwidth]{%
		\includegraphics[width=0.32\textwidth]{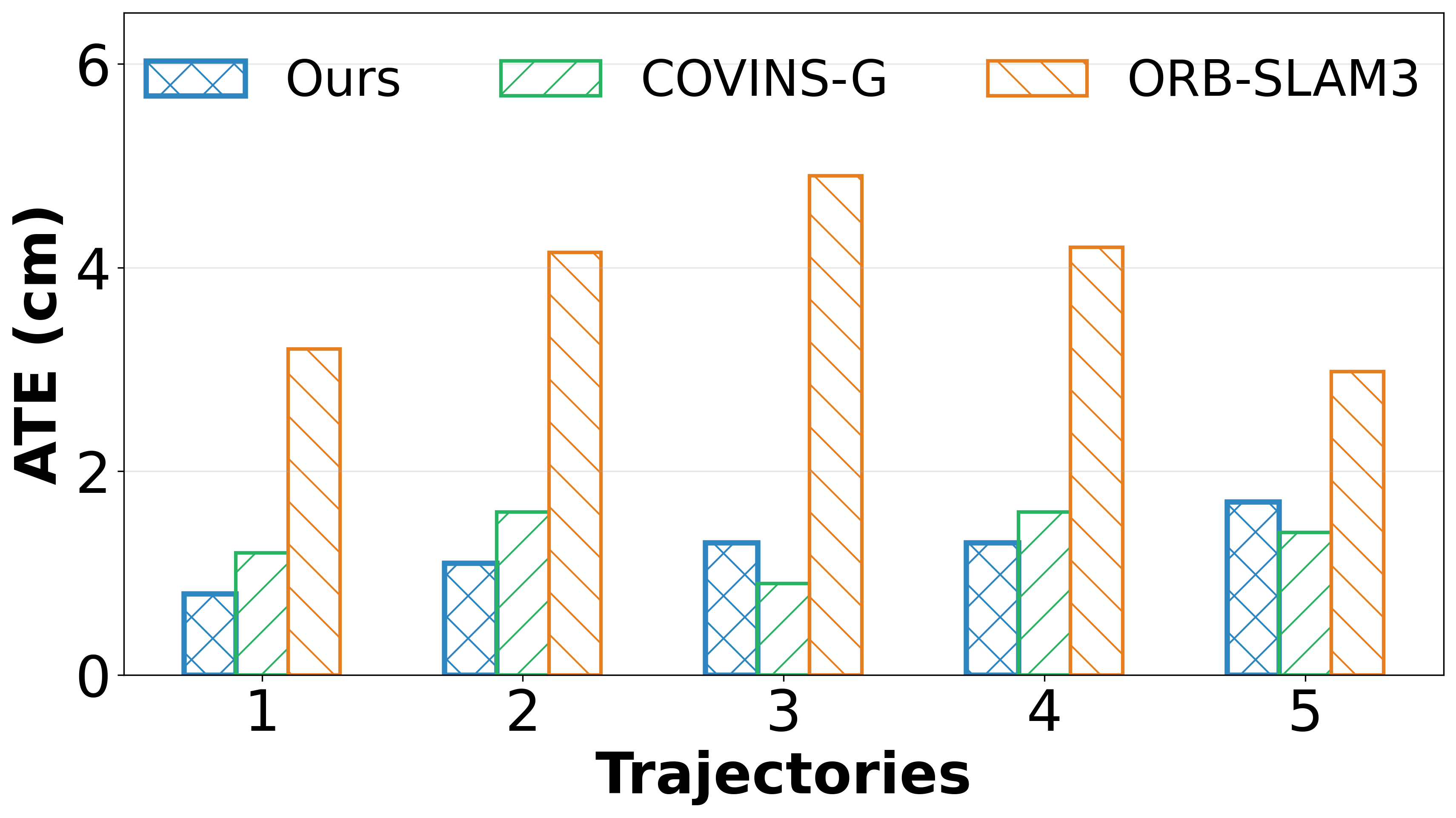}%
	}%
	\hfill%
	\subcaptionbox{Firestation (Emulation) robot.\label{fig:ate_firestation_robot}}[0.32\textwidth]{%
		\includegraphics[width=0.32\textwidth]{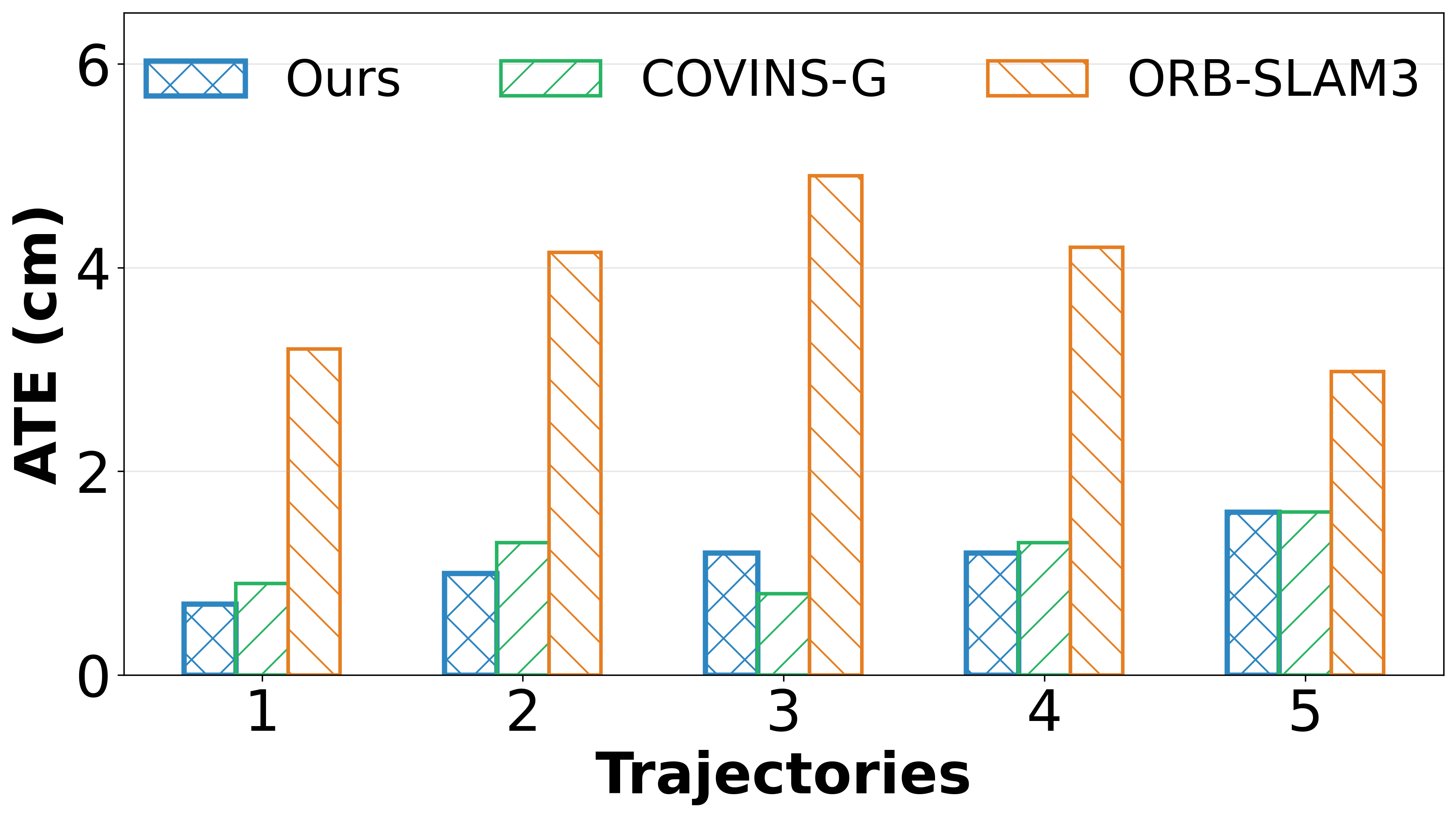}%
	}
	
	\caption{ATE comparison between \SysName, COVINS-G, and ORB-SLAM3 across three environments. Lower is better.}
	\label{fig:ate_full}
\end{figure*}

\begin{figure}[t]
    \begin{subfigure}{.48\columnwidth}
        \centering  
       \includegraphics[width=\columnwidth]{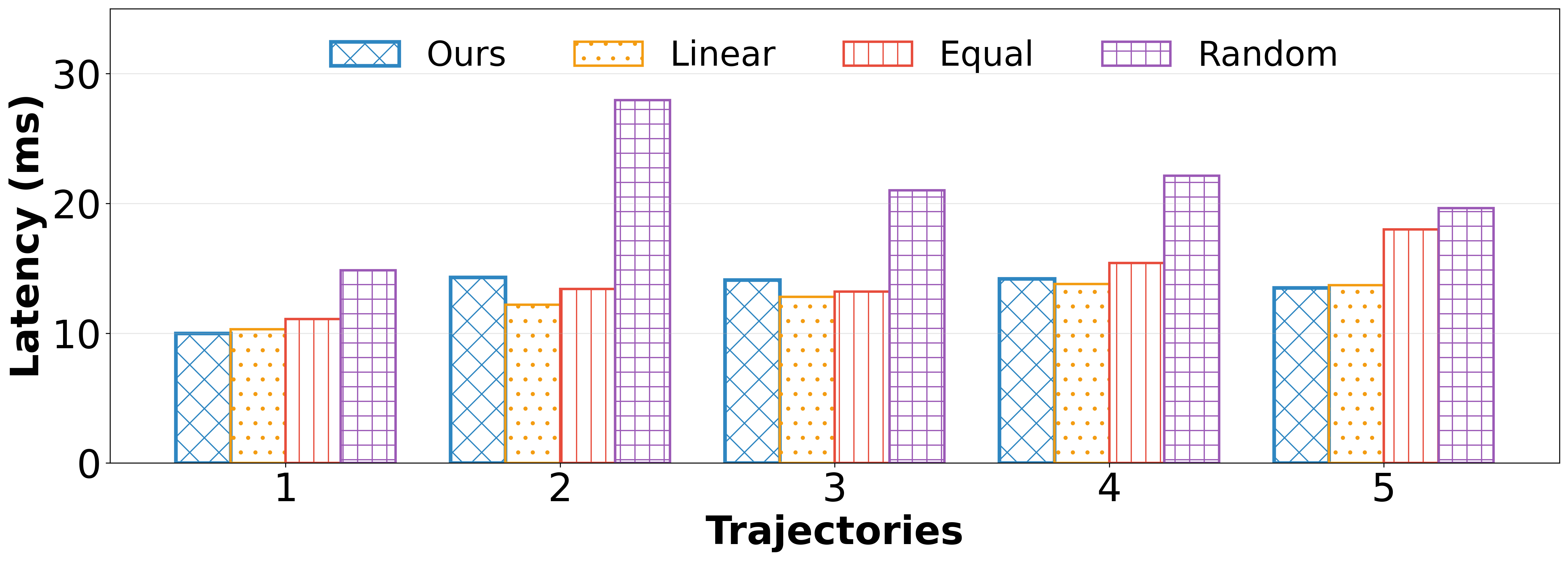}
    \caption{AR users latency}
    \label{fig_dr_protector} 
    \end{subfigure}
    \begin{subfigure}{.48\columnwidth}
        \centering
          \includegraphics[width=\columnwidth]{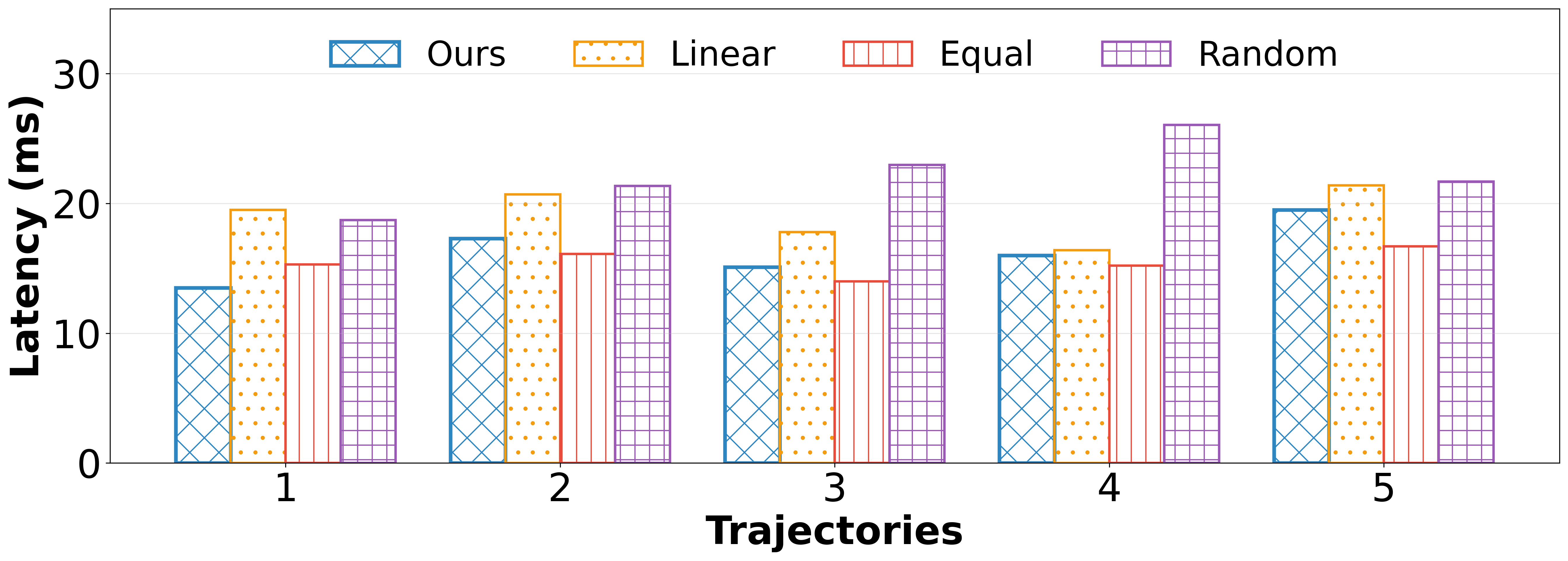}
        \caption{\editting{Robot latency}}
        \label{fig_rp_protector} 
    \end{subfigure}
    \begin{subfigure}{.48\columnwidth}
        \centering
          \includegraphics[width=\columnwidth]{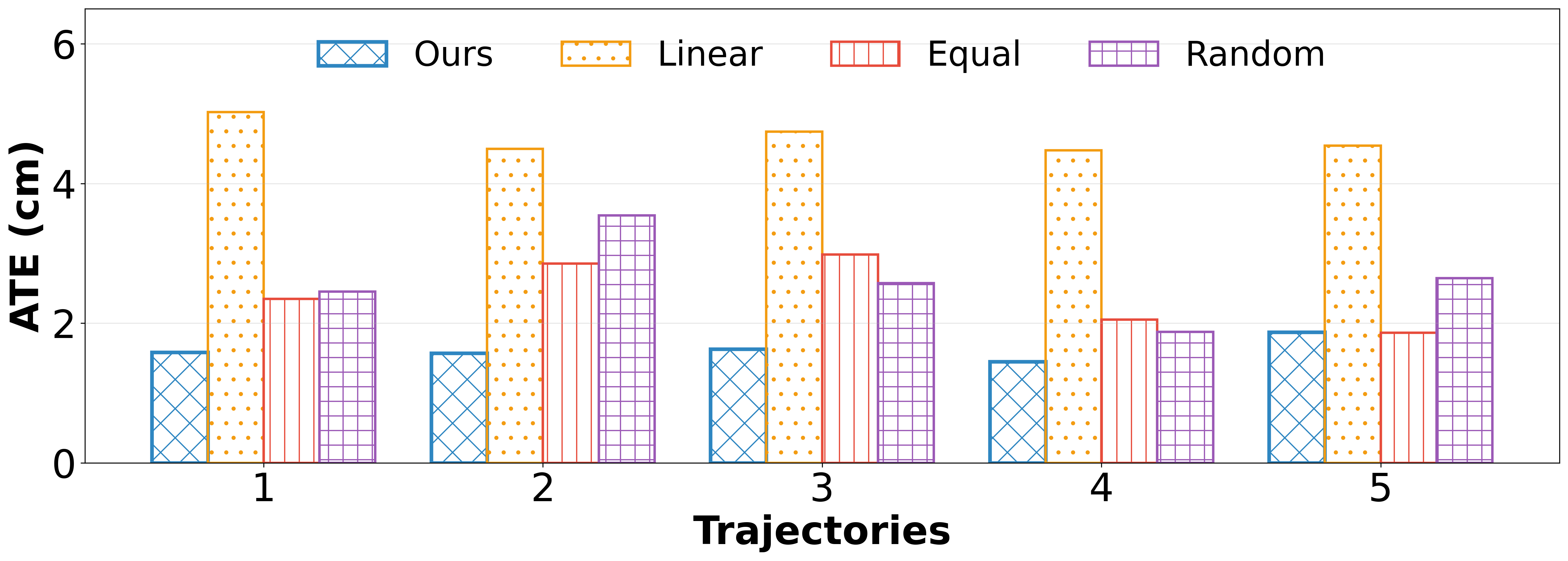}
        \caption{AR users accuracy}
        \label{fig_fp_protector} 
    \end{subfigure}
    \begin{subfigure}{.48\columnwidth}
        \centering
          \includegraphics[width=\columnwidth]{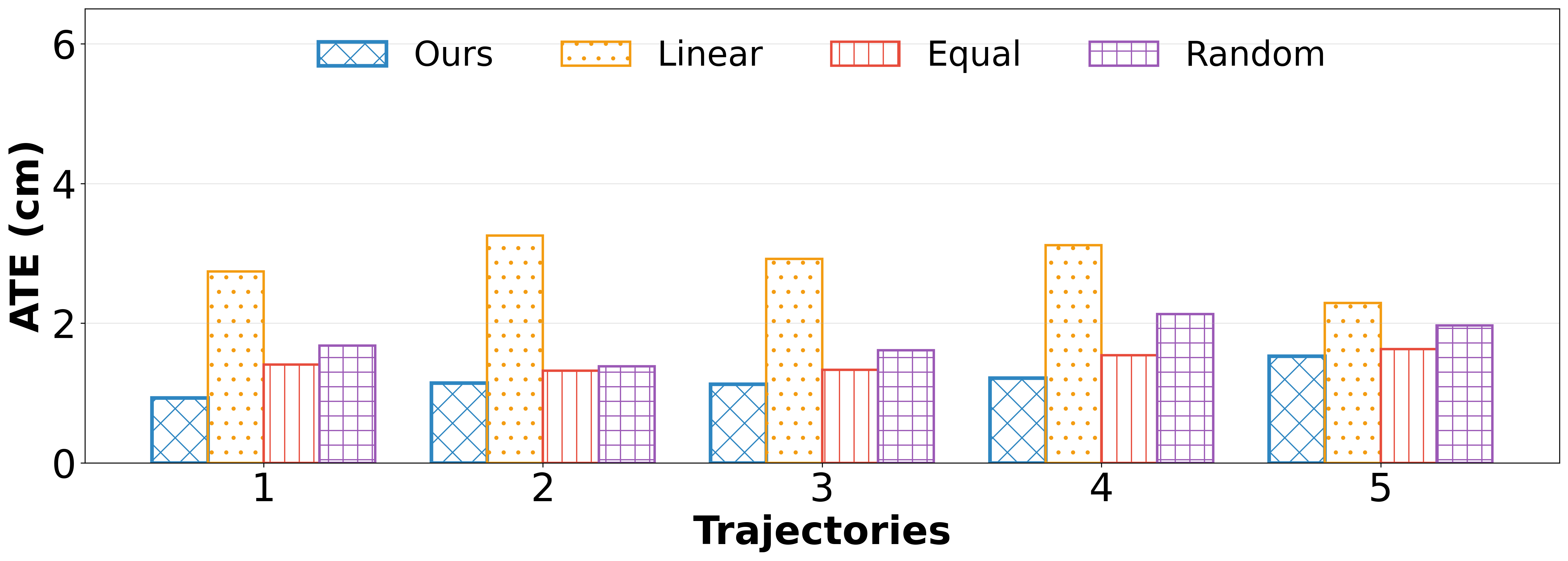}
        \caption{\editting{Robot accuracy}}
        \label{fig_rd_protector} 
    \end{subfigure}
    \vspace{-2.5mm}
    \caption{Ablation study results on field study trajectories. Linear is replacing PID scheduler with linear dropping, Equal is assigning equal QoE for heterogeneous agents. Random is setting non-tuned PID parameters. The lower the better.}
    \label{fig_diff_protector}
\end{figure}

\textbf{Impact of QoE.}
The results on both latency and accuracy comparisons are demonstrated in Figure~\ref{fig_diff_protector}. The Equal Baseline generates lower AR latency only on trajectories \textcircled{2} and \textcircled{3} and the advantage is marginal, but it gives higher latency for human agents on trajectories \textcircled{1}, \textcircled{4}, \textcircled{5}, especially on trajectory \textcircled{5} where it is 5 ms higher than \SysName. This is because trajectories \textcircled{2} and \textcircled{3} are shorter in length, holding less than 90 meters, where other trajectories are longer than 100 meters. On the accuracy perspective, converting to equal QoE results in increase for localization error on AR users in all cases, which is a result of higher investment in robot's performance. In addition, it is observed that aligning all agents' QoE models with the same as AR experience's QoE results in the system having limited care on accuracy but aggressively looking to drop latency. This results in decreased accuracy for all agents. The results provide the insight that when trajectories are short, QoE differences are addressed to a limited extent. Results suggest that the threshold for the scheduler to reflect the QoE differences and utilize it for distinguishing different agents is between 90 m to 100 m of total operation distance, but this distance is correlated to multiple factors, and the observation may only apply to the specific experiment setup. \editting{Concerning the performance of the robot, the trade-offs across baselines reveal distinct optimization behaviors. Equal Baseline achieves lower robot latency than \SysName in 4 out of 5 trajectories, consistent with its undifferentiated QoE that allocates communication resources without preference for AR users. However, this latency advantage comes at a consistent accuracy cost, where robot ATE under Equal Baseline is substantially higher than \SysName across all trajectories, reflecting the same pattern on users. Linear Baseline presents the opposite failure mode as it incurs the highest robot latency among all adaptive methods while also producing the worst robot ATE, suggesting that latency-proportional frame dropping without agent-awareness simultaneously harms both metrics for the robot. \SysName achieves the lowest robot ATE in all trajectories while keeping robot latency within an acceptable range, demonstrating that differentiated QoE optimization does not sacrifice robot localization quality in exchange for AR latency gains. }

\textbf{Impact of PID.}
The results on both latency and accuracy comparisons are demonstrated in Figure \ref{fig_diff_protector}. The latency performance of the Linear Baseline is on par with our proposed method for AR agents. On the human side's trajectories, \textcircled{2}, \textcircled{3}, \textcircled{4} see the Linear Baseline demonstrating lower latency than the proposed method with advantage of less than 3 ms. On trajectories \textcircled{1} and \textcircled{5}, the two methods have equivalent latency. However, linear dropping results in significantly higher (ATE), which is infeasible for the system's overall performance. Compared to the baseline of linear frame dropping, PID based frame dropping delivers significantly lower ATE in both human and robot agents in all trajectories, as demonstrated in Figure \ref{fig_diff_protector}. In all experiment sets, converting PID scheduler to linear dropping causes accuracy degradation. This is because the linear dropping mechanism has no awareness of visual environment, and it has a higher chance of removing informative visual frames, causing tracking to be inaccurate. 
On the other hand, Random Baseline has worse performance than both our method and the Linear Baseline in latency, but has performance between the two in ATE. This shows that without informed tuning, PID scheduler could remain awareness of visual environment, but cannot effectively address users' latency concerns.

\textbf{Impact of map merging.}
The effect of map merging redundancy of our method compared to COVINS-G is demonstrated in Table \ref{tab:map_merge_comparison}. Compared to the baseline COVINS-G, we have removed an average of 43.72 percent of merges averaged across the 5 trajectories. The percentage of removal is also relevant to the length of trajectories, with longer trajectories experiencing more removal. These removals cast limited impact on accuracy, as suggested in Figure~\ref{fig:ate_full}, which validates that our removal of redundancy is effective and accurate.

\begin{table}[t]
\centering
\begin{minipage}[t]{0.49\columnwidth}
    \centering
    \caption{Map Merges per Meter}
    \vspace{-3mm}
    \label{tab:map_merge_comparison}
    \small
    \setlength{\tabcolsep}{3pt}
    \begin{tabular}{cccc}
    \toprule
    \textbf{Traj.} & \textbf{COVINS-G} & \textbf{\SysName} & \textbf{Improvement (\%)} \\
    \midrule
    \textcircled{1} & 0.13 & 0.06 & 57.18 \\
    \textcircled{2} & 0.14 & 0.09 & 33.28 \\
    \textcircled{3} & 0.10 & 0.07 & 28.51 \\
    \textcircled{4} & 0.09 & 0.05 & 45.42 \\
    \textcircled{5} & 0.09 & 0.04 & 55.59 \\
    \midrule
    \textbf{Avg.} & \textbf{0.11} & \textbf{0.06} & \textbf{43.72} \\
    \bottomrule
    \end{tabular}
\end{minipage}%
\hfill
\begin{minipage}[t]{0.49\columnwidth}
    \centering
    \caption{Error Predictor Comparison}
    \vspace{-3mm}
    \label{tab:ablation_dnn}
    \small
    \setlength{\tabcolsep}{3pt}
    \begin{tabular}{lcc}
    \toprule
    \textbf{Method} & \textbf{RMSE (cm)} & \textbf{Time (ms)} \\
    \midrule
    \textbf{\SysName DNN} & \textbf{0.43} & \textbf{8.2}  \\
    Random Forest & 0.88 & 63.2 \\
    Linear Regression & 9.57 & 2.2\\
    MLP & 0.85 & 14.4\\
    Contrastive CNN & 0.67 & 19.6\\
    \bottomrule
    \end{tabular}
\end{minipage}
\vspace{-2mm}
\end{table}

\textbf{Impact of DNN frame evaluator.}
To evaluate the effectiveness of our design choice of using the DNN frame evaluator, we implement four distinct approaches across the same dataset partition using consistent evaluation protocols for estimating SLAM error from visual information:
\emph{1) Linear Regression:} a linear regression model using only the 8 scalar features extracted from the SLAM system (number of keypoints, loop closures, tracking quality metrics, etc.). \emph{2) Random Forest:} Random forest using the same features as the neural network, combining 1280-dimensional MobileNetV2 image features, 16-dimensional PSD features, and 16-dimensional processed scalar features, resulting in a 1312-dimensional feature vector. The model employs 70 trees with default scikit-learn parameters, trained on the extracted feature representations. \emph{3) Multi-Layer Perceptron (MLP):} MLP architecture utilizing the same multi-modal feature extraction as our proposed method but replacing the final prediction layer with a deeper fully connected layer.
\emph{4) Contrastive-style CNN:} a similar architecture to our proposed method but focusing solely on feature extraction capabilities, using a simpler final prediction head compared to our proposed approach. Training is conducted for 20 epochs with optimization settings identical to our proposed approach. The DNN evaluation is performed on two metrics: \emph{Root Mean Squared Error (RMSE)} measures the accuracy of models' estimation, and \emph{Process Time} measures the time elapsed for a model to predict one input.

Table~\ref{tab:ablation_dnn} presents the quantitative comparison of all evaluated approaches across accuracy and efficiency metrics. Our proposed method achieves the lowest RMSE of 0.43 cm, representing a 35\% improvement over the next-best performing approach (Contrastive-style CNN at 0.67 cm). While Linear Regression offers the fastest inference time (2.2 ms), its prediction accuracy is poor (9.57 cm RMSE), confirming that multi-modal information integration is essential for accurate SLAM error prediction. The Random Forest and MLP approaches achieve similar accuracy levels (0.88 cm and 0.85 cm respectively), but both underperform compared to our proposed architecture. The Random Forest approach suffers from slow inference (63.2 ms), making it unsuitable for real-time applications despite acceptable accuracy. In comparison, our proposed method achieves the second-fastest processing time (8.2 ms) while maintaining high accuracy.

\begin{figure}[t]
    \begin{subfigure}{.2\columnwidth}
       \includegraphics[width=\columnwidth]{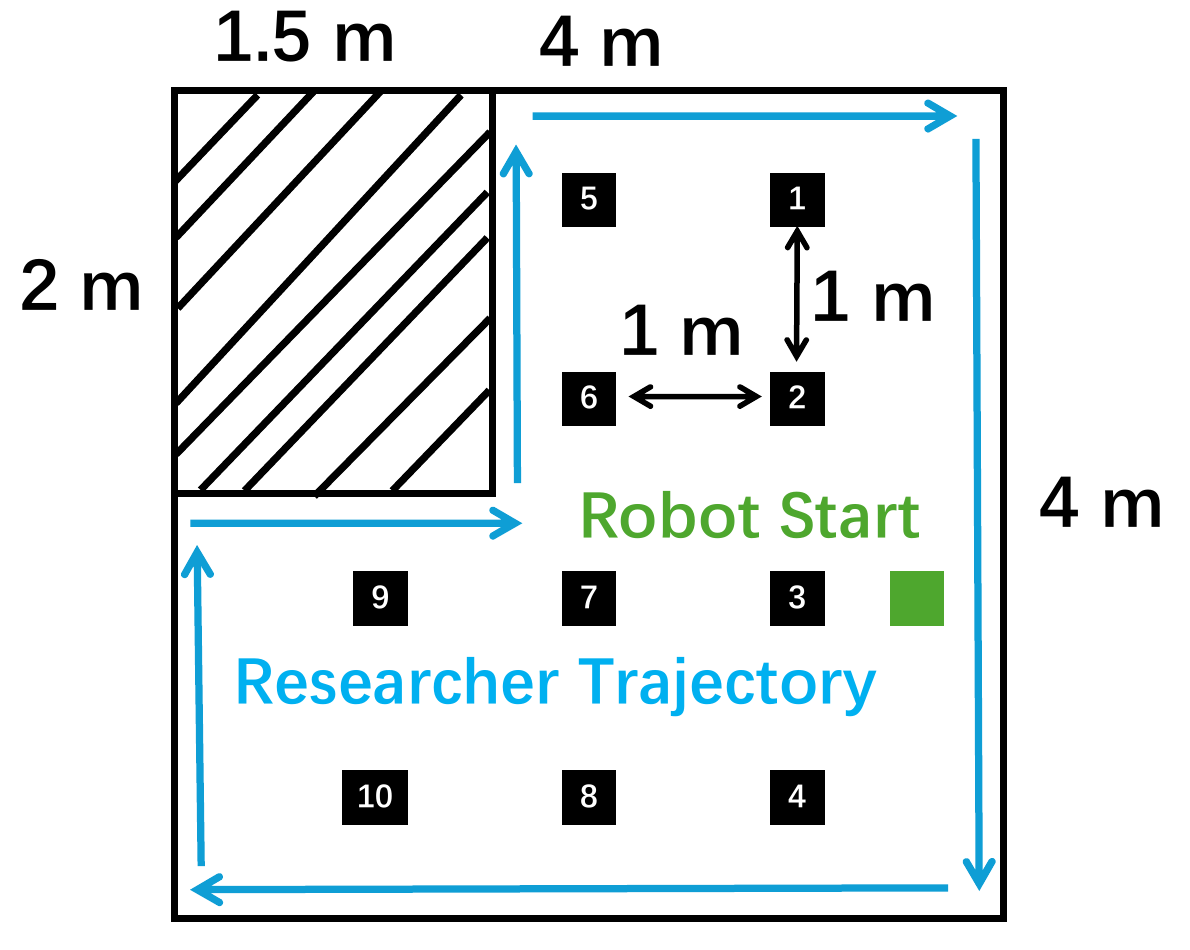}
    \caption{Experiment environment layout.}
    \label{fig:user-study-setup} 
    \end{subfigure}
    \hfill
    \begin{subfigure}{.2\columnwidth}
          \includegraphics[width=\columnwidth]{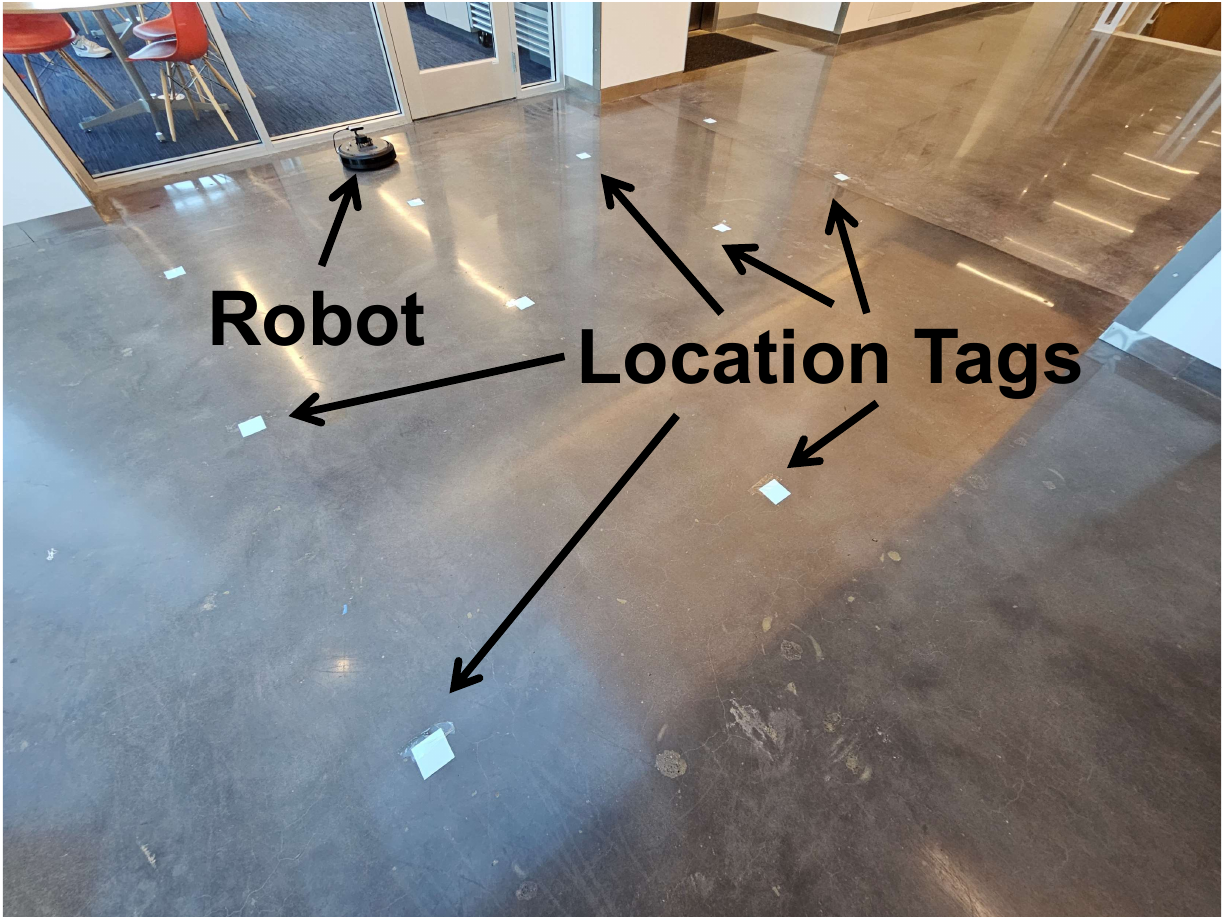}
        \caption{User study real environment.}
        \label{fig:user-study-picture} 
    \end{subfigure}
    \hfill
        \begin{subfigure}{.5\columnwidth}
        \includegraphics[width=\columnwidth]{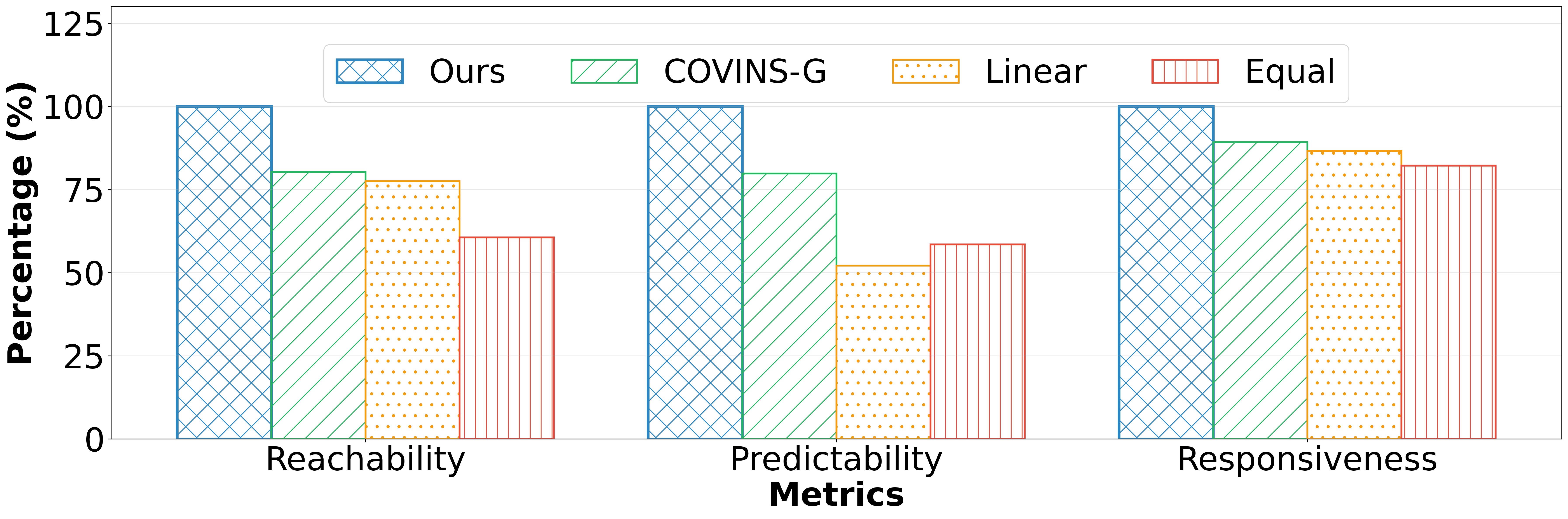}
        \caption{\editting{Results of the user study, normalized to percentages for comparison. Higher is better.}}
        \label{fig:user-study-result}
        \end{subfigure}
    
    \vspace{-3mm}
    \caption{User study setup.}
    \label{fig:user-study}
    \vspace{-3mm}
\end{figure}

\subsection{User Study}

To evaluate the effectiveness of \SysName in the context of human experience, \editting{we recruit 20 participants (7 female, 11 male, and 2 non-binary, aged between 18 and 51) through our university mailing lists, social networks, and advertising boards in our department building. Among the participants, 10 reported high familiarity with AR, 3 reported low familiarity, and 7 reported no prior experience. For robot familiarity, 4 participants reported high familiarity, 2 reported low familiarity, and 14 reported no prior experience.} The user study is approved by an Institutional Review Board (IRB).

\textbf{Setup.}
The user study is conducted in an open space contained in a 4-meter by 4-meter square area, as demonstrated in Figure \ref{fig:user-study}. The researcher and a participant act as two AR user agents working in this area. The participant wears the composed Meta Quest 3 headset (cf. Section \ref{sec_implementation}) and uses our custom-developed application to control the robot's movement. The researcher traverses a fixed trajectory around the user study area at a steady speed of 0.5 meters per second, demonstrated as "Researcher Trajectory" in Figure \ref{fig:user-study}.

\textbf{Task design.} The full study process contains four trials. For each trial, one of four solutions runs: Ours, COVINS-G (cf. Section \ref{sec_performance_field}), Linear Baseline, and Equal Baseline (cf. Section \ref{sec_ablation}). The ordering of the four solutions is randomly assigned and differs between users. Both the researcher and the participant are unaware of the underlying algorithm for each trial. During each trial of the study, the participant is asked to traverse the 10 designated spots in Figure \ref{fig:user-study} in a random order. Once the participant approaches one of the spots, they are asked to stop and press the controller. A virtual cube is then generated at the spot to signal that a command has been sent to the robot, and the robot approaches the user. Once the robot stops moving, the human user moves to another spot, until all ten spots are covered.

\textbf{Results.}
After each trial, the participant is asked the following two questions on a \textit{10-point Likert scale (1 for not at all; 10 for absolutely)}:
\begin{itemize}
    \item[-] Q1: \emph{How predictable is the movement of the robot?}
    \item[-] Q2: \emph{How responsive is the robot to your command?}
\end{itemize}

The users are instructed to rate predictability as 1 if the robot's movement is completely random, and 10 if the robot's movement completely aligns with their expectations; rate the responsiveness to 1 if the response of the robot is slow enough to cause irritation, and 10 if the response is instant and not noticeable. After all four trials are completed, the participant is asked the following questions:
\begin{itemize}
    \item[-] Q3: \emph{Among all four trials, which one do you prefer most?}
    \item[-] Q4: \emph{Among all four trials, which one do you prefer least?}
\end{itemize}

Based on objective measurements and the answers to the four questions, we conduct evaluation on the following metrics.

\emph{Reachability.} An objective metric regarding the task completion rate of the robot, where it describes how well the robot performs in the task of reaching a designated location. For each command of location reaching, it either succeeds or fails, and \emph{Reachability} describes the rate of success. \editting{Over the course of the entire experiment, our method achieves the highest task completion rate among all conditions, as demonstrated in Figure~\ref{fig:user-study-result}. \SysName achieves relative improvements of 24.6\%, 29.1\%, and 65.1\% compared to COVINS-G, Linear Baseline, and Equal Baseline. This result validates that our method delivers the best applicability with the highest completion rate.}

\emph{Predictability.} A subjective metric measuring perception of the movement trajectory of the robot. This metric describes how well the robot's movement aligns with the user's anticipation. \editting{As demonstrated in Figure~\ref{fig:user-study-result}, COVINS-G, Linear Baseline, and Equal Baseline achieve 79.79\%, 52.13\%, and 58.51\% of our predictability score, respectively, corresponding to relative improvements of 25.3\%, 91.8\%, and 70.9\% in favor of \SysName.} \editting{Notably, Equal Baseline ranks above Linear Baseline on this metric, reversing the ordering observed in Reachability and Responsiveness. This inversion is consistent with the finding in Section~\ref{sec_ablation} that linear frame dropping, while reducing communication overhead, does so without regard for the informativeness of the discarded frames, thereby degrading tracking accuracy. The resulting spatial inconsistencies manifest perceptually as less predictable robot trajectories. Equal Baseline, by contrast, maintains more uniform tracking quality across agents, which translates to more coherent robot motion as perceived by users, even though it does not optimize for latency. This result further demonstrates our proposed method's capability in addressing human users' spatial cognition in addition to the completion of assigned tasks.}

\emph{Responsiveness.} A subjective metric measuring perception of the time the robot takes to initiate a movement. This term describes how quickly the robot responds to the user's commands. \editting{As demonstrated in Figure~\ref{fig:user-study-result}, \SysName improved by 12.1\%, 15.4\%, and 21.7\% compared to the COVINS-G, Linear Baseline, and Equal Baseline. The comparatively smaller margin on this metric across all conditions reflects that responsiveness is partially attributable to hardware-level robot actuation delays shared across all software conditions. Nonetheless, \SysName consistently achieves the highest rating, demonstrating its advantage under low-latency operation.}

\emph{Preference.} \editting{A summary of preferences from all 20 participants. 13 participants (65\%) identify our method as most preferred, while no participant selects it as least preferred, yielding a net preference score of $+13$. COVINS-G receives 4 most-preferred and 5 least-preferred selections (net $-1$). Equal Baseline receives 3 most-preferred and 6 least-preferred selections (net $-3$). Linear Baseline receives no most-preferred selections and 9 least-preferred selections (net $-9$). The consistent ranking across all preference indicators confirms that \SysName is the most preferred solution, and its absence from any participant's least-preferred selections indicate broad acceptability across the user population.}

\editting{
\textbf{Statistical Analysis.}
To assess whether the observed differences across methods are statistically reliable, we conduct two complementary hypothesis tests on each of the three metrics. Because the same participants complete all four conditions, a within-subjects design is employed throughout. The \textit{Friedman test}~\cite{Friedman1937} is selected as the primary omnibus test as it is appropriate for the ordinal structure of Likert-scale responses. The \textit{repeated-measures ANOVA} is applied as a parametric complement, providing higher statistical power when distributional assumptions are approximately satisfied, and its convergence with the Friedman results strengthens the overall conclusions on all three metrics.}

\editting{
The Friedman test yields significant omnibus effects for all three metrics: Predictability ($\chi^2 = 19.62$, $p < 0.005$), Responsiveness ($\chi^2 = 13.24$, $p < 0.05$), and Reachability ($\chi^2 = 24.44$, $p < 0.005$). Repeated-measures ANOVA corroborates these findings: Predictability ($F(3, 57) = 8.76$, $p < 0.005$), Responsiveness ($F(3, 57) = 3.06$, $p < 0.05$), and Reachability ($F(3, 57) = 13.05$, $p < 0.005$). The agreement between the non-parametric and parametric tests across all three metrics indicates that the observed performance ordering is robust to the choice of statistical assumption. Reachability exhibits the strongest overall effect, consistent with its status as an objective binary outcome measure, while Responsiveness shows a smaller but still significant effect, reflecting the partial contribution of hardware actuation delays shared across all conditions. Together, these results confirm that the performance differences reported in Figure~\ref{fig:user-study-result} reflect genuine systematic differences attributable to the choice of method, with \SysName yielding the best performance across all metrics.}

\section{Discussion}

\editting{The evaluation results collectively demonstrate that \SysName achieves consistent latency and accuracy advantages over all baselines across diverse environments and agent configurations, with gains most pronounced under the operationally realistic conditions of longer trajectories, feature-sparse scenes, and sustained multi-agent interaction. The user study corroborates these objective findings at the perceptual level, with statistically significant improvements in predictability, responsiveness, and reachability across 20 participants. While the participant pool was recruited from a single institution and may not fully capture the demographic diversity of target deployment contexts, the effect sizes and statistical consistency across both parametric and non-parametric tests suggest that the observed advantages reflect genuine system-level differences rather than population-specific responses. These results motivate a closer examination of the design trade-offs underlying \SysName and the conditions under which its assumptions hold, which the remainder of this section addresses.}

\editting{\textbf{Robot latency.}
A deliberate consequence of the QoE prioritization strategy is an average increase in robot latency of 2.1 to 4.6 ms, as shown in Figure~\ref{fig:latency_full}. The latency remains within the tolerance for human--robot interaction tasks since tolerance is comparatively higher in domestic and logistic robot assistance scenarios that we target, as reported by prior works~\cite{shiwacomm,yangdelay,humanrobodelay}. Furthermore, certain deployment contexts may impose much stricter latency requirements on the robot itself. For high-speed industrial robots where latency degradation is intolerable, the QoE formulation should be inverted to treat the robot as the latency-critical agent. For multi-purpose robots, the QoE model could vary per operational mode. In the case of tightly coupled closed-loop control, a promising direction is to dynamically partition the AR and robot control loops based on their instantaneous QoE states. Given that these robots are not widely available, we leave such framework formalization and validation for future research.}

\editting{
\textbf{PID Transferability.} 
The parameters of the PID scheduler are currently tuned for the environments evaluated, since our application scenarios are focused on domestic and logistics assistance that only require fine-tuning once per environment. \SysName can be extended to settings with substantially different visual or communication characteristics that may benefit from adaptive gain scheduling~\cite{LeithLeithead2000} or reinforcement learning-based parameter tuning~\cite{Lakhani2022stability}. These advancements will further facilitate \SysName applications across a wide range of environments.}

\editting{\textbf{Practical Deployment.}
Currently, we mount external cameras on the Meta Quest 3 to obtain visual captures and IMU data, since privacy considerations on commercial headsets restrict direct access to integrated cameras and IMU data. The external camera configuration is consistent with established practice in AR tracking research~\cite{outcamera1, outcamera2}, and the compute requirements of \SysName are met by modestly provisioned hardware. Looking forward, \SysName's core mechanisms are designed to be decoupled from the specific sensor configuration, which reduces migration effort as open sensor access becomes available across headsets including Magic Leap~2~\cite{MagicLeapSensors2025} and Samsung Galaxy XR~\cite{SamsungGalaxyXR2025}. Together, these factors suggest that \SysName is well-positioned to support collaborative AR applications including AR-guided assembly, domestic cleaning assistance, and first-response coordination as device ecosystems mature.}

\editting{\textbf{Large-scale environment.}
The present work targets small-scale indoor environments, and no outdoor evaluation has been performed. Some AR-HRC applications, such as warehouse coordination and search-and-rescue support, may require operation across larger and more open environments. Potential solutions include adjusting overlap threshold to account for sparser co-visibility in larger spaces, retraining error estimation model to account for more diverse lighting and motion, and adjusting buffer window size to reflect the lower frequency of agent viewpoint overlap in a larger environment. We will explore them in our future work.}

\section{Conclusion}
This paper presented \SysName, a system that enables multi-agent SLAM to address critical latency challenges in AR-powered human-robot collaboration. We build a first-of-its-kind QoE model for HRC agents to quantify the impact of various factors introduced by a SLAM-based tracking service. Building upon it, \SysName employs a PID scheduler to adaptively adjust transmission priorities and meet diverse QoE requirements across HRC agents. We further reduce edge computation by removing visual redundancy in shared human-robot workspaces. Through extensive real-world evaluations and user studies using commercial AR headsets and robots, \SysName achieved latency well below the critical 20 ms threshold while maintaining good tracking performance for both users and robot collaborators. Its insights extend beyond SLAM: any edge-based multi-agent system serving heterogeneous clients can benefit from quality-of-experience differentiation and context-aware data fusion.
\begin{acks}

This work was supported in part by NSF grants CSR-2312760, CNS-2112562, and IIS-2231975, NSF CAREER Award IIS-2046072, NSF NAIAD Award 2332744, a CISCO Research Award, a Meta Research
Award, Defense Advanced Research Projects Agency Young Faculty Award HR0011-24-1-0001, and the Army Research Laboratory under Cooperative Agreement Number W911NF-23-2-0224. The views and conclusions contained in this document are those of the authors and should not be interpreted as representing the official policies, either expressed or implied, of the Defense Advanced Research Projects Agency, the Army Research Laboratory, or the U.S. Government. This paper has been approved for public release; distribution is unlimited. No official endorsement
should be inferred. The U.S. Government is authorized to reproduce and distribute reprints for Government purposes notwithstanding any copyright notation herein. The authors would also like to gratefully acknowledge the assistance of Dhun Pandya from North Carolina School of Science and Mathematics on this work.

\end{acks}

\bibliographystyle{ACM-Reference-Format}
\bibliography{reference}

\end{document}